\documentclass[a4paper,11pt]{article}
\pdfoutput=1 

\usepackage{jheppub} 
               
\usepackage{besphysics}
\usepackage[T1]{fontenc} 
\usepackage{siunitx}
\usepackage{rotating}
\usepackage{booktabs}
\usepackage[capitalize]{cleveref}
\usepackage{multirow}
\usepackage{overpic}
\usepackage{lineno}
\usepackage{cancel}

\title{\boldmath Measurement of the $D \to \kmthreepi$ and $D \to \kmpipio$ coherence factors and average strong-phase differences in quantum-correlated \\ ${D\bar{D}}$ decays}

\collaboration{BESIII Collaboration}

\author{
\begin{small}
\begin{center}
\normalfont{
M.~Ablikim$^{1}$, M.~N.~Achasov$^{10,c}$, P.~Adlarson$^{67}$, S. ~Ahmed$^{15}$, M.~Albrecht$^{4}$, R.~Aliberti$^{28}$, A.~Amoroso$^{66A,66C}$, M.~R.~An$^{32}$, Q.~An$^{63,49}$, X.~H.~Bai$^{57}$
% ...
}
\\
\vspace{0.2cm}
(BESIII Collaboration)\\
\vspace{0.2cm} {\it
$^{1}$ Institute of High Energy Physics, Beijing 100049, People's Republic of China\\
% ...
$^{4}$ Bochum Ruhr-University, D-44780 Bochum, Germany\\
% ...
$^{10}$ G.I. Budker Institute of Nuclear Physics SB RAS (BINP), Novosibirsk 630090, Russia\\
% ...
$^{15}$ Helmholtz Institute Mainz, Johann-Joachim-Becher-Weg 45, D-55099 Mainz, Germany\\
% ...
$^{28}$ Johannes Gutenberg University of Mainz, Johann-Joachim-Becher-Weg 45, D-55099 Mainz, Germany\\
% ...
$^{32}$ Liaoning Normal University, Dalian 116029, People's Republic of China\\
% ...
$^{49}$ State Key Laboratory of Particle Detection and Electronics, Beijing 100049, Hefei 230026, People's Republic of China\\
% ...
$^{57}$ University of Jinan, Jinan 250022, People's Republic of China\\
% ...
$^{63}$ University of Science and Technology of China, Hefei 230026, People's Republic of China\\
% ...
$^{66}$ University of Turin and INFN, (A)University of Turin, I-10125, Turin, Italy; (B)University of Eastern Piedmont, I-15121, Alessandria, Italy; (C)INFN, I-10125, Turin, Italy\\
$^{67}$ Uppsala University, Box 516, SE-75120 Uppsala, Sweden\\
% ...
$^{c}$ Also at the Novosibirsk State University, Novosibirsk, 630090, Russia\\
% ...
}\end{center}

\vspace{0.4cm}
\end{small}}

\abstract{The decays $D\to \kmthreepi$ and $D \to \kmpipio$ are studied in a sample of quantum-correlated $D\bar{D}$ pairs produced through the process $e^+e^- \to \psi(3770) \to D\bar{D}$, exploiting a data set collected by the BESIII experiment that corresponds to an integrated luminosity of 2.93\,fb$^{-1}$.  Here $D$ indicates a quantum superposition of a $D^0$ and a $\bar{D}^0$ meson.  By reconstructing one neutral charm meson in a signal decay, and the other in the same or a different final state, observables are measured that contain information on the coherence factors and average strong-phase differences of each of the signal modes.  These parameters are critical inputs in the measurement of the angle $\gamma$ of the Unitarity Triangle in $B^- \to DK^-$ decays at the LHCb and Belle II experiments.  The coherence factors are determined to be $R_{K3\pi}=0.52^{+0.12}_{-0.10}$ and $R_{K\pi\pi^0}=0.78 \pm 0.04$, with values for the average strong-phase differences that are $\delta_D^{K3\pi}=\left(167^{+31}_{-19}\right)^\circ$ and  $\delta_D^{K\pi\pi^0}=\left(196^{+14}_{-15}\right)^\circ$, where the uncertainties include both statistical and systematic contributions.  The analysis is re-performed in four bins of the phase-space of the $D \to \kmthreepi$ to yield results that will allow for a more sensitive measurement of $\gamma$ with this mode, to which the BESIII inputs will contribute an uncertainty of around 6$^\circ$.}

\begin{document} 
\maketitle
\flushbottom

\section{Introduction}
\label{sec:intro}

In the Standard Model, $C\!P$ violation is described by the irreducible complex phase of the Cabibbo-Kobayashi-Maskawa (CKM) quark-mixing matrix~\cite{Cabibbo,Kobayashi:1973fv}.  This matrix can be represented geometrically by the Unitarity Triangle in the complex plane, with angles $\alpha$, $\beta$ and $\gamma$ (also denoted $\phi_2$, $\phi_1$ and $\phi_3$). The angle $\gamma$, equal to $\arg(-V_{us}{V_{ub}}^*/V_{cs}{V_{cb}}^*)$ at ${\cal{O}}(\lambda^4)\sim 10^{-3}$, is of particular interest, as it can be measured with negligible theoretical uncertainty in decays of the class $B^- \to D K^-$, where $D$ represents a superposition of $\Dz$ and $\bar{D}^{0}$ mesons reconstructed in a final state common to both.  The $\gamma$ sensitivity arises from the phase difference between the $b\to u$ and $b \to c$ amplitudes, which manifests itself in an interference term in the partial width.
These decays are mediated by tree-level amplitudes and thus their rates are not expected to deviate from the Standard Model prediction. 
In contrast, other observables that can be used to infer $\gamma$ involve loop-level amplitudes, which are sensitive to non-Standard-Model processes.  Comparison between the two sets of measurements therefore provides a powerful method to probe for new physics beyond the Standard Model.

An important category of charm decays with which to perform $B^- \to DK^-$ measurements is $D \to K^- n \pi$, where $n \pi$ signifies $n\ge 1$ pions~\cite{ADS0,ADS}\footnote{Throughout the discussion charge conjugation is implicit, but it should be noted that both charge configurations $B^- \to (K^- n\pi)_D K^-$ and $B^- \to (K^+ n\pi)_D K^-$ may be exploited in the measurement of $\gamma$.}.
Here the final state can be accessed either by the Cabibbo-favoured (CF) amplitudes of a $D^0$ decay, or the doubly Cabibbo-suppressed (DCS) amplitudes of a $\bar{D}^{0}$ decay, or either of these amplitudes in conjunction with those responsible for $D^0$-$\bar{D}^{0}$ oscillations.
In the case of final states where $n\ge 2$, such as $D \to \kmthreepi$ or $D \to \kmpipio$, the interference term involving $\gamma$ is also dependent on the variation of the $D^0$ and $\bar{D}^{0}$ decay amplitudes over the multi-body phase space.  Inclusive measurements that integrate over this phase space can be interpreted in terms of $\gamma$ provided that three charm hadronic parameters are known.  Generalising for the decay $D \to S$, where $S$ denotes a $K^-n\pi$ final state, these parameters are the coherence factor $R_S$, and the amplitude ratio  $r_D^{S}$ and  $C\!P$-conserving  strong-phase difference  $\delta_D^{S}$ between the CF and DCS amplitudes averaged over phase space~\cite{Atwood:2003mj}:
 \begin{equation}
 R_{S}e^{-i\delta_{D}^{S}}=\frac{\int {\cal A}^{\star}_{S}(\textbf{x}){\cal  A}_{\bar{S}}(\textbf{x})\rm d\textbf{x}}{A_{S}A_{\bar{S}}} {\;\;\;{\rm and}\;\;\;}
  r^{S}_{D} =A_{\bar{S}}/A_{S}.
  \label{eq:rds}
 \end{equation}
Here ${\cal A}_{S}(\textbf{x})$ is the decay amplitude of $\Dz \to S$ at a point in multi-body phase space described by parameters $\textbf{x}$ and $A^2_{S}=\int |{\cal A}_{S}(\textbf{x})|^2{\rm d}\textbf{x}$, with equivalent definitions for the $D^0 \to \bar{S}$ amplitude.  (The amplitudes are normalised such that, in the absence of oscillations, $A^2_{S}$ corresponds to the branching ratio of the $\Dz$ meson into mode $S$.)
Maximum (zero) interference in the term sensitive to $\gamma$ occurs in the limiting case $R_{S} = 1$ $(0)$. The parameter $r_D^{S}$ is of the order $\lambda^2 \approx 0.06$ for $D\to K^-n\pi$ decays.

Strong constraints exist on the values of $r_D^S$ from measurements of branching fractions.
The coherence factors and strong-phase differences may be determined through decays of quantum-correlated $D\bar{D}$ pairs produced in $e^+e^-$ collisions at the $\psi(3770)$ resonance. The decay rate of a $D$ meson into the final state of interest is modified from the uncorrelated expectation if the other meson is observed to decay into a $C\!P$ eigenstate, or indeed any final state that is not flavour specific.  These modifications are dependent on the hadronic parameters, and so observables measured in $\psi(3770)$ decays can be used to determine $R_D$ and $\delta_D^S$. 
This strategy has been pursued on a data set corresponding to an integrated luminosity of $0.81~{\rm fb^{-1}}$ collected by the CLEO-c detector~\cite{Lowery:2009id,Libby:2014rea,Evans:2016tlp}.  Complementary information on the hadronic parameters may be obtained from measurements of $D^0$-$\bar{D}^{0}$ oscillations above charm threshold~\cite{Harnew:2013wea}.  Here the coherence factor dampens the oscillation signal
with respect to that which occurs for the two-body mode $D \to K^-\pi^+$.  Such a measurement has been performed by the LHCb collaboration for $D \to \kmthreepi$~\cite{Aaij:2016rhq}, which, when combined with the results from the CLEO-c data set, yields $R_{K3\pi} = 0.43^{+0.17}_{-0.13}$, $\delta_D^{K3\pi}=\left(128^{+28}_{-17}\right)^\circ$,~\footnote{In this paper strong-phase differences and expressions are given in the convention $C\!P |D^0\rangle = |\bar{D}^{0}\rangle$.} and $r_D^{K3\pi} = (5.49 \pm 0.06) \times 10^{-2}$.  The equivalent parameters for $D \to \kmpipio$ are determined to be $R_{K\pi\pi^0} = 0.81 \pm 0.06$, $\delta_D^{K\pi\pi^0}= \left(198^{+14}_{-15}\right)^\circ$ and $r_D^{K\pi\pi^0} =  (4.47 \pm 0.12) \times 10^{-2}$~\cite{Evans:2016tlp}.

Measurements of $C\!P$ asymmetries and associated observables  in $B^- \to DK^-$ decays using these modes have been performed by LHCb,
with data collected during Run 1 of the LHC~\cite{Aaij:2015jna,Aaij:2016oso,Aaij:2019tmn}, and by Belle~\cite{Nayak:2013tgg}.  
The measurements are interpreted using the
reported values of the hadronic parameters to obtain constraints on the value of $\gamma$ and related parameters of the $b$-hadron decay. In these interpretations the uncertainties associated with the knowledge of the $D$ hadronic parameters are currently smaller than those associated with the finite size of the $b$-decay samples, but this will no longer be the case with larger Run 2 data set now under analysis at LHCb, and the data sets that both LHCb and Belle II expect to collect over the coming years~\cite{Malde:2223391,Bediaga:2018lhg,Kou:2018nap}.   Therefore, more precise measurements of the $D$ hadronic parameters are essential to enable improved knowledge of the  angle $\gamma$.  Such measurements, when obtained at charm threshold, will also be valuable for interpretation of $D^0$-$\bar{D}^{0}$ oscillation measurements performed at higher energies.

It is noteworthy that the coherence factor of $D\to \kmthreepi$ is low, which is an attribute that dilutes its sensitivity to $\gamma$ in a $B^- \to DK^-$ analysis.  In order to ameliorate this problem, it is advantageous to partition the phase space of the decay into regions of higher individual coherence, with well-separated values of average strong-phase difference, 
A binning scheme to achieve this goal has recently been proposed, based on input from $D^0 \to \kmthreepi$ and $D^0 \to \kpthreepi$ amplitude models constructed by LHCb~\cite{Aaij:2017kbo}.
Interpreting the $b$-decay properties within these bins requires corresponding binned information for the $D$ hadronic parameters. Initial measurements of these binned parameters has been performed using the CLEO-c data set, but these are rather imprecise~\cite{Evans:2019wza}.  Once more, improved measurements are desirable.

In this paper measurements are reported of the coherence factor and average strong-phase differences of $D \to \kmthreepi$ and $D \to \kmpipio$ decays, integrating over the phase space of both modes.
Results are also presented in bins of $D \to \kmthreepi$ phase space. The analysis exploits a data set equivalent to an integrated luminosity of $2.93\,{\rm fb^{-1}}$~\cite{Ablikim:2014gna,Ablikim:2015orh} delivered by the BEPCII collider and collected by the BESIII experiment in $e^+e^-$ collisions at a collision energy of 3773~MeV.  
The definition of the measured observables and their relationship to the underlying physics parameters is given in Sec.~\ref{sec:formalism}. The BESIII detector is described in Sec.~\ref{sec:detector}, together with information concerning the Monte Carlo simulation samples used. The event selection is presented in Sec.~\ref{sec:selection}.  The values of the measured observables, the fit to the hadronic parameters, and the assignment of systematic uncertainties are discussed in Sec.~\ref{sec:obs}. Section~\ref{sec:gamma}
assesses the impact of the results on measurements of $\gamma$ in $B^- \to DK^-$ decays, and a summary is given in Sec.~\ref{sec:summary}.
\section{Formalism and measurement strategy}
\label{sec:formalism}

Consider a $D\bar{D}$ pair produced in $e^+e^-$ collisions at the $\psi(3770)$ resonance, where one charm meson decays into the signal mode $S$, either $\kmthreepi$ or $\kmpipio$, and the other decays into a tagging mode, denoted $T$.  The two mesons are produced in a $C$-odd eigenstate and their decays are quantum correlated, with a rate given by
\begin{eqnarray}
\Gamma(S|T) & =  & 
\int\int{|\mathcal{A}_{S}(\mathbf{x})\mathcal{A}_{\bar{T}}(\mathbf{y})-\mathcal{A}_{\bar{S}} (\mathbf{x})\mathcal{A}_T(\mathbf {y})|^{2} {\rm d}\mathbf{x}{\rm d}\mathbf{y}} \; \nonumber \\
	   & = & [ A_{S}^{2}A_{\bar{T}}^{2}+A_{\bar{S}}^2 A_T^2 - \nonumber \\
	   &  & 2R_{S} R_T A_{S} A_{\bar{S}}A_{T}A_{\bar{T}}\cos{(\delta_D^{T}-\delta_D^{S})} ] \nonumber \\
	   & = & A_{S}^{2}A_{T}^{2}\,  [ (r_D^S)^{2} + (r_D^T)^2 -  2R_S R_T r_D^S r_D^T \cos{(\delta_D^{T}-\delta_D^{S})} ] \;. \label{eq:general_ST}
\end{eqnarray}
Note that in general the tagging mode may be a multi-body decay, and therefore has its own coherence factor $R_T$, average amplitude ratio $r_D^T$ and strong-phase difference $\delta_D^T$.  Throughout the discussion it is assumed that $C\!P$ violation can be neglected in the charm system.  In addition, Eq.~\ref{eq:general_ST} omits terms of order ${\cal{O}}(x^2,y^2)\sim 10^{-5}$  associated with $D^0$-$\bar{D}^{0}$ oscillations, where $x$ and $y$ are the usual mixing parameters~\cite{Xing:1996pn}. These are safe approximations at the current  level of experimental precision.

When both $S$ and $T$ are reconstructed the event is said to be {\it double-tagged}. Three separate classes of tag are employed in the analysis: $C\!P$ tags, like-sign tags and $\kspipi$ tags.  These are described for the global measurement, integrated over the full phase space of the signal decay, and then discussed for the case where the $D\to \kmthreepi$ phase space is binned. 

\subsection{$C\!P$ tags}

In the case when the tagging mode is a $C\!P$ eigenstate, then $r^T_D=1$, $R_T = 1$,  and $\delta^T_D =0$ or $\pi$, and Eq.~\ref{eq:general_ST} becomes
\begin{equation}
   \Gamma(S|C\!P) =  A_{S}^{2}A_{C\!P}^{2}\,  \left( 1 + (r_D^S)^{2}  -  2 \lambda R_S r_D^S  \cos{\delta_D^{S}} \right),
   \label{eq:cptag}
\end{equation}
with $C\!P$ indicating a specific mode with eigenvalue $\lambda$ ($\lambda = +1$ when $\delta_D^T=0$ and $-1$ when $\delta_D^T=\pi$).  Some self-conjugate decay modes are not pure $C\!P$ eigenstates, but are known to be predominantly $C\!P$ even or $C\!P$ odd.  Then Eq.~\ref{eq:cptag} continues to apply with the substitution $\lambda = (2F_+^T - 1)$, where $F_+^T$ is the $C\!P$-even fraction of the decay.  The most striking known example is $D\to\pipipio$, where $F_+^{\pi\pi\pi^o}$ is in excess of 95\%~\cite{Nayak:2014tea,Malde:2015mha}.

It is useful to define the observables $\rho^S_{C\!P\pm}$, which are the ratios of the number of $C\!P$-tagged signal events to the number expected in the absence of quantum-correlations:
\begin{equation}
\rho_{C\!P\pm}^{S} \equiv \frac{ N(S|C\!P)+N(\bar{S}|C\!P)} 
{2N_{D\bar{D}}\,\br{}\left( D^{0}\to C\!P \right) \left[\br{}\left(D^{0}\to S\right)+\br{}\left( D^{0}\to \bar{S}\right) \right]}. 
\label{eq:rho_cp}
\end{equation}
Here $N(S|C\!P)$ 
($N(\bar{S}|C\!P)$) is the number of decays of channel $S$ ($\bar{S}$) tagged with a $C\!P$ eigenstate after efficiency correction, $N_{D\bar{D}}$ is the produced number of neutral $D\bar{D}$ pairs in the sample and $\br{}(D^0 \to X)$ is the branching fraction of a $D^0$ meson into state $X$. The observable $\rho^S_{C\!P+}$ ($\rho^S_{C\!P-}$) corresponds to the case when the $C\!P$ tag is even (odd).

In relating the squared amplitudes to branching fractions it is necessary to include corrections arising from $D^0$-$\bar{D}^{0}$ oscillations.
Initially keeping terms to ${\cal{O}}\left(r_D^2, x^2,y^2 \right)$, the below relations apply:
\begin{eqnarray}
\br{}(D^0\to S) &=& A^2_S\left(1 - R_Sr^S_D(y\cos\delta^S_D + x\sin\delta^S_D) + (y^2-x^2)/2 \right), \label{eq:BRCF}\\
\br{}(D^0\to \bar{S}) &=& A^2_{{S}}\left((r_D^S)^2 - R_S r_D^S (y \cos \delta^S_D - x \sin\delta^S_D) + (x^2 + y^2)/2 \right), \label{eq:BRDCS}\\ 
\br{}(D^0 \to C\!P) &=& A^2_{C\!P}\left(1 - \lambda y + y^2 \right), \label{eq:BRCP}
\end{eqnarray}
where it is noted that both $x$ and $y$ are of magnitude $\sim r^S_D/10$~\cite{Amhis:2019ckw}. 
It thus follows that 
\begin{equation}
\begin{aligned}
\rho _ {C\!P\pm} ^ { S } = 
\frac{\left( 1 + (r_{D}^{S})^2 \mp 2 r_{D}^{S} R_{S} \cos \delta_{D}^{S}  \right)}{\left(1  \mp y + (r_D^S)^2 -2r_D^SR_Sy\cos\delta^S_D  \right)}.
%
\end{aligned}
\label{eq:rho_cp_the}
\end{equation}
This and subsequent expressions are now written to ${\cal{O}}\left(r_D^2, x,y \right)$.
In order to combine the information from $\rho^S_{C\!P+}$ and $\rho^S_{C\!P-}$ into a single, $C\!P$-invariant, parameter, it is convenient to define
\begin{equation}
\Delta_{C\!P}^S \equiv \pm\left( \rho^S_{C\!P\pm} - 1 \right),
\label{eq:deltacp}
\end{equation}
so that
\begin{equation}
\Delta_{C\!P}^S = y -2 r_D^S R_S \cos \delta_D^S.
\label{eq:deltacp_th}
\end{equation}

In practice, the precision of the measurement of $\rho_{C\!P\pm}^S$ with many $C\!P$ eigenstates is limited by the uncertainty on the branching fraction of the tagging mode.  This problem can be largely circumvented by re-expressing $\rho_{C\!P}^S$ in a manner which involves $N(K^\mp\pi^\pm|C\!P)$, the efficiency-corrected yield of $D\to K^\mp\pi^\pm$ decays tagged by the same $C\!P$ eigenstate:
\begin{equation}
\rho_{C\!P\pm}^S = \frac{N(S|C\!P)+N(\bar{S}|C\!P)}{N(K^-\pi^+|C\!P)+N(K^+\pi^-|C\!P)}\cdot
\frac{\br{}(D^0\to K^-\pi^+) + \br{}(D^0 \to K^+\pi^-)}{\br{}(D^0\to S) + \br{}(D^0 \to \bar{S})}\cdot
\rho_{C\!P\pm}^{K\pi}.
\label{eq:kpinorm}
\end{equation}
The values of $\rho_{C\!P\pm}^{K\pi}$ can be calculated  from Eq.~\ref{eq:rho_cp_the} with a precision of $0.5\%$.

\subsection{Like-sign  tags}

Consider the case where the tagging mode is also a flavoured final state of the category $K^- n \pi$, for example $K^-\pi^+$, $\kmthreepi$ or $\kmpipio$, and the charge of the kaon in the tagging mode is the same as that of the signal.    

In the case where the signal and tagging mode are identical Eq.~\ref{eq:general_ST} reduces to
\begin{equation}
   \Gamma(S|S) =  A_{S}^{2}A_{\bar{S}}^{2}\,  [ 1  -   R_S^2 ].
   \label{eq:lstag}
\end{equation}
Again, an observable is constructed that expresses the ratio of the number of efficiency-corrected double-tagged events, $N(S|S)$ and $N(\bar{S}|\bar{S})$, to the expectation without quantum correlations:
\begin{equation}
\rho^S_{LS}  \equiv  \frac{N(S|S)+N(\bar{S}|\bar{S})}{2N_{D\bar{D}}\,\br{}( D^{0}\to S)\br{}( D^{0}\to \bar{S})}. \label{eq:rho_ls} 
\end{equation}
Since quantum-correlation effects are negligible in opposite-sign double tags, it is experimentally advantageous to determine this observable through the equivalent expression:
\begin{equation}
\rho^S_{LS} = \frac{N(S|S)+N(\bar{S}|\bar{S})}{2N(S|\bar{S})\left(\br{}(D^0 \to \bar{S})/\br{}(D^0 \to S)\right)}. \label{eq:rho_ls2}
\end{equation}
Consideration of Eq.~\ref{eq:general_ST}, Eq.~\ref{eq:lstag} and the relationship between the squared amplitudes and branching ratios leads to 
\begin{equation}
\rho _ { L S } ^ { S } =
\frac{ \left( 1 -  R _ { S } \right) ^ { 2 } }{ 1  - R_{S} \left((y/r_D^S)\cos\delta_{D}^{S}-(x/r_D^S)\sin\delta_{D}^{S}\right) + (x^2 + y^2)/(2[r_D^S]^2)},
\label{eq:rho_ls2_th}
\end{equation}
from which it may be seen  that this observable has high sensitivity to the coherence factor.

When the tag is the two-body mode $D \to K^-\pi^+$ then $R_T = 1$ and the other parameters $r_D^{K\pi}$ and $\delta_D^{K\pi}$ are well known from measurements of $D^0$-$\bar{D}^{0}$ oscillations. The case $S=\kmthreepi$ and $T=\kmpipio$ carries simultaneous information on both multi-body signal modes of interest.  The quantum-correlation effects in these double tag can be studied through the observables
\begin{equation}
\rho^S_{T,LS} \equiv
\frac{ N(S|T)+N(\bar{S}|\bar{T})} 
{2N_{D\bar{D}}\,\left( \br{}( D^{0}\to S) \br{}(D^{0}\to \bar{T})+
\br{}( D^{0}\to \bar{S}) \br{}(D^{0}\to {T}) \right)},
\label{eq:rho_lsx}
\end{equation}
where $T \ne S$. Again, it is beneficial to take advantage of the yields of opposite-sign double tags and to evaluate the equivalent expression
\begin{equation}
\rho^S_{T,LS} =
\frac{ N(S|T)+N(\bar{S}|\bar{T})} 
{\left(N(S|\bar{T})+N(\bar{S}|T)\right)\left(
\frac{\br{}(D^0 \to \bar{T})}{\br{}(D^0 \to T)} +\frac{\br{}(D^0 \to \bar{S})}{\br{}(D^0 \to S)}\right)}.
\label{eq:rho_lsx2}
\end{equation}
It follows from Eq.~\ref{eq:general_ST}, Eq.~\ref{eq:rho_lsx} and the branching-ratio relations that
\begin{eqnarray}
\rho_{T, LS}^{S} &=& \left( 1 + (r_{D}^{S}/r_D^T)^2 - 2 (r_{D}^ {S}/r_{D}^{T}) R_{S}R_T \cos ( \delta_{D}^{T} - \delta_{D}^{S} )\right) / \nonumber \\
& & \left( 1 + (r_D^S/r_D^T)^2 - R_T( [y/r_D^T]\cos\delta_D^T -
x/r_D^T]\sin\delta_D^T
)\, - \right. \nonumber \\
& & \left. R_S (  [yr_D^S/(r_D^T)^2]\cos\delta_D^S -
[xr_D^S/(r_D^T)^2]\sin\delta_D^S) + (x^2 + y^2)/(r_D^T)^2 \right).
\end{eqnarray}

\subsection{$\kspipi$ tags}

The self-conjugate decay mode $D\to \kspipi$ has been extensively studied at charm threshold, and measurements have been performed by both the CLEO and BESIII Collaborations~\cite{Libby:2010nu,Ablikim:2020yif,Ablikim:2020lpk} of the strong-phase variation over the phase space of the three-body final state.

\begin{figure}
    \centering
    \includegraphics[width=.48\textwidth]{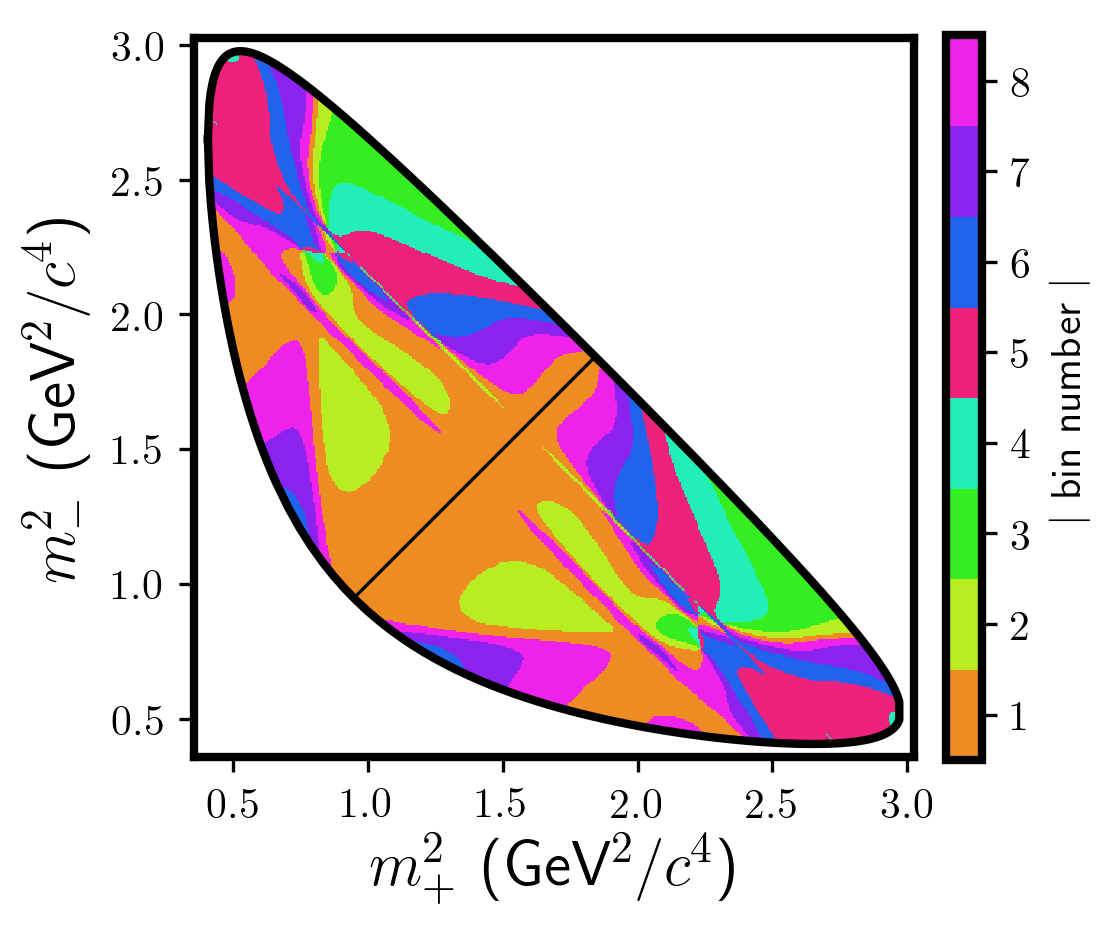}
    \caption{Dalitz plot of $D \to \kspipi$ decays showing the equal-$\Delta \delta_D$ binning scheme.}
    \label{fig:kspipidalitz}
\end{figure}

The Dalitz plot of the decay has axes corresponding to the squared invariant masses $m_-^2 = m(\ks\pi^-)^2$ and $m_+^2= m(\ks \pi^+)^2$ for each $\ks$ and pion combination.
Eight pairs of bins are defined, symmetric about the line $m_-^2 = m_+^2$, such that 
the bin number changes sign under the exchange $(m_-^2,m_+^2) \leftrightarrow (m_+^2,m_-^2)$.  The bins are labelled $-8$ to 8 (excluding 0), with the positive bins lying in the region $m_+^2>m_-^2$. The strong-phase difference
between symmetric points in the Dalitz plot is given by $\Delta \delta_D^{\nkspipi} \equiv \delta_D^{\nkspipi}(m_+^2,m_-^2) - \delta_D^{\nkspipi}(m_-^2,m_+^2)$.  
The bin boundaries are chosen such that each bin spans an equal range in $\Delta \delta_D^{{\nkspipi}}$ (the `equal-$\Delta \delta_D$ binning scheme'), as shown in Fig.~\ref{fig:kspipidalitz} where the variation in  $\Delta \delta_D^{\nkspipi}$ is assumed to follow that predicted by an amplitude model~\cite{Aubert:2008bd}. 

Measurements performed with quantum-correlated $D\bar{D}$ pairs determine the actual strong-phase difference within each Dalitz bin.  More precisely, what are measured are $c_i$, the cosine, and $s_i$, the sine of the strong-phase difference weighted by the $D$-decay amplitude in bin $i$:
\begin{align}
c_i &\equiv \frac{\int_{i} {\rm d}{m_+^2} \, {\rm d}{m_-^2} \, |{\cal{A}}_{\nkspipi}({m_+^2},{m_-^2})| |{\cal{A}}_{\nkspipi}({m_-^2},{m_+^2})| \cos
\Delta \delta_D^{\nkspipi}({m_+^2},{m_-^2})}
{\sqrt{\int_{i} {\rm d}{m_+^2} \, {\rm d}{m_-^2} \, |{\cal{A}}_{\nkspipi}({m_+^2},{m_-^2})|^2 \int_{i} {\rm d}{m_+^2} \, {\rm d}{m_-^2} \, |{\cal{A}}_{\nkspipi}({m_-^2},{m_+^2})|^2}}\,,
\label{eq:ci}
\end{align}
with an analogous expression for $s_i$.

When employing $D\to\kspipi$ as a tag mode it is also necessary to know $K_i$, which is the probability of a single $D^0$ decay occurring in bin $i$ with no requirement on the decay of the other charm meson in the event:
\begin{equation}
\label{eq:ki}
K_i = \frac{\int_{i} {\rm d}{m_+^2} {\rm d}{m_-^2} |{\cal A}_{\nkspipi}(m_+^2,m_-^2)|^2 }{\sum_j \int_{j} {\rm d}{m_+^2} {\rm d}{m_-^2} |{\cal A}_{\nkspipi}(m_+^2,m_-^2  )|^2},
\end{equation}
where the sum in the denominator is over all bins.
This quantity may be measured in flavour-tagged decays, either at charm threshold or at higher energies (although in the latter case small corrections must be applied to account for the effects of $D^0$-$\bar{D}^{0}$ oscillations~\cite{Bondar:2010qs}).

Events where one meson decays to $K^- n\pi$ ($K^+ n\pi$) and the other to $\ks\pi^+\pi^-$ are labelled with a negative (positive) bin number if $m_-^2<m_+^2$ ($m_-^2>m_+^2$).
With these definitions, it can be shown from Eq.~\ref{eq:general_ST} that, with uniform acceptance over the Dalitz plot, the yield of double-tagged events  is given by
\begin{equation}\label{eq_kspipi}
\begin{aligned}
 Y^{S}_{i} =  H \left( K _ { i } + \left( r _ { D } ^ {S} \right) ^ { 2 } K _ { - i } -   2 r _ { D } ^ {S} R_S \sqrt { K _ { i } K _ { - i } } \left[ c _ { i } \cos \delta _ { D } ^ {S} - s _ { i } \sin \delta _ { D } ^ {S} \right] \right),
\end{aligned}
\end{equation}
where $H$ is a bin-independent normalisation factor. 

In principle, the channel $D \to K^0_L \pi^+\pi^-$ may be employed as a tagging mode in an analogous manner.  However, the $K_i$ factors for this decay are less well known than those for $D \to \ks \pi^+\pi^-$, and limit the sensitivity of this tag.  Hence only the $D \to \ks\pi^+\pi^-$ tag is used in the current analysis.

\subsection{Binning the ${D\to \kmthreepi}$ phase space}

In Ref.~\cite{Evans:2019wza} it is demonstrated how the sensitivity to $\gamma$ in $B^- \to DK^-$, $D\to \kmthreepi$ decays can be enhanced by dividing the phase space of the $D$ decay into bins, each with its own coherence factor and strong-phase difference.  A binning scheme is proposed based on the resonance sub-structure of the CF and DCS decays as modelled in the LHCb study reported in Ref.~\cite{Aaij:2017kbo}. The phase space is divided into four bins, each of which span a variation in the strong-phase difference between the $D^0$ and $\bar{D}^{0}$ decay amplitudes.  The boundaries are chosen to equalise the product of the integrated CF amplitude and integrated DCS amplitude between bins, which is a metric that is shown to optimise the $\gamma$ sensitivity.  In this binning scheme a small region of phase space is excluded where the invariant mass of $\pi^+\pi^-$ pairs lies close to the mass of the $\ks$ meson.  This is done in order to exclude background from $D \to \ks K^-\pi^+$ decays, and, according to the amplitude models, rejects $5\%$ of the signal.  The phase space of $D\to\kmthreepi$ decays is five-dimensional, and therefore there is no convenient way to visualise the bins. However, the bin in which each decay lies can be unambiguously assigned through knowledge of the four-momenta of the final-state particles.

The observables discussed  may be determined within each bin in an identical manner to the global case, giving four measurements each for $\rho^{S}_{C\!P\pm}$, $\rho^{S}_{F,LS}$ and $Y^{S}_i$.  For $\rho^{K3\pi}_{LS}$, where 
the signal and tag decays are both $D\to \kmthreepi$, ten distinct measurements can be performed.
\section{Detector and simulation samples}
\label{sec:detector}

The BESIII detector is a magnetic
spectrometer~\cite{Ablikim:2009aa} located at the Beijing Electron
Positron Collider (BEPCII)~\cite{Yu:2016cof}. The
cylindrical core of the BESIII detector consists of a helium-based
 multilayer drift chamber (MDC), a plastic scintillator time-of-flight
system (TOF), and a CsI(Tl) electromagnetic calorimeter (EMC),
which are all enclosed in a superconducting solenoidal magnet
providing a $1.0\,$T magnetic field. The solenoid is supported by an
octagonal flux-return yoke with resistive plate counter muon-identifier modules interleaved with steel. The acceptance of
charged particles and photons is 93\% of the $4\pi$ solid angle. The
charged-particle momentum resolution at $1\,{\rm GeV}/c$ is
$0.5\%$, and the specific energy loss (${\rm d}E/{\rm d}x$) resolution is $6\%$ for electrons
from Bhabha scattering. The EMC measures photon ~~energies with a
resolution of $2.5\%$ ($5\%$) at $1$\,GeV in the barrel (end-cap)
region. The time resolution of the TOF barrel section is $68\,$ps, while
that of the end cap is $110\,$ps.

Simulated samples produced with the {\sc
geant4}-based~\cite{geant4} Monte Carlo  package, which
includes the geometric description of the BESIII detector and the
detector response, are used to determine the detection efficiencies
and to estimate the backgrounds. The simulation includes the beam-energy spread of 0.97\,MeV and initial-state radiation (ISR) in the $e^+e^-$
annihilations modelled with the generator {\sc
kkmc}~\cite{ref:kkmc}. The inclusive Monte Carlo samples consist of the production of $D^0\bar{D}^0$ and $D^+D^-$ pairs from decays of the $\psi(3770)$,
decays of the $\psi(3770)$ to charmonia or light hadrons, the ISR
production of the $J/\psi$ and $\psi(3686)$ states, and the
continuum processes incorporated in {\sc kkmc}~\cite{ref:kkmc}. The equivalent integrated luminosity of the inclusive Monte Carlo samples is about ten times that of the data. The known decay modes are modelled with {\sc
evtgen}~\cite{Lange:2001uf,Rong_Gang_2008}
using branching fractions taken from the
Particle Data Group~\cite{Zyla:2020zbs}, and the remaining unknown decays
from the charmonium states with {\sc
lundcharm}~\cite{PhysRevD.62.034003,
YANGRui-Ling:61301}. The final-state radiation (FSR)
from charged final-state particles is incorporated with the {\sc
photos} package~\cite{RICHTERWAS1993163}. The signal processes are generated separately taking the spin-matrix elements into account in {\sc evtgen}. 
The signal mode $D \to \kmpipio$ is generated with a resonant sub-structure which matches that of the CF process reported in Ref.~\cite{Zyla:2020zbs}; for the decay $D \to \kmthreepi$ the simulated events are re-weighted so that the invariant-mass distributions agree with those produced by the CF model of Ref.~\cite{Aaij:2017kbo}.
Sample sizes of 200,000 events are simulated for each class of double tag, apart from the case where the tagging mode is $D\to \kspipi$, for which 400,000 events are produced for each signal channel. 
\section{Event selection and yield determination}
\label{sec:selection}

\subsection{Selection of double-tagged events}

In order to determine the observables introduced in Sec.~\ref{sec:formalism}, double-tagged  samples are collected by reconstructing one charm meson in its decay to a signal mode, and the other charm meson in its decay to a tag mode, which is classed as either flavour, $C\!P$, or self-conjugate.  Flavour tags can be divided into like sign, where the kaon is of the same sign as the signal mode, or opposite sign.  The latter category is used for normalisation purposes.  Six categories of $C\!P$-even tags are reconstructed, including the quasi-eigenstate mode $D\to \pipipio$, and six $C\!P$-odd tags.
Events containing both $D\to \kmpi$ and $C\!P$ tags are also reconstructed for normalisation purposes.
The self-conjugate category contains one tag mode $D \to \kspipi$.  The full list of tags used in the analysis is given in Table~\ref{tab:taglist}. Events containing a $K^0_L$ meson are reconstructed using a missing-mass technique, discussed below.  All other double-tagged events are fully reconstructed. $C\!P$-violation and matter-interaction effects within the neutral-kaon system are not considered because their impact on the measurements is negligible in comparison to the experimental sensitivity.
Those mesons that decay within the detector are reconstructed through the modes: $\ks \to \pi^+\pi^-$, $\pi^0\to\gamma\gamma$, $\eta\to\gamma\gamma$ and $\pi^+\pi^-\pi^0$, $\omega\to\pi^+\pi^-\pi^0$, $\eta'\to\pi^+\pi^-\eta$ and $\gamma\pi^+\pi^-$, and $\phi \to K^+K^-$.

\begin{table}[!ht]
\caption{Summary of tag modes selected against the signal decays $D \to \kmthreepi$ and $D \to \kmpipio$.  The $C\!P$ tags are also reconstructed against the decay $D \to \kmpi$.}
\label{tab:taglist}
\begin{center}
\begin{tabular}{llccc}
\toprule
\multirow{2}{*}{Flavour} & Like sign  &$\kmthreepi$, $\kmpipio$, $\kmpi$\\
					    & Opposite sign  &$\kpthreepi$, $\kppipio$, $\kppi$\\
\multirow{2}{*}{$C\!P$} & Even &$\kk$, $\pipi$, $\kspiopio$, $\klpio$, $\klomega$, $\pipipio$\\
& Odd &$\kspio$, $\kseta$, $\ksomega$, $\ksetap$, $\ksphi$, $\klpiopio$\\
\multicolumn{2}{l}{Self-conjugate}  & $\kspipi$\\
\bottomrule
\end{tabular}
\end{center}
\end{table}


Selected charged tracks must satisfy $|\cos\theta| < 0.93$, where $\theta$ is the polar angle with respect to the beam axis. The distance of closest approach of the track to the interaction point (IP) is required to be less than $10\,$cm in the beam direction and less than $1\,$cm in the plane perpendicular to the beam, except for tracks from $\ks$ candidates where the closest approach to the IP is required to be within $20\,$cm along the beam direction. Separation of charged kaons from charged pions is implemented by combining the d$E$/d$x$ measurement in the MDC and the time-of-flight information from the TOF.  
This information is used to calculate the probabilities $P_K$ and $P_\pi$ for the $K$ and $\pi$ hypothesis, respectively, 
and the track is labelled a $K$ ($\pi$) candidate if $P_K>P_\pi$ ($P_\pi>P_K$).

Photon candidates are selected from showers deposited in the EMC crystals, with energies larger than $25\,$MeV in the barrel ($|\cos{\theta}|<0.8$) and $50\,$MeV in the end cap ($0.86<|\cos{\theta}|<0.92$). In order to suppress fake photons from beam background or electronic noise, the shower clusters are required to be within $[0, 700]\,$ns of the start time of the event. Furthermore, the photon candidates are required to be at least $20^{\circ}$ away from any charged tracks to eliminate fake photons caused by the interactions of hadrons in the EMC.

Candidate $\ks$ mesons are reconstructed from pairs of tracks with opposite charge. To improve efficiency no particle-identification requirements are imposed on these tracks. 
The $\ks$ candidate is required to satisfy the flight-significance criterion $L/\sigma_L>2$, where $L$ is its flight distance obtained from a fit to the vertex of the track pair, and $\sigma_L$ the uncertainty on this quantity.
In addition, a constrained vertex fit is performed for each candidate, and the resulting invariant mass is required to lie within [0.487, 0.511]\,\gev/$c^{2}$.

In forming $\pi^0$ ($\eta$) candidates with pairs of photons it is required that the di-photon invariant mass lies within 
[0.115, 0.150] ([0.480, 0.580]) \gev/$c^{2}$, 
and that at least one photon candidate is found in the barrel region, where the energy resolution is best. To improve momentum resolution, a kinematic fit is performed, where the reconstructed $\pi^0$ $(\eta)$ mass is constrained to the nominal value~\cite{Zyla:2020zbs}, and the resulting four-vector is used in the later steps of the analysis.  When reconstructing $\eta\to\pi^+\pi^-\pi^0$ decays, the invariant mass of the $\pi^+\pi^-\pi^0$ combination is required to lie within [0.530, 0.565]\,\gev/$c^{2}$. Likewise, 
mass windows of [0.750, 0.820]\,\gev/$c^{2}$, [0.940, 0.970] \gev/$c^{2}$ and [0.940, 0.976]\,\gev/$c^{2}$ are imposed for the decays $\omega \to {\pi^+\pi^-\pi^0}$, $\eta' \to {\pi^+\pi^-\gamma}$ and $\eta' \to {\pi^+\pi^-\eta(\gamma\gamma)}$, respectively.

To suppress combinatorial background, the energy difference, $\Delta{E} = E_{D} - \sqrt{s}/2$ is required to be within $\pm3\sigma_{\Delta E}$ around the $\Delta E$ peak, where $\sigma_{\Delta E}$ is the $\Delta E$ resolution and $E_{D}$ is the reconstructed energy of a $D$ candidate in the rest frame of the initial $e^+e^-$ collision.
The resolution varies between decay modes, and the allowed interval in $\Delta{E}$ ranges from [$-$0.018, 0.017]\,\gev/$c^{2}$ for $D \to \kmthreepi$ to [$-$0.069, 0.044]\,\gev/$c^{2}$ for $D \to \kspio$. To suppress background from cosmic and Bhabha events in the tag modes $D \to K^+K^-$, $\pi^+\pi^-$ and $\kmpi$, the event is required to have two charged tracks with a TOF time difference less than $5\,$ns and that neither track is identified as an electron or a muon.
As discussed below in Sec.~\ref{sec:bckpeak}, a $\ks$ veto based on flight distance is applied to $D \to \pipipio$ candidates to suppress background from $D \to \kspio$, and similarly to
$D\to \kmthreepi$ candidates, to reduce contamination from $D \to \ks K^-\pi^+$ decays. 
If there are multiple double-tagged candidates in one event, the combination with the average reconstructed invariant mass lying closest to the nominal $D^0$ mass is chosen. Around 10\% of selected events contain more than one candidate.

\begin{figure}
    \centering
    \includegraphics[width=.9\textwidth]{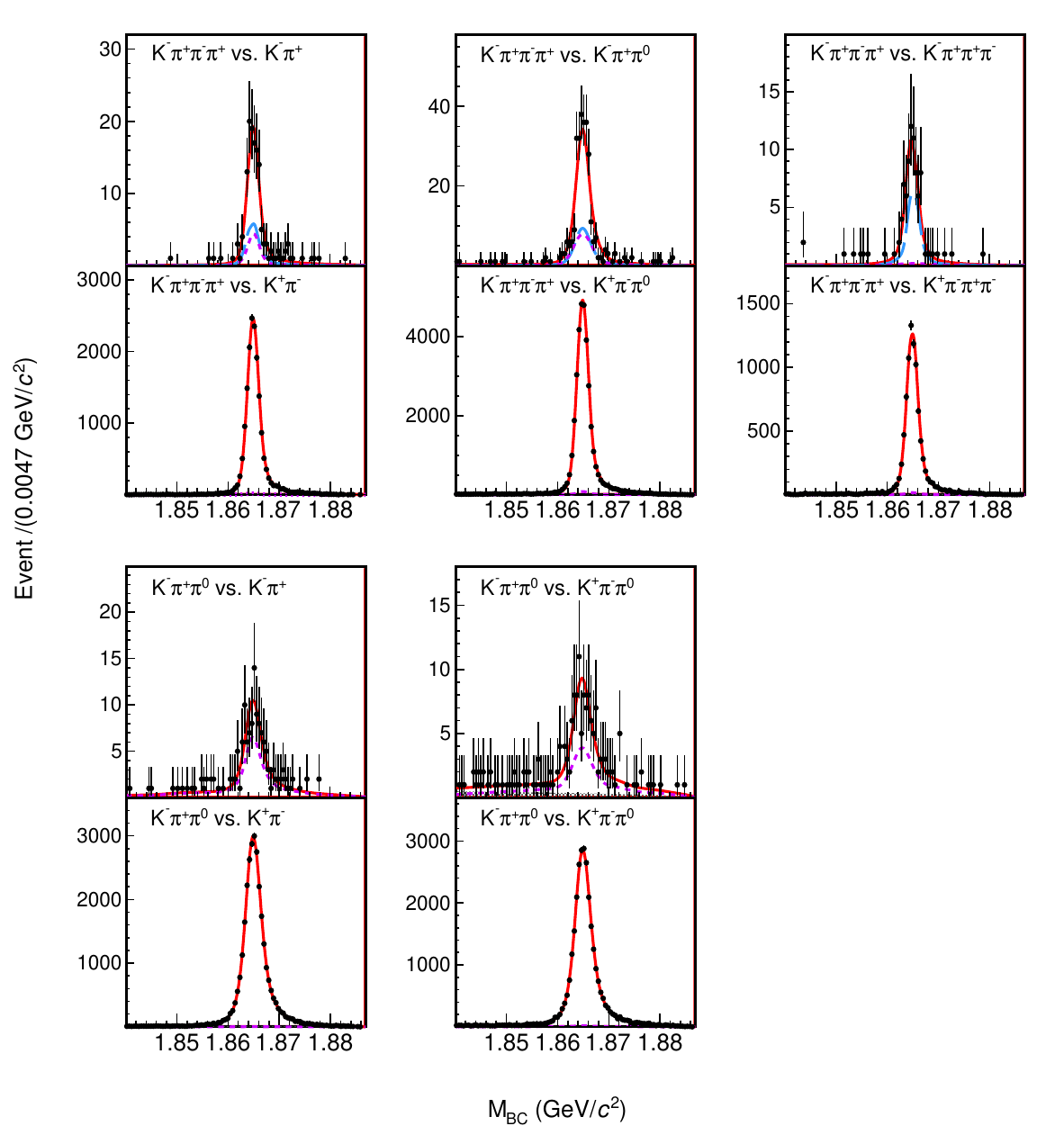}
    \caption{$M_{\rm BC}$ distributions  for the like-sign and opposite-sign flavour tags. The points with error bars are data; the red line indicates the total fit; the long-dashed azure line is the $D \to K^0_S K^-\pi^+$ background; the dashed purple line shows the other peaking-background contributions, which in the like-sign case are saturated by doubly misidentified opposite-sign events; the shaded region, which is at a very low level, represents the combinatorial background.}
    \label{fig:mbcexamples1}
\end{figure}

\begin{figure}
    \centering
    \includegraphics[width=.9\textwidth]{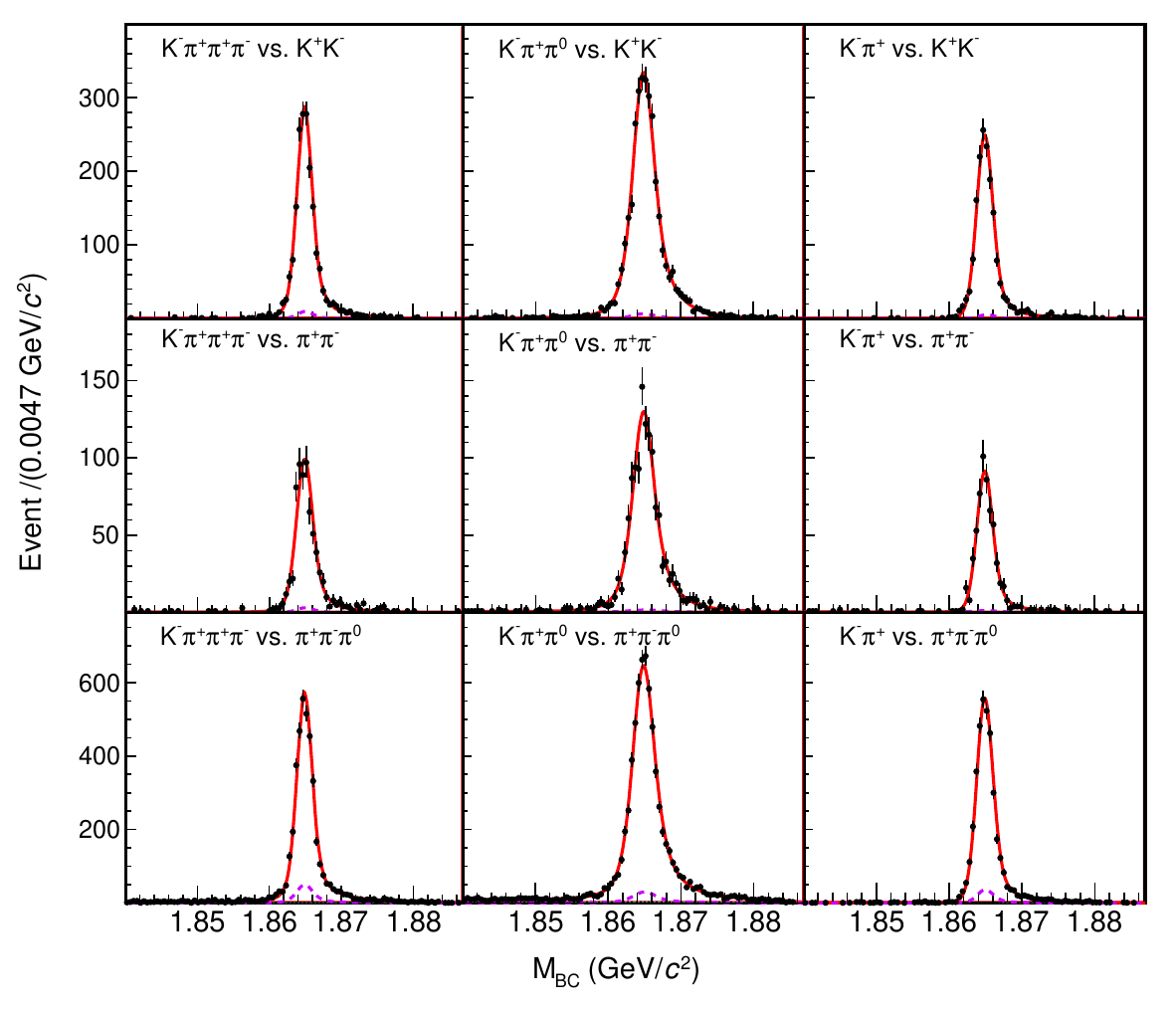}
    \caption{$M_{\rm BC}$ distributions for a selection of $C\!P$-tagged decays.
     The points with error bars are data; the red line indicates the total fit;  the  dashed purple line shows the  peaking-background contributions; the combinatorial-background contribution is at too low a level to be visible.}
    \label{fig:mbcexamples2}
\end{figure}

For each fully reconstructed $D$ candidate the beam-constrained mass, $M_{\rm BC}$, is calculated in order to provide optimal separation between signal and background:
\begin{equation}
   M_{\rm BC} = \sqrt{(\sqrt{s}/2)^2/c^4-|{\mathbf{p_{D}}}|^2/c^2},
   \label{eq:mbc}
\end{equation}
where $\mathbf{p_{D}}$ is the momentum of the $D$ candidate  in the rest frame of the initial $e^+e^-$ collision.   Figures~\ref{fig:mbcexamples1} and~\ref{fig:mbcexamples2} show $M_{\rm BC}$ distributions of the signal decay for a selection of double tags. The distributions for the other double tags may be found in Appendix~\ref{sec:appendix0}.

\begin{figure}
    \centering
    \includegraphics[width=.9\textwidth]{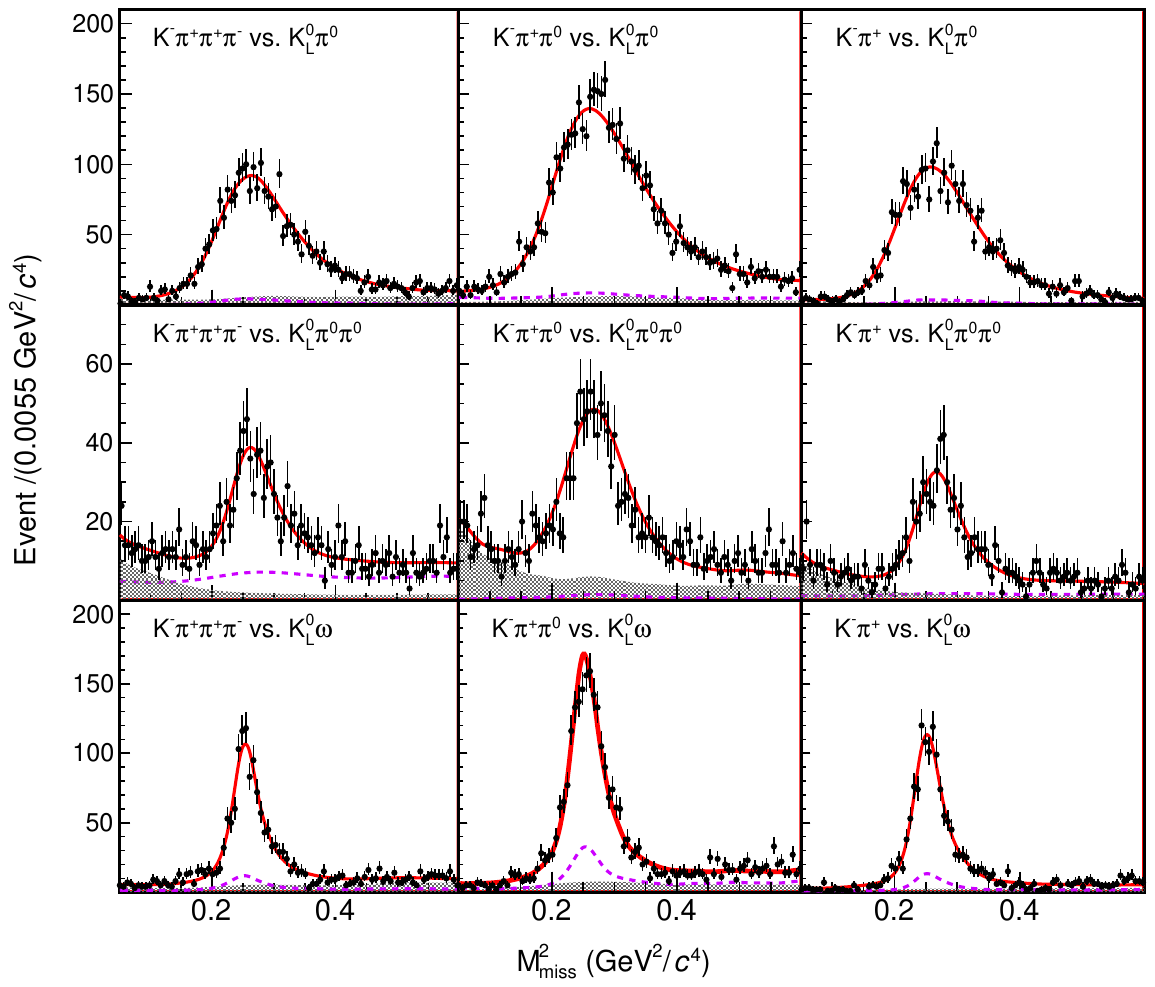}
    \caption{$M_{\rm miss}^2$ distributions for double tags containing a $K^0_L$ meson.  The points with error bars are data;   the red line indicates the total fit; 
    the dashed purple line shows the peaking-background contribution; the shaded region represents the combinatorial background.
    }
    \label{fig:mmexamples}
\end{figure}

Double-tagged events where the $C\!P$-tag mode involves a $K^0_L$ meson cannot be fully reconstructed.  Instead, these events are selected using a missing-mass technique.  First the signal mode is reconstructed, and its momentum, $\mathbf{p_S}$, is measured in the centre-of-mass frame of the $e^+e^-$ collision. If more than one candidate is found, that one with the smallest value of $|\Delta E|$ is retained. Then the total energy, $E_{X}$, and momentum, $\mathbf {p_{X}}$, of the charged particles and $\pi^0$ candidates not associated with the signal mode are determined.  This information allows the missing-mass squared,
\begin{equation}
   M_{\rm miss}^2 = (\sqrt{s}/2-E_{X})^2/c^4-|\mathbf{p_{S}}+\mathbf{p_{X}}|^2/c^2,
   \label{eq:mm2}
\end{equation}
to be calculated, which is expected to peak at the squared mass of the $K^0_{L}$ meson for the $C\!P$ tags under consideration.
To suppress contamination from $\ks \to \pi^0\pi^0$ decays and other backgrounds, events are rejected that contain surplus $\pi^0$ candidates, surplus charged tracks, any $\eta \to \gamma \gamma$ candidates, or multiple $\pi^0$ candidates that share common showers.  Figure~\ref{fig:mmexamples} shows $M_{\rm miss}^2$ distributions for those double tags containing a $K^0_{L}$ meson.

\subsection{Signal yield determination and consideration of peaking backgrounds}
\label{sec:bckpeak}

The signal yields for the  double tags are determined from an extended unbinned maximum-likelihood fit to the $M_{\rm BC}$ distributions of the tag decay in the case of the fully reconstructed events, and to the $M_{\rm miss}^2$ distributions for the events containing a $K^0_L$ candidate. (An alternative approach, where a two-dimensional fit is performed to the $M_{\rm BC}$ distributions of both the signal and tag decays in fully reconstructed events, is found to give very similar results, and is further considered when discussing the assignment of systematic uncertainties in Sec.~\ref{sec:syst}.) The signal shape is modelled from fits to Monte Carlo simulation using the {\sc RooKeysPdf} one-dimensional kernel estimator~\cite{ROOT}, and is convolved with a Gaussian function to account for differences between the resolution in data and simulation, and whose width and mean are free parameters in the fit. The combinatorial background of the $M_{\rm BC}$ distribution is described with an ARGUS function~\cite{ALBRECHT1990278}, and that of the $M_{\rm miss}^2$ distribution with a {\sc RooKeysPdf}-determined shape taken from simulation.
In general the contribution from background that peaks under the signal region is fixed according to what is found in the Monte Carlo simulation, with a shape determined with {\sc RooKeysPdf} from the same source.  Most of these contributions are at a low level in the fully reconstructed events, summing to typically less than 5\%, 10\% and 10\% in the flavour-tagged, $C\!P$-tagged and $D\to \kspipi$-tagged samples, respectively. However some are significantly larger, or require individual treatment, as is explained below.  All fits converge with satisfactory residuals.

The two singly Cabibbo-suppressed decays $D \to \ks K^\mp \pi^\pm$ give rise to the same final states as the signal modes $D \to K^\mp \pi^\pm \pi^\pm \pi^\mp$, and lead to significant contamination in the like-sign samples for the global measurement. In the binned measurement these decays are rejected by explicitly excluding the $\ks$ region from the phase space that is analysed.  In the global analysis a mass veto is undesirable, as it would prevent the measurement being representative of all phase space.  Instead an alternative $\ks$ veto is applied, in which the two pairs of oppositely charged pions in the decay of the signal candidate are fitted to a common vertex in turn, and their flight significance is determined.  Events are rejected when this quantity exceeds a value of two for either combination, after which the net selection efficiency for the background events is ${\cal{O}}(1\%)$. (This requirement is also imposed in the binned analysis, in addition to the mass veto.) The residual contamination is at a similar level to the signal itself in the affected like-sign samples.   Its exact contribution is determined by assuming the measured branching fractions~\cite{Zyla:2020zbs} and the efficiencies found in Monte Carlo simulation, and also correcting for quantum-correlation effects not present in the simulation.  For the latter calculation the hadronic parameters of the background decay are taken from Ref.~\cite{Insler:2012pm}, and those of the other decays contributing to the double tags from Refs.~\cite{Evans:2016tlp,Amhis:2019ckw}.  This background accounts for around 1\% of events in  the samples containing $C\!P$ and $D\to \kspipi$ tags, and is negligible for the opposite-sign flavour-tagged events.

\begin{figure}
    \centering
    \includegraphics[width=.48\textwidth]{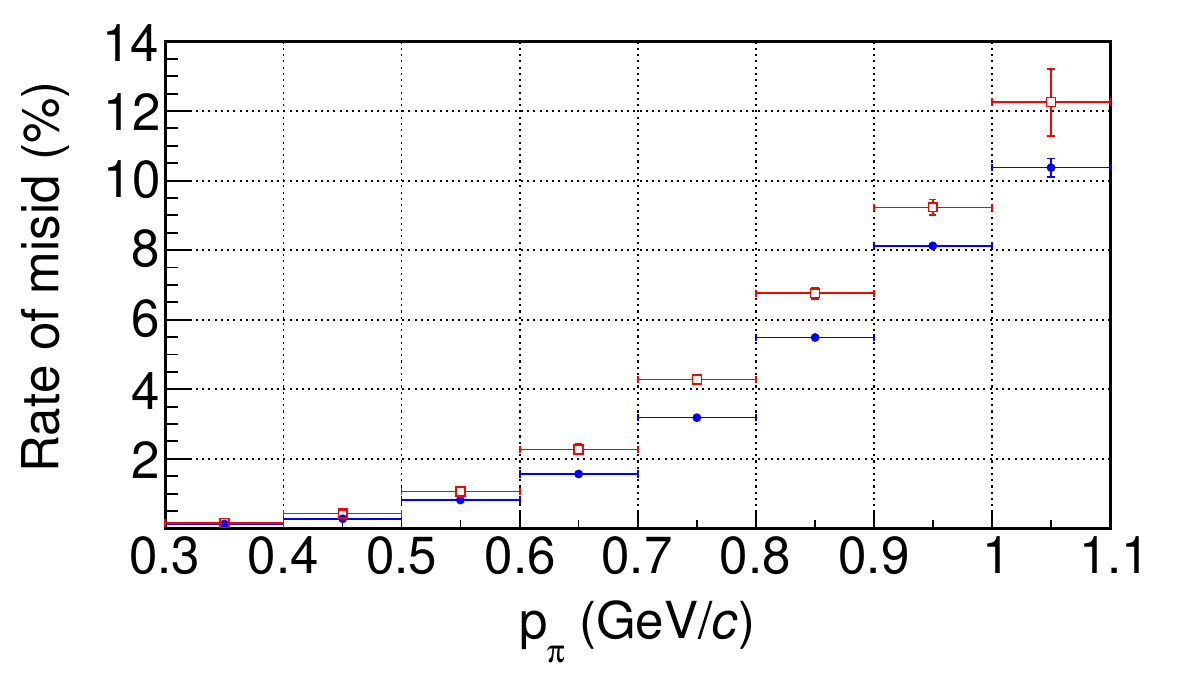}
    \includegraphics[width=.48\textwidth]{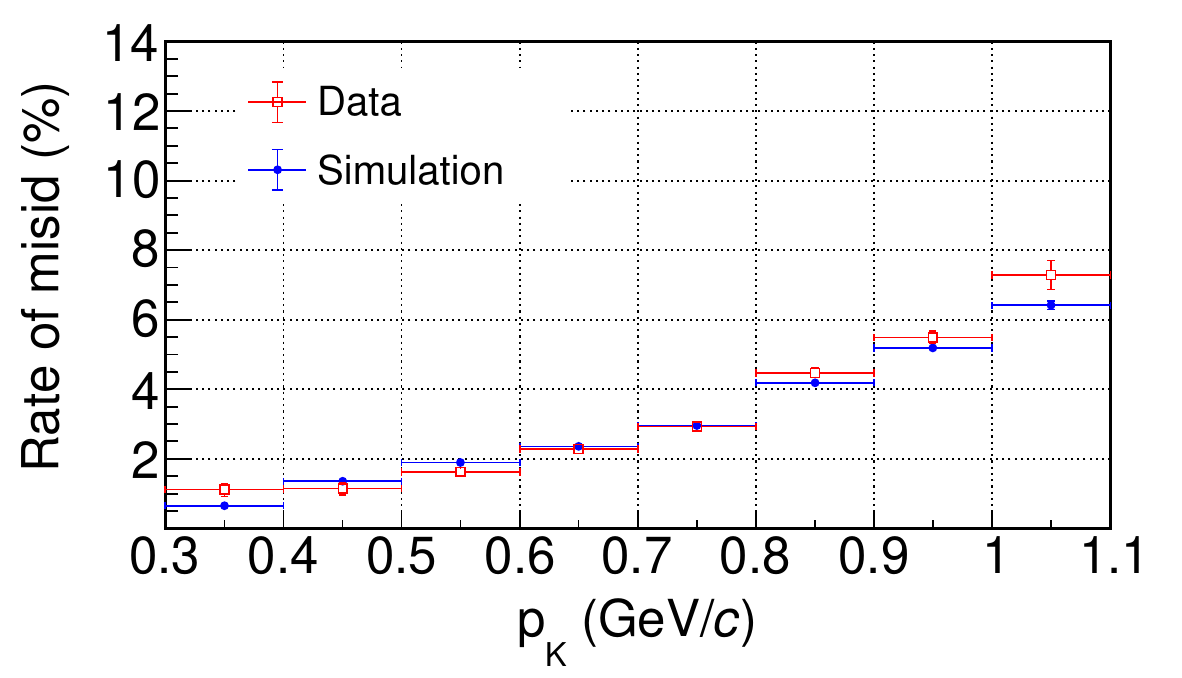}
    \caption{Probability of $\pi \to K$ (left) and $K \to \pi$ (right) misidentification in bins of momentum, as measured in data and Monte Carlo simulation.}
    \label{fig:misid}
\end{figure}

Another important source of background in the like-sign samples arises from opposite-sign events in which both a kaon and pion of opposite charge are misidentified.  In order to calibrate the level of this contamination, the rate of misidentification is measured in bins of momentum for a sample of opposite-sign events that is selected with no particle-identification requirements on one of the kaon or pion candidates.
This sample includes all the double-tag categories listed in Table~\ref{tab:taglist}.
The results in data are similar to those found in simulation, as can be seen in Fig.~\ref{fig:misid}.  Those differences that are observed are applied as corrections in the simulation. The amount of this background depends on the multiplicity of the final states, because of the momentum dependence of the particle identification.  It is approximately double the signal size in the $\kmpipio$ vs.~$\kmpi$ sample, and an order of magnitude smaller than the signal contribution in the $\kmthreepi$ vs. $\kmthreepi$ sample.

The decay $D\to \kspio$ is a dangerous background for $D \to \pipipio$ tags, as the two modes have opposite $C\!P$ eigenvalues.
Therefore a $\ks$ veto is applied based on the flight significance and identical to that imposed for the $D \to \kmthreepi$ selection.
The residual background is corrected for quantum correlations, as is the low level of $D \to \pipipio$ contamination in the $D\to \kspio$ sample.
The $\ks \pi^+\pi^-\pi^0$ final state that is used to construct the $C\!P$-odd $D\to \ks \omega$ and $D \to \ks \eta$  tags has a non-resonant component that is determined to be at the 20\% and 10\% level, respectively.  
Studies reported in Ref.~\cite{K:2017qxf} suggest that this background is mildly $C\!P$ odd.  
Non-resonant $D \to \ks \pi^+\pi^-\pi^0$ and $D \to K^0_{L} \pi^+\pi^-\pi^0$ decays comprise around 30\% of the $D \to K^0_L \omega$ sample, and are expected to be $C\!P$ neutral.  These estimates, obtained from simulation, are corroborated by fits to the $\pi^+\pi^-\pi^0$ mass spectrum in data, indicating that any quantum-correlation effects are here negligible.
The $D \to \klpiopio$ sample suffers background from $D \to \pi^0\pi^0\pi^0$ decays, which generates a structure in the $M_{\rm miss}^2$ spectrum that peaks at low values.  The contribution from this component is a free parameter in the fit,  with its shape fixed from simulation studies.

The signal yields for the flavour-tagged and the $D \to \kspipi$ tagged events are given in 
Table~\ref{tab:flavouryields}.  Those for the $C\!P$ tags may be found in Table~\ref{tab:cpyields}, together with the selection efficiencies that are required to determine the $\rho^{K3\pi}_{C\!P\pm}$ and $\rho^{K\pi\pi^0}_{C\!P\pm}$ observables.

\begin{table}[!ht]
\caption{Signal yields for the flavour-tagged and $D \to \kspipi$ tagged events. The uncertainties are statistical only.}\label{tab:flavouryields}
\begin{center}
\begin{tabular}{lrr}
\toprule
Tag  & $\kmthreepi$ & $\kmpipio$ \\
\midrule
$\kppi$     	&17368$\pm$\,\,136 & 
29462$\pm$180 
\\
$\kppipio$  	&33734$\pm$\,\,192 &
28672$\pm$182
  \\
$\kpthreepi$ &8602$\pm$\;\;\;97    & 
/ \hspace*{0.5cm}
\\
 & 	& \\
 $\kmpi$ 	& 63.0$\pm$13.9 	  &33.0$\pm$\;\,9.6 \\ 
 $\kmpipio$  &128.8$\pm$22.0    &53.5$\pm$11.2  \\
$\kmthreepi$ &41.0$\pm$11.2      &
/ \hspace*{0.5cm}
\\ 
& & \\
$\kspipi$ & 4927.5$\pm$73.5 &  8647$\pm$101 
\\ 
\bottomrule
\end{tabular}
\end{center}
\end{table}

\begin{table}[!ht]
\caption{Signal yields for the $C\!P$-tagged events. The uncertainties are statistical only. Also shown are the selection efficiencies as determined from Monte Carlo simulation, all of which have a statistical uncertainty of $0.1\%$ or less.}\label{tab:cpyields}
\begin{center}
\begin{tabular}{lrr rr rr} \toprule
Tag & \multicolumn{2}{c}{$\kmthreepi$} & \multicolumn{2}{c}{$\kmpipio$} & \multicolumn{2}{c}{$\kmpi$} \\
& Yield\hspace*{0.65cm} & Eff. (\%) & Yield\hspace*{0.65cm} & Eff. (\%) & Yield\hspace*{0.65cm} & Eff. (\%) \\
\midrule
$\kk$ 		 &1849.0$\pm$44.4 & 23.3\;\; &3261.3$\pm$58.5 & 24.1\;\;\ &1646.4$\pm$41.5 & 43.2\;\;			 \\
$\pipi$		 &684.6$\pm$26.8 & 24.4\;\; &1273.1$\pm$36.8 &  25.7\;\;& 592.4$\pm$25.0 & 46.5\;\; 
\\
$\pipipio$ 	 &3539.0$\pm$67.8 & 12.6\;\; &6269.1$\pm$88.5 & 13.5\;\; &3646.9$\pm$62.7 &
28.2\;\; \\
$\kspiopio$  &727.5$\pm$29.8 & 5.8\;\; &1346.9$\pm$41.1   &  5.7\;\; &803.6$\pm$29.7	&	12.3\;\;
\\
$\klpio$ 	 &2293.3$\pm$61.3 & 11.5\;\; &4203.2$\pm$71.1 & 11.2\;\; &2589.9$\pm$59.5	&
25.1\;\; \\
$\klomega$ 	 &1041.7$\pm$61.8 & 4.5\;\; & 1543.6$\pm$53.9 & 4.2\;\; &1137.2$\pm$40.0 & 10.7\;\;	
 \\ 
 \\
$\kspio$ 			     &1958.4$\pm$47.8 & 14.2\;\; &3553.8$\pm$63.4 &14.7\;\;  &1697.1$\pm$42.2	& 28.4\;\;
\\
${\kseta}(\gamma\gamma)$  &289.3$\pm$18.9 & 12.2\;\; & 472.2$\pm$23.8 & 11.9\;\; & 230.2$\pm$16.2 &
25.0\;\; \\
${\kseta}(\pi\pi\pi^0)$   &67.5$\pm$\;\,8.9 	& 6.3\;\; & 122.1$\pm$12.2 & 6.4\;\;  &65.6$\pm$\;\,8.5 &
13.0\;\; \\
${\ksetap}(\gamma\pi\pi)$ &  220.4$\pm$17.9 & 7.6\;\; & 445.0$\pm$23.5 & 8.0\;\; & 220.0$\pm$15.8 &
15.8\;\;\\
${\ksetap}(\pi\pi\eta)$   &104.1$\pm$10.7 &4.6\;\; &159.2$\pm$12.9 & 4.7\;\; &94.6$\pm$\,\;9.8 &
10.1\;\; \\
${\ksomega}$ &728.8$\pm$34.1 & 5.8\;\; &1153.4$\pm$39.8 & 5.6\;\; & 643.4$\pm$27.8 & 12.1\;\;
\\
$\ksphi$ 				 &187.0$\pm$16.5 & 5.1\;\; &349.2$\pm$20.3 & 5.6\;\; &
180.9$\pm$13.5 & 10.4\;\;
\\ 
$\klpiopio$ 				 &651.4$\pm$47.8 & 2.8\;\;& 1113.1$\pm$40.4 &2.4\;\; & 639.7$\pm$40.3 & 4.9\;\;
\\
\bottomrule
\end{tabular}
\end{center}
\end{table}

In order to improve the resolution of the reconstructed position of each decay in phase space, a kinematic fit is applied to the $D \to \kmthreepi$ candidates with the $D^0$ mass imposed as a constraint.  The events are partitioned into the four bins defined in Ref.~\cite{Evans:2019wza}, and the same procedure is followed as above, to determine the signal yields in each bin.  By way of example, the signal yields of $D \to \kmthreepi$ tagged with $D \to \kspio$ are $497.5 \pm 23.2$, $415.0 \pm 21.0$, $400.1 \pm 20.2$ and $578.6 \pm 24.8$ for bins 1 to 4, respectively.

\subsection{Efficiency corrected double-tagged yields in $D \to \kspipi$ bins}

The $D \to \kspipi$ tagged events are divided into 16 bins of phase space of the tag decay. Again, a mass-constrained fit is applied to improve the resolution of the position of the decay in phase space.  Monte Carlo studies indicate that the correct bin is assigned 86\% of the time for the positive bins, and 94\% of the time for the negative bins.
The signal yields are determined in each bin and efficiency corrections are applied.
These corrections account for both the relative efficiency variation bin-to-bin, which lies within a range of $\pm 15\%$, and for migration between bins.  The resulting signal yields $Y^{K3\pi}_i$ and $Y^{K\pi\pi^0}_i$ are listed in Table~\ref{tab:kspipiglobal}.  This procedure is repeated for the binned $D \to \kmthreepi$ analysis to measure the signal yields $Y^{K3\pi,j}_i$ presented in Table~\ref{tab:kspipibins}.

\begin{table}[!htb]
    \caption{Efficiency corrected signal yields $Y^{K3\pi}_i$ and $Y^{K\pi\pi^0}_i$ in bin $i$ of $D \to \kspipi$ phase space. The efficiencies integrated over all bins are 12.9$\%$ and 13.6$\%$ for $D \to \kmthreepi$ and $D \to \kmpipio$ decays, respectively.}
    \label{tab:kspipiglobal}
    \centering
    \begin{tabular}{lrrlrr}\toprule
Bin    &$\kmthreepi$ &$\kmpipio$\hspace*{0.0cm}  & \;\,Bin     &$\kmthreepi$ &$\kmpipio$\hspace*{0.0cm}\\
\midrule
%

1     	&7011$\pm$\;303 &11095$\pm$395 &\;\;$-$1	 &3126$\pm\;$178 &5328$\pm$\;237\\
2  		&3630$\pm$\;220 &5427$\pm$290 &\;\;$-$2 	 &807.4$\pm$93.4 &1377$\pm$\;136 \\
3 		&2632$\pm\;$160 &4254$\pm$215 &\;\;$-$3 	 &681.4$\pm$79.8   &1302$\pm$\;116\\
4		&816.7$\pm$90.9 &1505$\pm$126 &\;\;$-$4 	 &701.7$\pm$82.8 &1067$\pm$\;111\\
5 	    &2978$\pm\;$171 &5127$\pm$237 &\;\;$-$5 	 &1933$\pm$\;136 &3145$\pm$\;184\\ 
6  		&2020$\pm$\;163  &3272$\pm$219 &\;\;$-$6 	 &454.8$\pm$82.2 &510$\pm$\;103\\
7 		&4782$\pm$\;245 &7657$\pm$295 &\;\;$-$7   &545.3$\pm$86.6	&990$\pm$\;127 \\ 
8      	&5029$\pm$\;274 &8162$\pm$374 &\;\;$-$8   &956$\pm$\;109 &1822$\pm$\;154 \\
\bottomrule
    \end{tabular}
\end{table}

\begin{table}[!htb]
\caption{Efficiency corrected signal yields $Y^{K3\pi}_i$ in bin $i$ of $D \to \kspipi$ phase space (rows) and each bin of $D \to \kmthreepi$ phase space (columns).} \label{tab:kspipibins}
    \centering
\begin{tabular}{rrrrr}    \toprule
\multicolumn{1}{l}{Bin} & \multicolumn{1}{c}{1} & \multicolumn{1}{c}{2} & \multicolumn{1}{c}{3}& \multicolumn{1}{c}{4}          \\
\midrule
1     	&1646$\pm$\;155 & 1415$\pm$\;134 &1688$\pm$\;158 &1963$\pm$\;168 \\
2  		&711$\pm$\;105 &715$\pm$\;102 &943$\pm$\;118 &1013$\pm$\;122  \\
3 		&735.2$\pm$88.9 &402.6$\pm$65.2 &522.7$\pm$76.2 &800.7$\pm$92.1 \\
4		&162.4$\pm$39.7 &161.8$\pm$44.1 &158.1$\pm$43.1 &309.1$\pm$53.5 \\
5 	    &738.5$\pm$86.7 &580.8$\pm$75.2 &560.8$\pm$78.5 &956.1$\pm$99.9  \\ 
6  		&430.9$\pm$77.3 &343.7$\pm$69.8 &523.7$\pm$84.1 &612.8$\pm$89.2 \\
7 		&1322$\pm$\;131 &1103$\pm$\;115 &926$\pm$\;117 &1135$\pm$\;117 \\ 
8      	&1313$\pm$\;145 &1048$\pm$\;126 &962$\pm$\;132 &1472$\pm$\;154  \\
$-$1	 &884.7$\pm$96.7 &706.9$\pm$81.4 &608.7$\pm$82.3 &641.3$\pm$89.5\\
$-$2 	 &107.7$\pm$40.3 &150.0$\pm$39.7 &288.8$\pm$56.2 & 165.8$\pm$44.8\\
$-$3 	 &128.0$\pm$35.4  &83.1$\pm$26.8 &182.7$\pm$41.0 & 211.3$\pm$44.2 \\
$-$4 	 &185.8$\pm$39.7  &167.0$\pm$40.9 &100.5$\pm$35.4 &207.7$\pm$46.7 \\
$-$5 	 &562.6$\pm$76.6 &410.1$\pm$64.3 &309.2$\pm$54.6 &550.5$\pm$74.3\\
$-$6 	 &142.4$\pm$43.5 &128.8$\pm$35.1  &43.9$\pm$29.5 &97.6$\pm$38.5\\
$-$7   &99.6	$\pm$36.3	  &164.6$\pm$43.3 &85.4$\pm$36.7 &189.9$\pm$51.4\\
$-$8   &213.5$\pm$51.2  &278.2$\pm$54.4 &225.3$\pm$49.6 &170.9$\pm$52.0\\
\bottomrule
\end{tabular}
\end{table}  
\section{Measurement of the observables and fit to the hadronic parameters}
\label{sec:obs}

A global analysis is performed, in which the $D\to\kmthreepi$ and $D \to \kmpipio$ phase spaces are considered inclusively.  In addition, a binned analysis is executed, where the $D\to\kmthreepi$ phase space is partitioned according to the four-bin scheme defined in Ref.~\cite{Evans:2019wza}.

The determination of the $C\!P$-tag and opposite-sign observables, and their interpretation in terms of the hadronic parameters, requires knowledge of branching fractions, ratios of branching fractions, and other parameters.  The values used, and their sources, are given in Table~\ref{tab:inputs}.

\begin{table}[!h]
\caption{Input parameters used in the determination of the observables and hadronic parameters.}
\begin{center}
\begin{tabular}{lllc}
\toprule
Input parameter \hspace{15mm} & Value &Reference\\
\midrule
$\br{}(\Dz\to \kmthreepi)$ & (8.23$\pm$0.14)$\%$ &\hspace*{0.5cm}\cite{2018Tanabashi}\\
$\br{}(\Dz\to \kpthreepi)/\br{}(\Dz\to \kmthreepi)$ & (3.22$\pm$0.05)$\times 10^{-3}$ &\hspace*{0.5cm}\cite{2018Tanabashi}\\
$\br{}(\Dz\to \kmpipio)$ & (14.4$\pm$0.5)$\%$ &\hspace*{0.5cm}\cite{2018Tanabashi}\\
$\br{}(\Dz\to \kppipio)/\br{}(\Dz\rightarrow \kmpipio)$ & (2.12$\pm$0.07)$\times 10^{-3}$ &\hspace*{0.5cm}\cite{2018Tanabashi}\\
$\br{}(\Dz\to K^-\pi)$  & (3.950$\pm$0.031)$\%$ &\hspace*{0.5cm}\cite{2018Tanabashi}\\
$\dkpi$ &$(192.1^{+8.6}_{-10.2})\si{\degree}$ &\hspace*{0.5cm}\cite{Amhis:2019ckw}\\
$(\rkpi)^2$ &(0.344$\pm$0.002)$\times 10^{-2}$
&\hspace*{0.5cm}\cite{Amhis:2019ckw}\\
$x$		   &($0.39^{+0.11}_{-0.12}$)$\%$
&\hspace*{0.5cm}\cite{Amhis:2019ckw}\\
$y$		   &$(0.651^{+0.063}_{-0.069})\%$ &\hspace*{0.5cm}\cite{Amhis:2019ckw}\\
$F_+^{\pi\pi\pi^0}$ &0.973$\pm$0.017 &\hspace*{0.5cm}\cite{Malde:2015mha}\\
\bottomrule
\end{tabular}
\end{center}
\label{tab:inputs}
\end{table}

When fitting the  signal yields  tagged by $D\to\kspipi$ decays in bins of phase space, $Y^{K3\pi}_i$ and $Y^{K\pi\pi^0}_i$,
to the expected distribution of events, expressed in Eq.~\ref{eq_kspipi}, 
it is necessary to know the strong-phase parameters $c_i$ and $s_i$, defined in Eq.~\ref{eq:ci}.  Values for these parameters are taken from the combined results of measurements performed by the BESIII~\cite{Ablikim:2020lpk,Ablikim:2020yif} and CLEO collaborations~\cite{Libby:2010nu}.  In addition, knowledge is required of the $K_i$ parameters, defined in Eq.~\ref{eq:ki}. Here, the most precise source of information comes from models fitted to the flavour-tagged yields of $D\to\kspipi$ decays at the BaBar and Belle experiments.  Predictions for $K_i'$ can be obtained from the models, where $K_i'= (1-\epsilon) K_i$ and $\epsilon= \sqrt{(K_{-i}/K_{i})}(yc_i - xs_i) + {\cal{O}}(x^2,y^2)$  is a small correction due to the presence of $D^0$-$\bar{D^0}$ oscillation effects. Table~\ref{tab:kspipifraction} lists the $K_i'$ results reported in Ref.~\cite{Adachi:2018jqe},  from which the $K_i$ values used in the current study are derived.  These results are considered to be reliable on account of the acceptable fit residuals found in that analysis.  The accompanying uncertainties are assigned from the statistical precision of the fit residuals in each bin, added in quadrature to
the root-mean-square of the variation of the predictions across this and three other models~\cite{Poluektov:2010wz,Aubert:2005iz,Aubert:2008bd}. 

\begin{table}[!ht]
\caption{The $K_i'$ parameters as calculated from the model of Ref.~\cite{2018Tanabashi}, equivalent to the fraction of flavour-tagged $D\to \kspipi$ decays falling in each bin of the Dalitz plot, including $D^0$-$\bar{D^0}$-oscillation effects, in the equal-$\Delta \delta_D$ binning scheme.}
\label{tab:kspipifraction}
\begin{center}
\begin{tabular}{cccc}
\toprule
 Bin		&{$K_i'$}         & Bin			     &{$K_i'$}\\
\midrule
1     	&$0.1734 \pm 0.0029$ & $-$1 			     & $0.0794 \pm 0.0023$		\\
2     	&$0.0876 \pm 0.0016$ & $-$2 			     & $0.0174 \pm 0.0002$			\\
3     	&$0.0692 \pm 0.0038$ & $-$3 			     & $0.0202 \pm 0.0002$			\\
4     	&$0.0255 \pm 0.0003$ & $-$4 			     & $0.0162 \pm 0.0008$			\\
5     	&$0.0850 \pm 0.0023$ & $-$5 			     & $0.0512 \pm 0.0013$		\\
6     	&$0.0592 \pm 0.0010$ & $-$6 			     & $0.0143 \pm 0.0009$			\\
7     	&$0.1269 \pm 0.0005$ & $-$7 			     & $0.0132 \pm 0.0003$			\\
8     	&$0.1338 \pm 0.0007$ & $-$8 			     & $0.0275 \pm 0.0026$			\\
\bottomrule
\end{tabular}
\end{center}
\end{table}

\subsection{Global analysis}
\subsubsection{Determination of the $C\!P$-tag and like-sign observables}

\begin{figure}[!ht]
\begin{center}
\includegraphics[width=.48\textwidth]{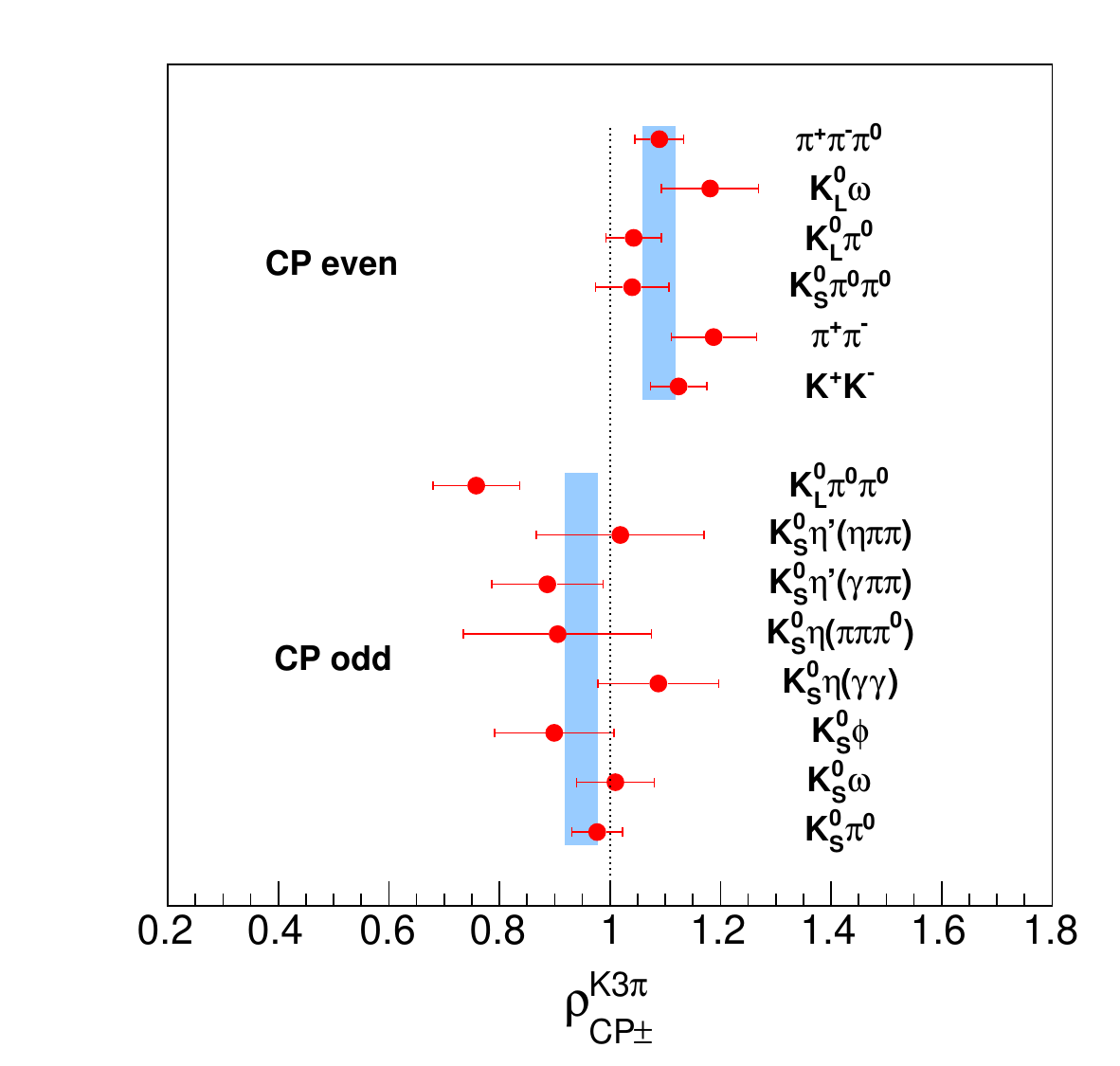}
\includegraphics[width=.48\textwidth]{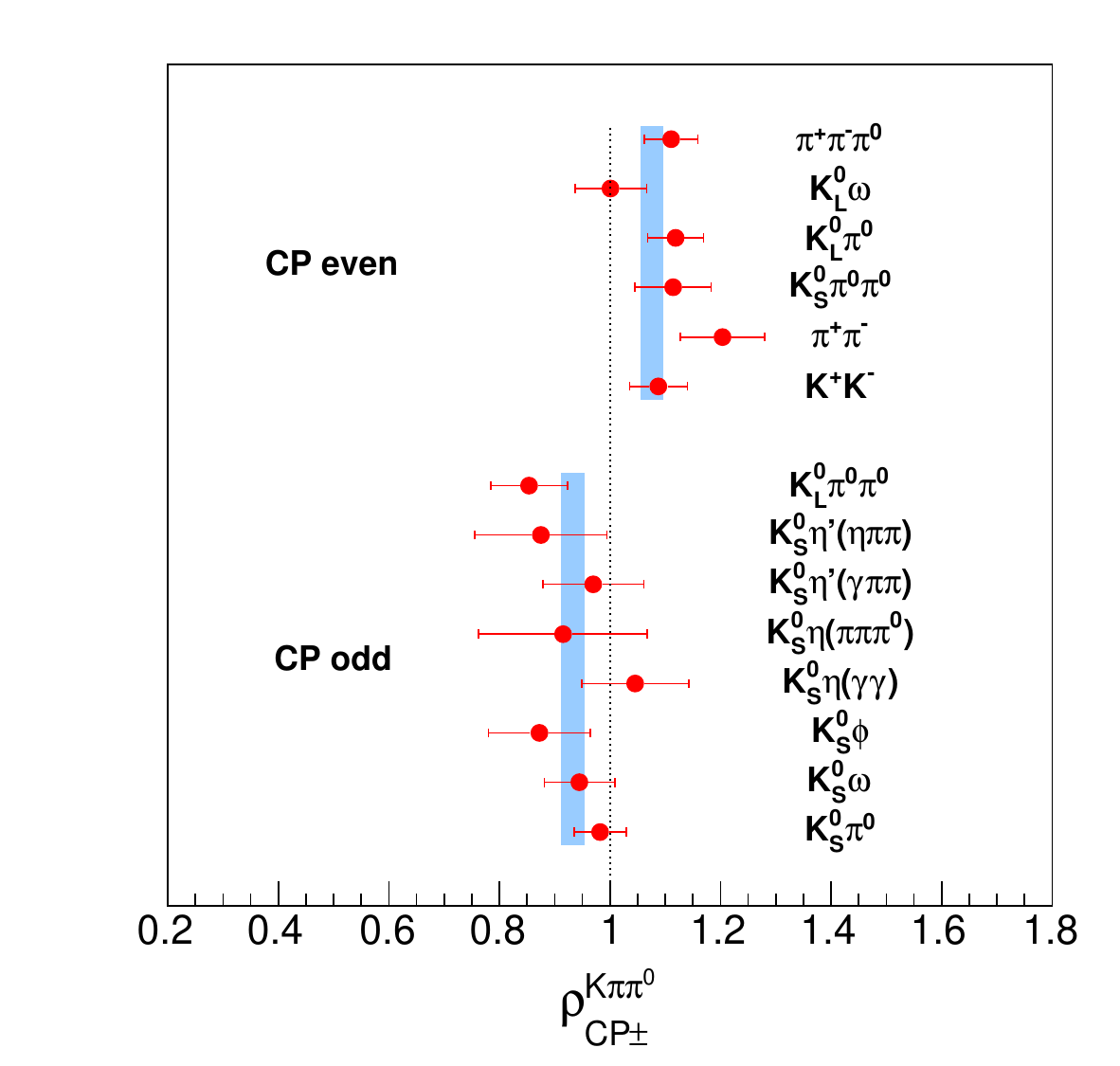}
\end{center}
{\caption{The $\rho^S_{C\!P+}$ and $\rho^S_{C\!P-}$ observables for (left) $S=\kmthreepi$ and (right) $S=\kmpipio$. The error bars indicate the sum in quadrature of the statistical and systematic uncertainties. The blue bands represent the $\pm$1$\sigma$ bound for the averaged results of each observable.}\label{fig:rhocp}}
\end{figure}

The $C\!P$-tag observables $\rho_{C\!P\pm}^{K3\pi}$ and $\rho_{C\!P\pm}^{K\pi\pi^0}$  are determined for each tag according to Eq.~\ref{eq:kpinorm}. They take as input the efficiency corrected $C\!P$-tagged signal and $D \to K^-\pi^+$ yields, listed in Table~\ref{tab:cpyields}, and the correction factors $\rho_{C\!P+}^{K\pi} = 1.119 \pm 0.005$ and $\rho_{C\!P-}^{K\pi} = 0.881 \pm 0.005$, which are calculated from knowledge of the external parameters $x$, $y$, $r_D^{\rm K\pi}$ and $\delta_D^{K\pi}$~\cite{Amhis:2019ckw}.  The results are given in Tables~\ref{tab:cp_k3pi} and~\ref{tab:cp_kpipi0} and are shown graphically in Fig.~\ref{fig:rhocp}. The results for each signal mode and class of tag are seen to be compatible and are therefore combined together to give a single, average, result for each observable.  Full account is taken of correlations between the systematic uncertainties.  Following Eq.~\ref{eq:deltacp} the parameters $\Delta^{K3\pi}_{C\!P}$ and  $\Delta^{K\pi\pi^0}_{C\!P}$ are also calculated, yielding the values shown in Table~\ref{tab:rhoresults}.

 \begin{table}[!htb]
 \caption{Results for  $\rho_{C\!P\pm}^{K3\pi}$ shown for individual tags, and averaged.  The uncertainties are statistical and systematic, respectively.}
\begin{center}
\begin{tabular}{lclc}
\toprule
 $C\!P$-even tag		&$\rho^{K3\pi}_{C\!P+}$    & $C\!P$-odd tag		&$\rho^{K3\pi}_{C\!P-}$  \\
\midrule
$\kk$    	&$1.124\pm 0.027\pm0.043$  &$\kspio$ 	&$0.977 \pm 0.024 \pm 0.038$ 			\\
$\pipi$  	&$1.188 \pm 0.047 \pm 0.061$ &${\kseta}(\gamma\gamma)$   & $1.088 \pm 0.071 \pm 0.083$ \\
$\pipipio$ & $1.089 \pm 0.022 \pm 0.038$  &${\kseta}(\pi\pi\pi^0)$   & $0.905 \pm 0.119 \pm 0.121$ \\
$\kspiopio$	& $1.040 \pm 0.043 \pm 0.050$ 				 &${\ksetap}(\gamma\pi\pi)$ & $0.887 \pm 0.072 \pm 0.071$  \\
$\klpio$  	& $1.043 \pm 0.028 \pm 0.041$ 	   &${\ksetap}(\pi\pi\eta)$   & $1.019 \pm 0.105 \pm 0.110$  \\ 
$\klomega$ & $1.181 \pm 0.070 \pm 0.053$   &${\ksomega}(\pi\pi\pi^0)$ & $1.010 \pm 0.047 \pm 0.053$ \\
 &      &$\ksphi$ 	 & $0.899 \pm 0.079 \pm 0.074$\\ 
      		 & 			  &$\klpiopio$ 				 & $0.758 \pm0.056 \pm 0.055$ \\
\midrule
Average & $1.088 \pm 0.013 \pm 0.026$ & 
& $0.948 \pm 0.017 \pm 0.025$  \\
\bottomrule
\end{tabular}
\end{center}
\label{tab:cp_k3pi}
\end{table}

\begin{table}[!htb]
 \caption{Results for  $\rho_{C\!P\pm}^{K\pi\pi^0}$ shown for individual tags, and averaged.  The uncertainties are statistical and systematic, respectively.}
\begin{center}
\begin{tabular}{lclc}
\toprule
$C\!P$-even tags		&$\rho^{K\pi\pi^0}_{C\!P+}$    & $C\!P$-odd tags		&$\rho^{K\pi\pi^0}_{C\!P-}$  \\
\midrule
$\kk$    	& $1.088\pm0.020\pm0.054$  &$\kspio$ 	&$0.982\pm 0.018\pm0.049$ 			\\
$\pipi$  	& $1.204 \pm 0.035 \pm 0.072$ &${\kseta}(\gamma\gamma)$   & $1.046 \pm 0.053\pm0.084$\\
$\pipipio$ & $1.111 \pm 0.016 \pm 0.052$  &${\kseta}(\pi\pi\pi^0)$   & $0.915 \pm 0.090 \pm 0.013$ \\
$\kspiopio$	& $1.114 \pm 0.034 \pm 0.065$ 				 &${\ksetap}(\gamma\pi\pi)$ & $0.970 \pm 0.051 \pm 0.079$  \\
$\klpio$  	& $1.119 \pm 0.019 \pm 0.053$ 	   &${\ksetap}(\pi\pi\eta)$   & $0.875 \pm 0.071 \pm 0.099$  \\ 
$\klomega$ & $1.001 \pm 0.035 \pm 0.060$   &${\ksomega}(\pi\pi\pi^0)$ & $0.945 \pm 0.033 \pm 0.059$ \\
 &      &$\ksphi$ 	 & $0.872 \pm 0.051 \pm 0.080$\\ 
      		 & 			  &$\klpiopio$ 				 & $0.854 \pm 0.031 \pm 0.066$\\
      		 \midrule
Average & $1.100 \pm 0.009 \pm 0.034$ & & $0.945 \pm 0.012 \pm 0.035$ \\
\bottomrule
\end{tabular}
\end{center}
\label{tab:cp_kpipi0}
\end{table}

The like-sign observables $\rho^{K3\pi}_{LS}$,  $\rho^{K3\pi}_{K\pi, LS}$, $\rho^{K3\pi}_{K\pi\pi^0, LS}$,  $\rho^{K\pi\pi^0}_{LS}$ and  $\rho^{K\pi\pi^0}_{K\pi, LS}$ are determined
according to  Eqs.~\ref{eq:rho_ls2} and~\ref{eq:rho_lsx2}.
from the yields of like-sign and opposite-sign double tags in Table~\ref{tab:flavouryields}. These results are also presented in Table~\ref{tab:rhoresults}.  The correlation matrix for all the observables may be found in Appendix~\ref{sec:appendix1}, accounting for both statistical and systematic contributions.

\begin{figure}[!ht]
\begin{center}
\includegraphics[width=.22\textwidth]{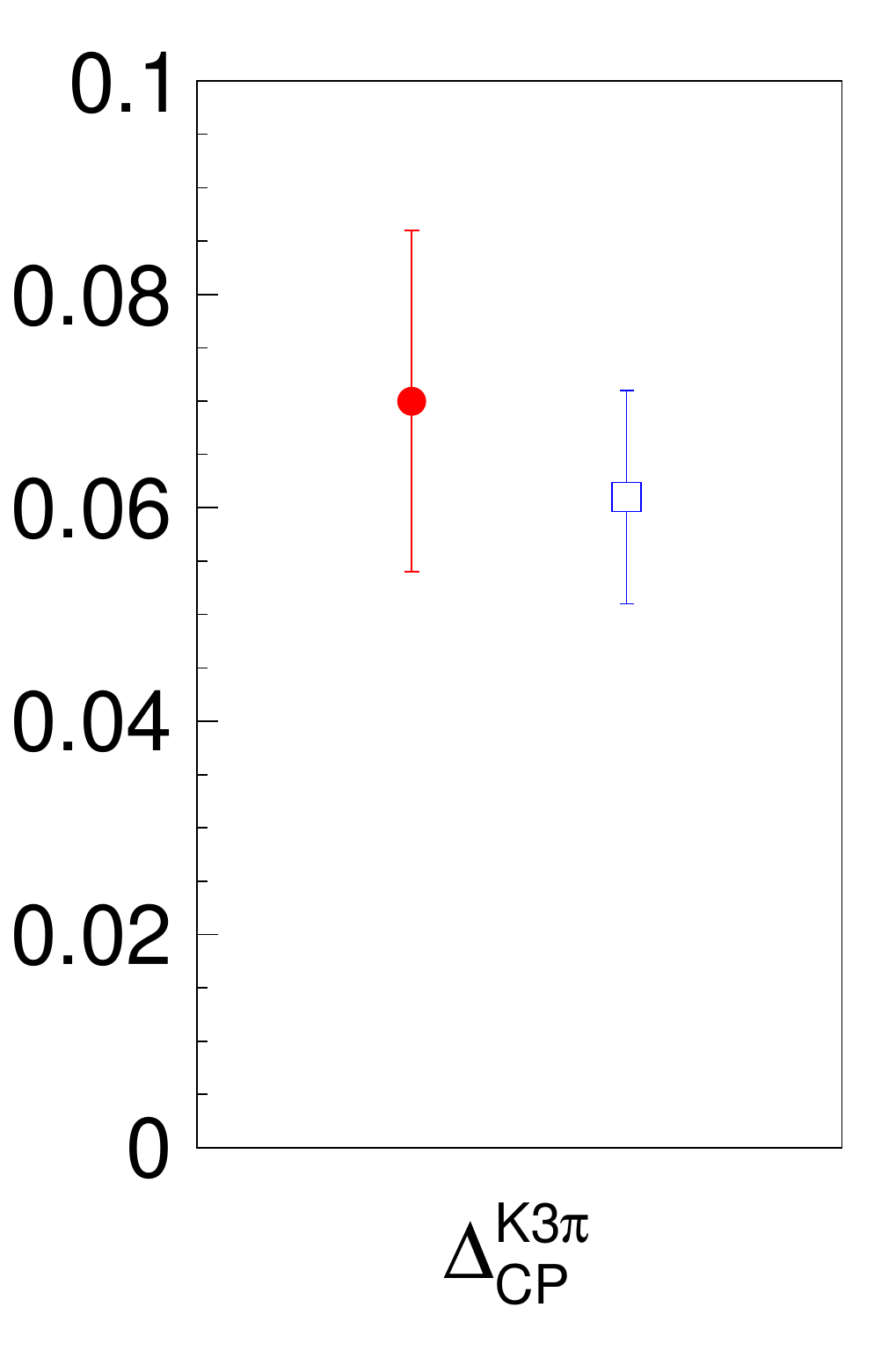}
\includegraphics[width=.22\textwidth]{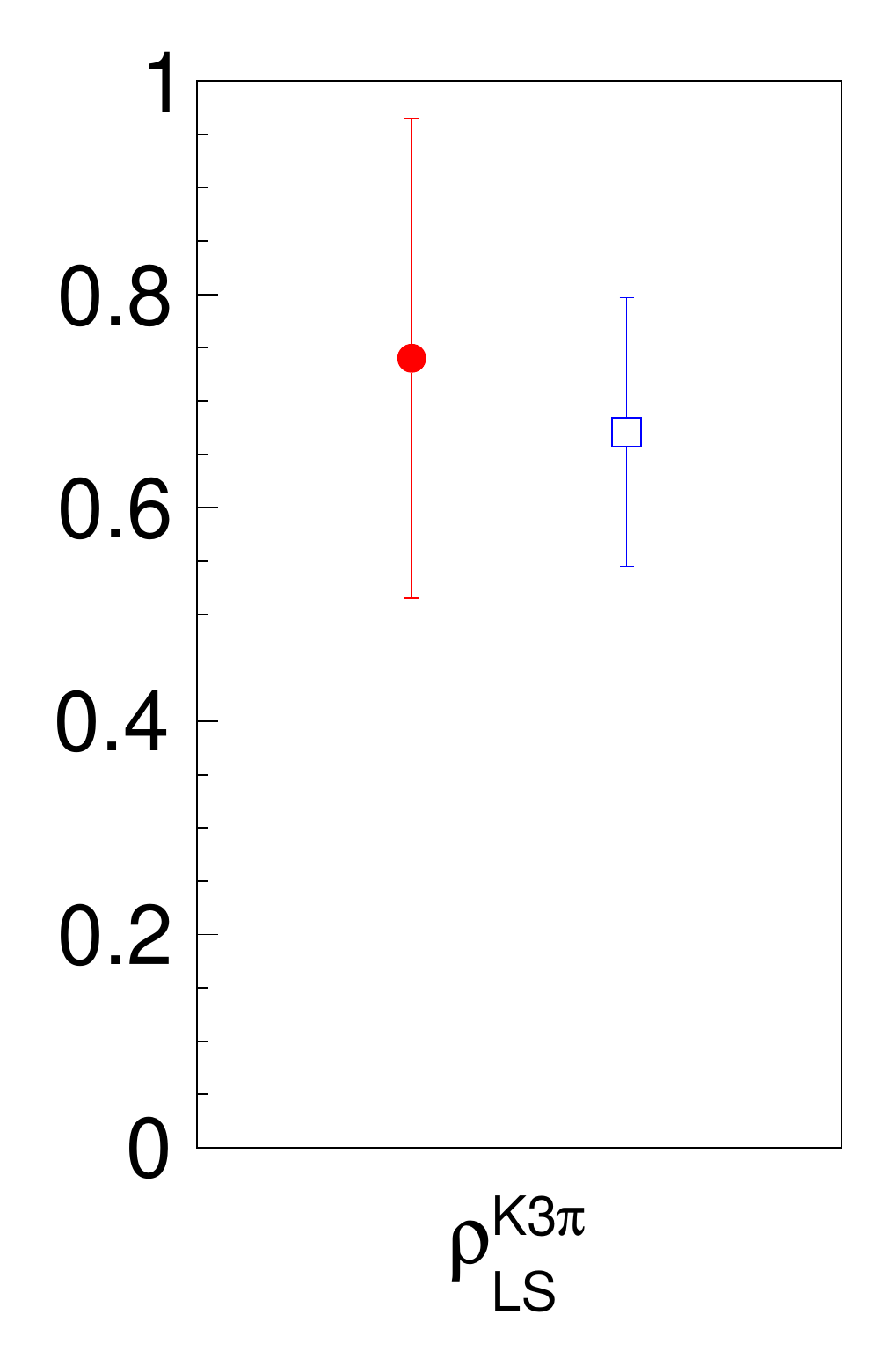}
\includegraphics[width=.22\textwidth]{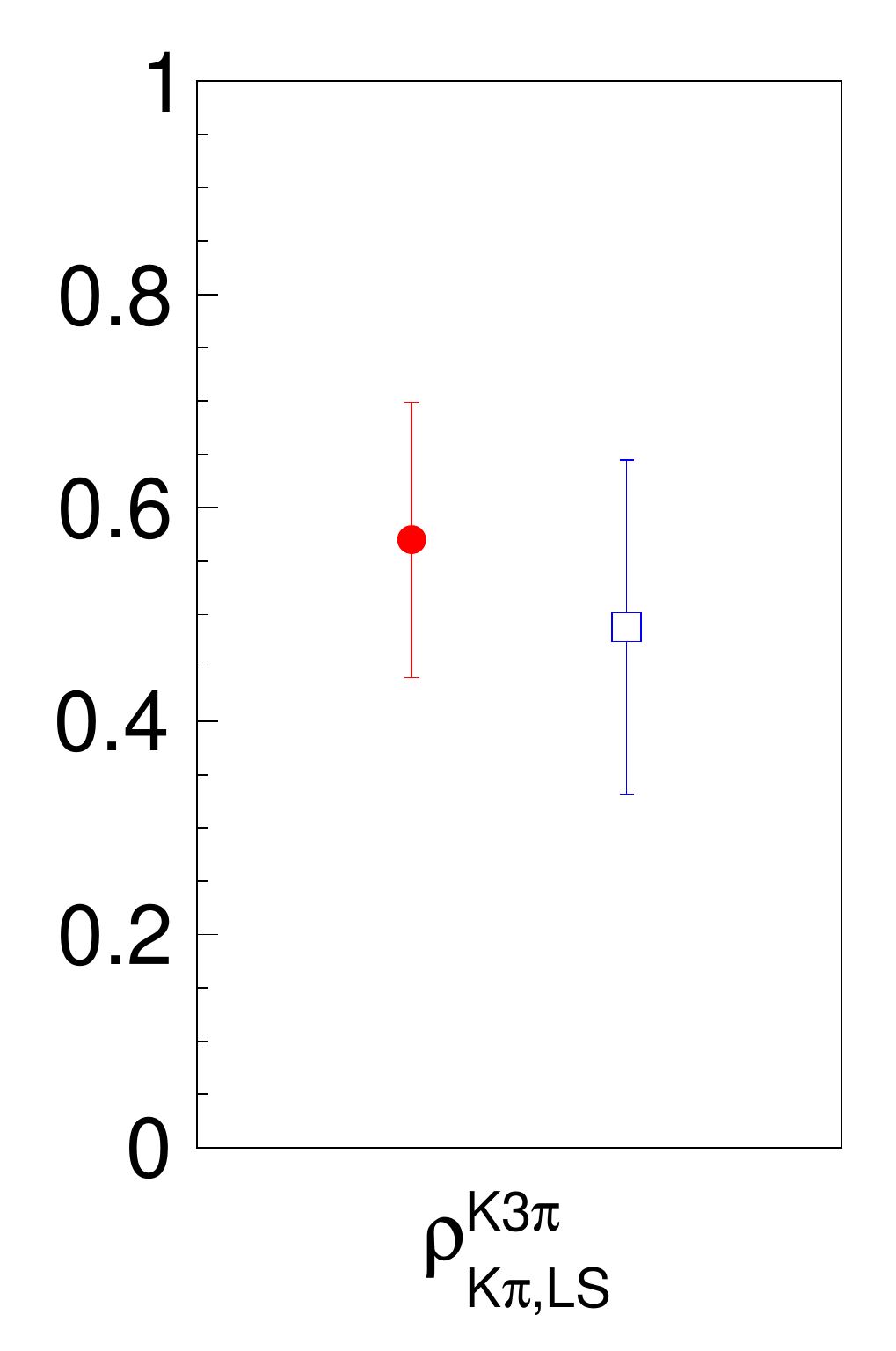}
\includegraphics[width=.22\textwidth]{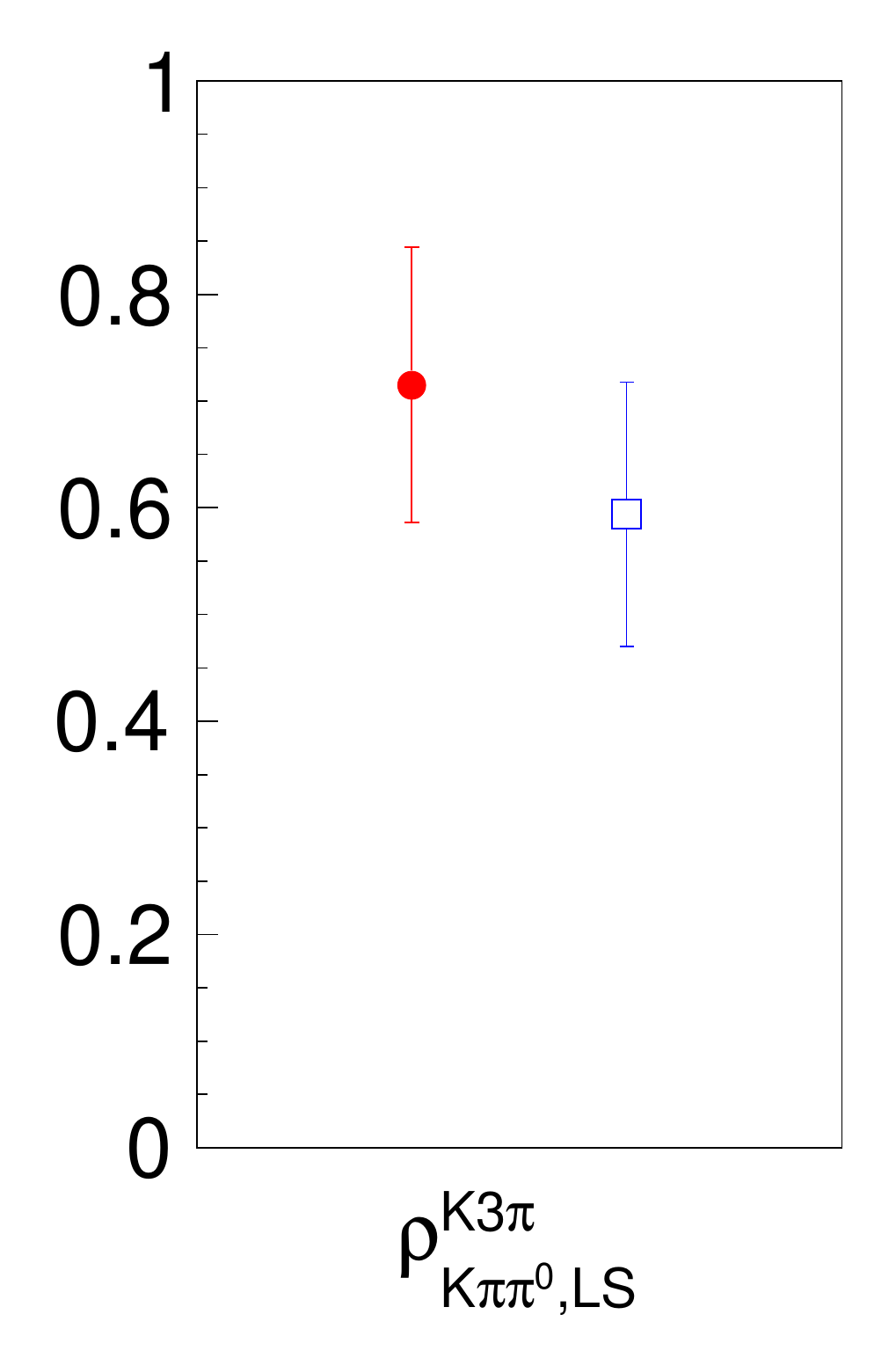}
\includegraphics[width=.22\textwidth]{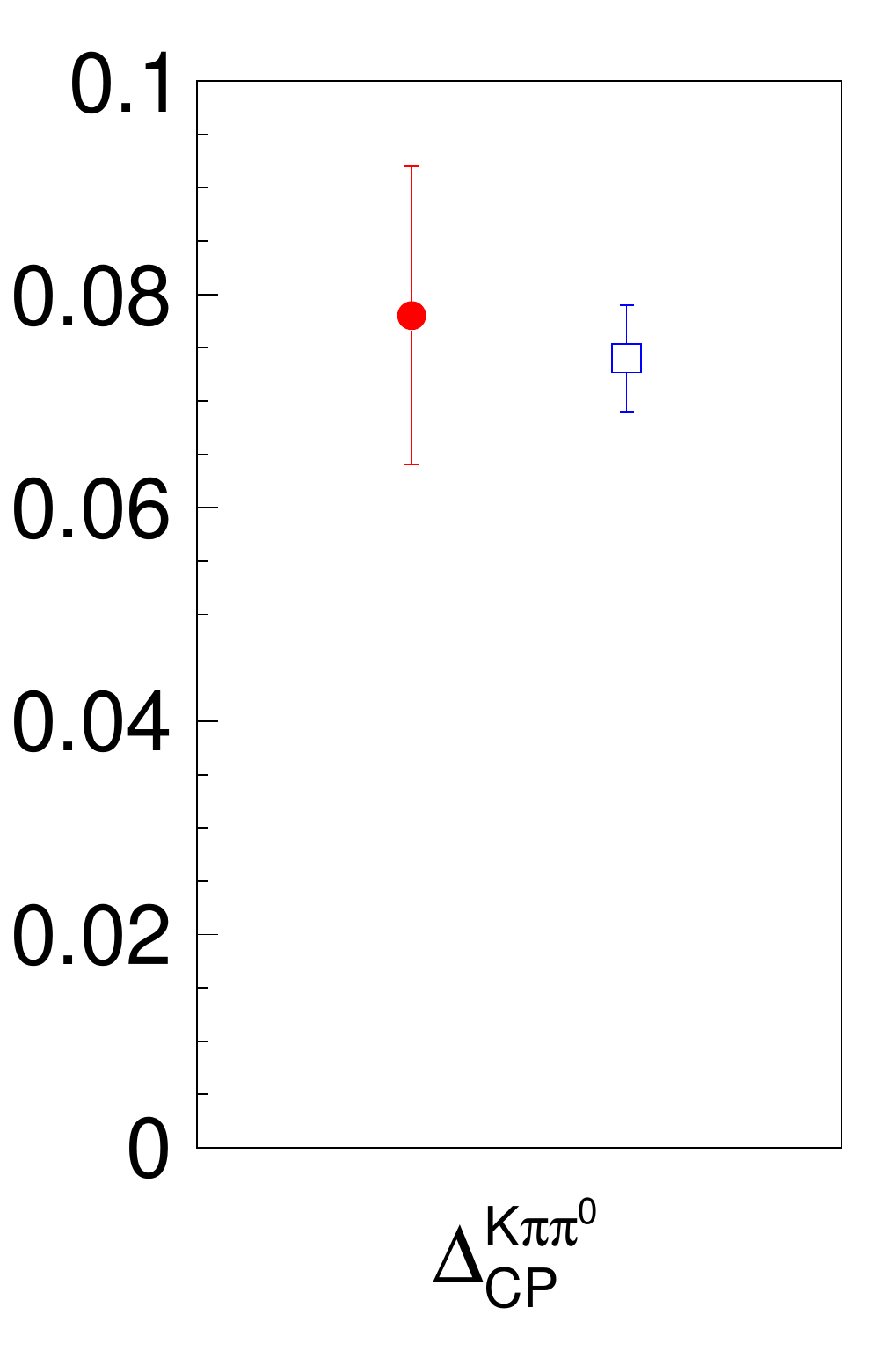}
\includegraphics[width=.22\textwidth]{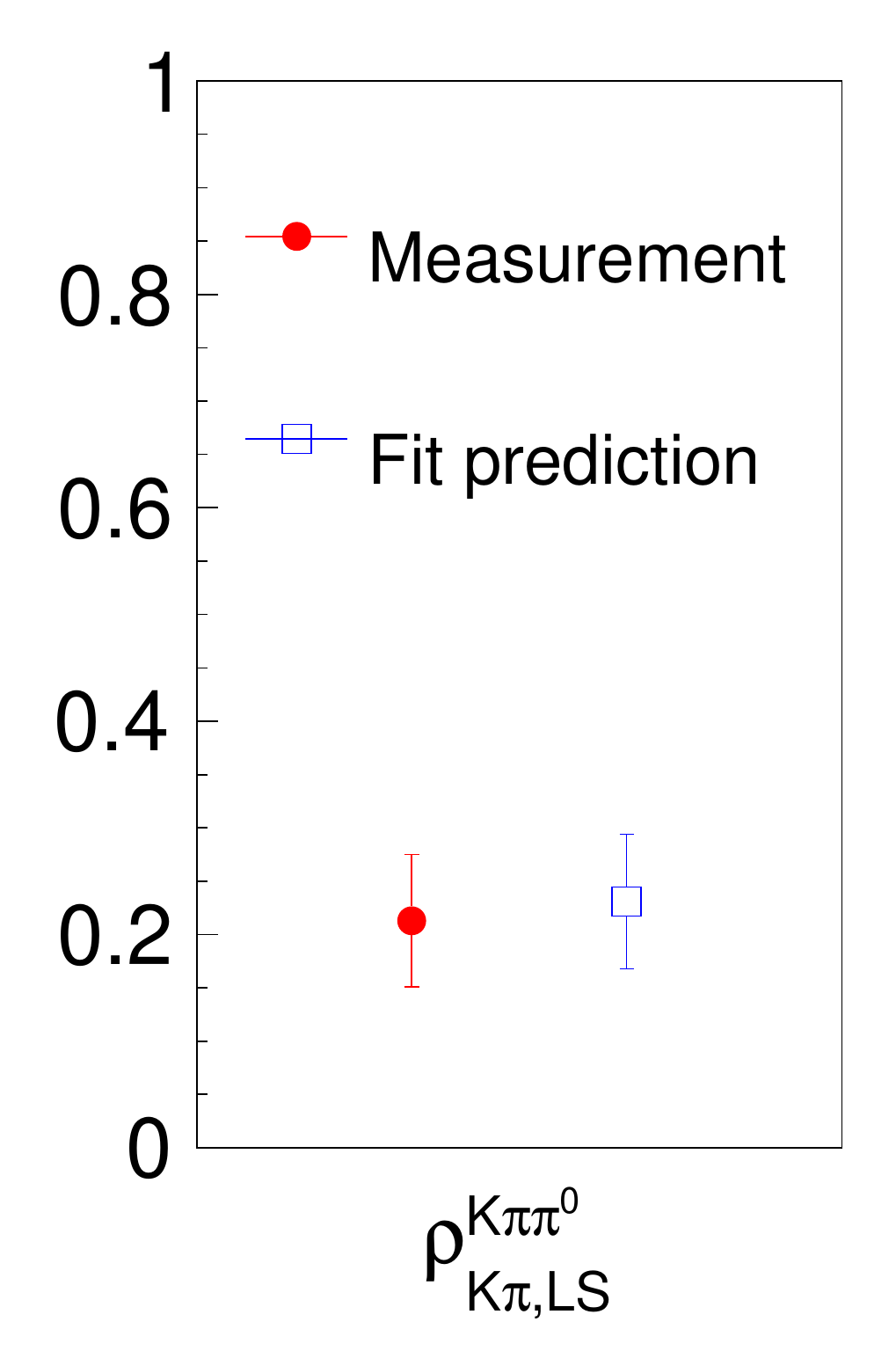}
\includegraphics[width=.22\textwidth]{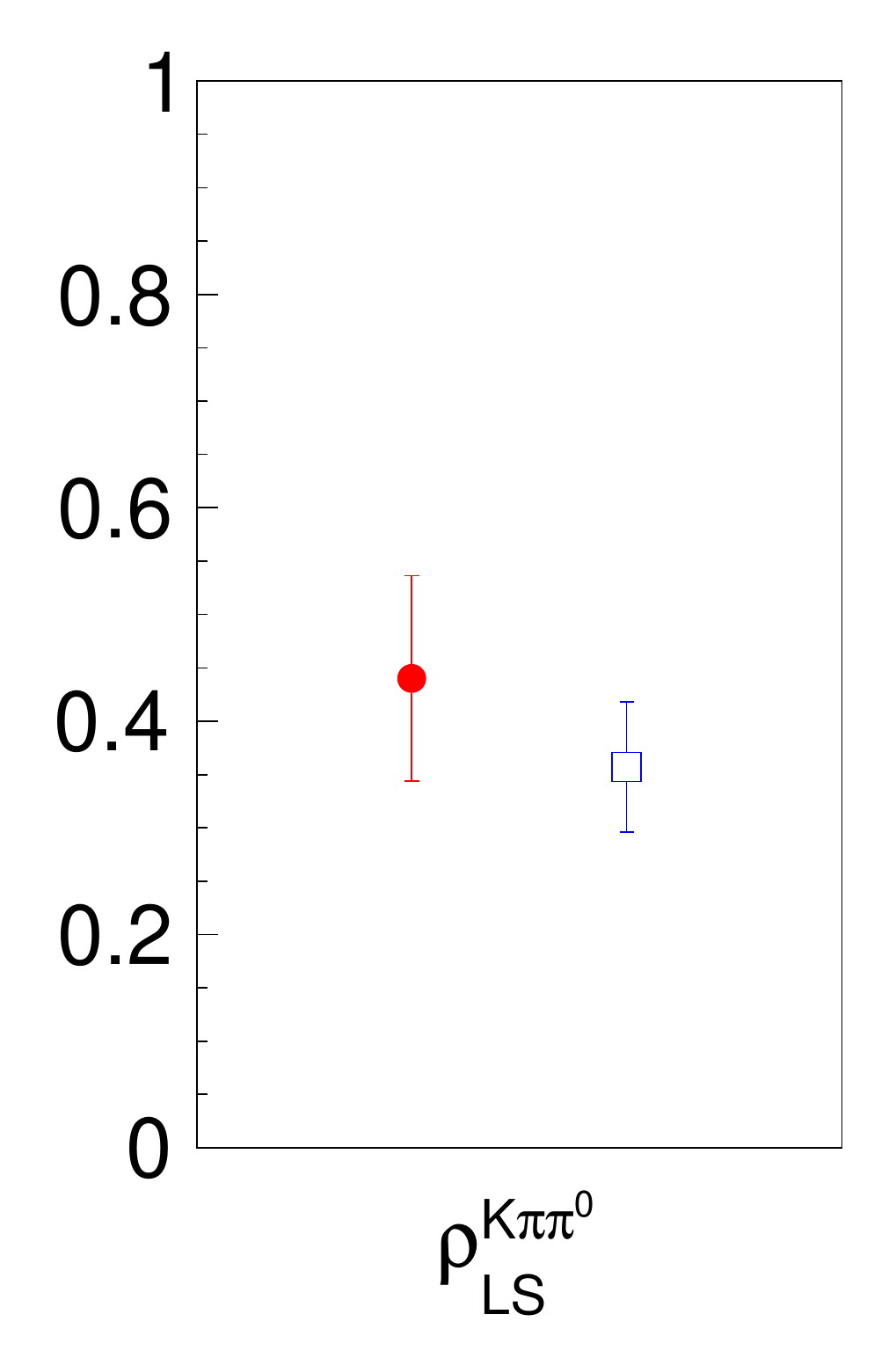}
\end{center}
{\caption{Results for the $C\!P$-tagged and like-sign global observables.  Also shown are  the predictions from the fit to these observables and the $\kspipi$ tags.}\label{fig:comparewithcleoc}}
\end{figure}

The observables are shown graphically in Fig.~\ref{fig:comparewithcleoc}.  The values of $\Delta^{K3\pi}_{C\!P}$ and $\Delta^{K\pi\pi^0}_{C\!P}$ are incompatible with zero,   and several of the like-sign observables are incompatible with unity, indicating the presence of significant effects from quantum correlations.  
The predictions from the global fit to the BESIII data, discussed below in Sec.~\ref{sec:globalhadronic}, are superimposed.

\begin{table}[!ht]
\caption{The measured $C\!P$-tagged and like-sign global observables for the modes $D \to \kmthreepi$ and $D \to \kmpipio$.  The first uncertainty is statistical and the second systematic.}\label{tab:rhoresults}
\begin{center}
\begin{tabular}{lc c lc}
\toprule
Observable  &  Value & & Observable & Value \\ \midrule
$\Delta^{K3\pi}_{C\!P}$ & $0.070 \pm 0.011 \pm 0.012$ & & $\Delta^{K\pi\pi^0}_{C\!P}$ & $0.078 \pm 0.007 \pm 0.012$ \\
$\rho^{K3\pi}_{LS}$           & $0.740 \pm 0.157 \pm 0.161$ & & $\rho^{K\pi\pi^0}_{LS}$ & $0.440 \pm 0.095 \pm 0.014$ \\
$\rho^{K3\pi}_{K\pi,LS}$      & $0.570 \pm 0.109 \pm 0.069$ & & $\rho^{K\pi\pi^0}_{K\pi,LS}$ & $0.213 \pm 0.062 \pm 0.004$ \\
$\rho^{K3\pi}_{K\pi\pi^0,LS}$ & $0.715 \pm 0.094 \pm 0.089$ & & & \\ 
\bottomrule
\end{tabular}
\end{center}
\end{table}

\subsubsection{Assignment of systematic uncertainties}
\label{sec:syst}

The systematic uncertainties reported in Table~\ref{tab:rhoresults} are estimated by repeatedly varying each input by a Gaussian distribution with a width set to the uncertainty of that component, and determining the shift in the central value. These shifts are then combined, taking account of any correlations that exist.

There are two uncertainties that are unique to the $\rho_{C\!P\pm}^{K3\pi}$ and $\rho_{C\!P\pm}^{K\pi\pi^0}$ observables. One arises from the limited size of the $C\!P$-tagged $D\to K^-\pi^+$ samples, and the other is associated with the selection efficiency.  The former is dominant for the individual measurements, while it is the selection-efficiency uncertainty that is more important in the evaluation of  $\Delta_{C\!P}^{K3\pi}$ and $\Delta_{C\!P}^{K\pi\pi^0}$. Consideration of Eq.~\ref{eq:kpinorm} indicates that it is necessary to know the ratio of selection efficiencies for the modes $D\to\kmthreepi$ and $D\to K^-\pi^+$, and $D \to \kmpipio$ and $D \to K^-\pi^+$.  These efficiencies are taken from Monte Carlo simulation, with uncertainties that are assigned from dedicated comparisons of data and Monte Carlo simulation~\cite{Ablikim:2018frk}. These uncertainties are $0.5\%$ for  the reconstruction of each charged track, $0.5\%$ for the identification of the pions and kaons, and $1\%$ for the $\pi^0$-reconstruction efficiency.
 These uncertainties are considered together with the smaller contribution arising from the finite size of the simulated data sets.

The like-sign observables are constructed by normalising the like-sign yields by the opposite-sign yields in the same final states, ensuring cancellation for most uncertainties associated with selection efficiencies.  However, the resonant sub-structure will in general be different for the like-sign events compared to the opposite-sign sample.  The consequence of this difference for $D \to \kmthreepi$ decays is estimated by re-weighting the Monte Carlo sample to the DCS model of LHCb, rather than the CF model~\cite{Aaij:2017kbo}, and re-evaluating the selection efficiency on $K^-\pi^+$-tagged signal events. A relative difference of 3\% is found, which is assigned as an uncertainty to all the like-sign $D\to\kmthreepi$ observables.  A similar study performed for $D\to\kmpipio$ decays, which makes use of the DCS model reported in~\cite{Aubert:2008zh}, leads to an uncertainty of $3.5\%$ for the corresponding observables associated with this mode.

Yields are determined from fits to the $M_{\rm BC}$ distribution of the signal decay.  In order to assign an uncertainty for this procedure, comparisons are made to the results when a two-dimensional fit is applied to the $M_{\rm BC}$ distributions of both the signal and tag decays, for the example the case where the tag mode is $D \to \kspio$.  Small differences are observed and uncertainties of $1.5\%$ and $0.8\%$ are assigned to $\rho^{K3\pi}_{C\!P\pm}$ and $\rho^{K\pi\pi0}_{C\!P\pm}$, respectively. No uncertainty is applied for the like-sign observables, as any bias cancels between the like-sign yields and the same-topology opposite-sign yields used for normalisation.

The most important source of background is the decay $D\to \ks K^-\pi^+$, particularly in the case of the like-sign double tags. The uncertainty on this background contribution is assigned by accounting for the knowledge of its coherence factor and mean strong-phase difference, measured to be $0.70 \pm 0.08$ and $(0.1 \pm 15.7)^\circ$, respectively~\cite{Insler:2012pm}. These inputs are required to correct for quantum-correlation in the decay rate.  Uncertainties on the branching fraction~\cite{2018Tanabashi} and the selection efficiency are also considered.
The uncertainty on this background component is the dominant systematic uncertainty for the like-sign tags.

A final source of uncertainty on the $C\!P$-tag and like-sign observables is the knowledge on the input parameters, which is taken from the measurements reported in Table~\ref{tab:inputs}.

When analysing the efficiency-corrected signal yields  tagged by $D\to\kspipi$ decays in bins of phase space, $Y^{K3\pi}_i$ and $Y^{K\pi\pi^0}_i$, it is only necessary to have control of the relative efficiency variation. This variation is less than 15\% and taken from Monte Carlo simulation.  It is assumed that any biases on these efficiency corrections are negligible. Small uncertainties are assigned  associated with the finite size of the simulation samples.  Fits to the expected distribution of events must take account of the uncertainties on the $D\to\kspipi$ strong-phase parameters, found in Ref.~\cite{Ablikim:2020yif}, and $K_i$ parameters, given in Table~\ref{tab:kspipifraction}.

\subsubsection{Fit to the hadronic parameters}
\label{sec:globalhadronic}

A $\chi^2$ fit is performed to the complete set of $C\!P$- (Tables~\ref{tab:cp_k3pi} and~\ref{tab:cp_kpipi0}), like-sign (Table~\ref{tab:rhoresults}) and $\kspipi$-tagged observables (Table~\ref{tab:kspipiglobal}) in order to determine the underlying physics parameters: $R_{K3\pi}$, $\delta_D^{K3\pi}$, $r_D^{K3\pi}$, $R_{K\pi\pi^0}$, $\delta_D^{K\pi\pi^0}$ and $r_D^{K\pi\pi^0}$.
In this fit each individual $C\!P$-tag result is entered as a separate measurement.  In total, therefore, there are 65 observables and six free fit parameters. 
The systematic uncertainties from external inputs are not included in the uncertainties of the observables. 
Instead, the auxiliary parameters $\delta_D^{K\pi}$, $r_D^{K\pi}$, $x$, $y$, $F_+^{\pi\pi\pi^0}$, $K_i$, $c_i$ and $s_i$ are also fitted, with Gaussian constraints introduced into the $\chi^2$ function according to their measured values and covariances.  When terms involve the ratio of branching fractions $\br{} (D^0 \to  K^+\pi^-)/\br{} (D^0 \to K^-\pi^+)$,
$\br{} (D^0 \to \kpthreepi) / \br{} (D^0 \to \kmthreepi)$ or $\br{} (D^0 \to \kppipio) / \br{} (D^0 \to \kmpipio)$, these ratios are replaced by the corresponding theoretical expressions (see Eqs.~\ref{eq:BRCF} and~\ref{eq:BRDCS}). Additional constraints are added to the $\chi^2$ function relating the measured values  
$\br{} (D^0 \to \kpthreepi) / \br{} (D^0 \to \kmthreepi)$ and $\br{} (D^0 \to \kppipio) / \br{} (D^0 \to \kmpipio)$ to the theoretical predictions. In summary, therefore, the function that is minimised is
\begin{equation}
\chi^2 = \chi^2_{C\!P} + \chi^2_{\rm LS} + \chi^2_{K^0_S \pi\pi} + \chi^2_{\rm aux}\, ,
\label{eq:chi2fn}
\end{equation}
where $\chi^2_{C\!P}$, $\chi^2_{\rm LS}$ and $\chi^2_{K^0_S \pi\pi}$ are the contributions from the $C\!P$ tags, like-sign tags and $\kspipi$ tags, respectively, and $\chi^2_{\rm aux}$ contains the constrained contributions from the auxiliary parameters and the ratios of branching fractions.
Studies performed with simulated data sets demonstrate that the fit is unbiased and returns correctly estimated uncertainties.

\begin{figure}[!htb]
    \centering
\includegraphics[width=.48\textwidth]{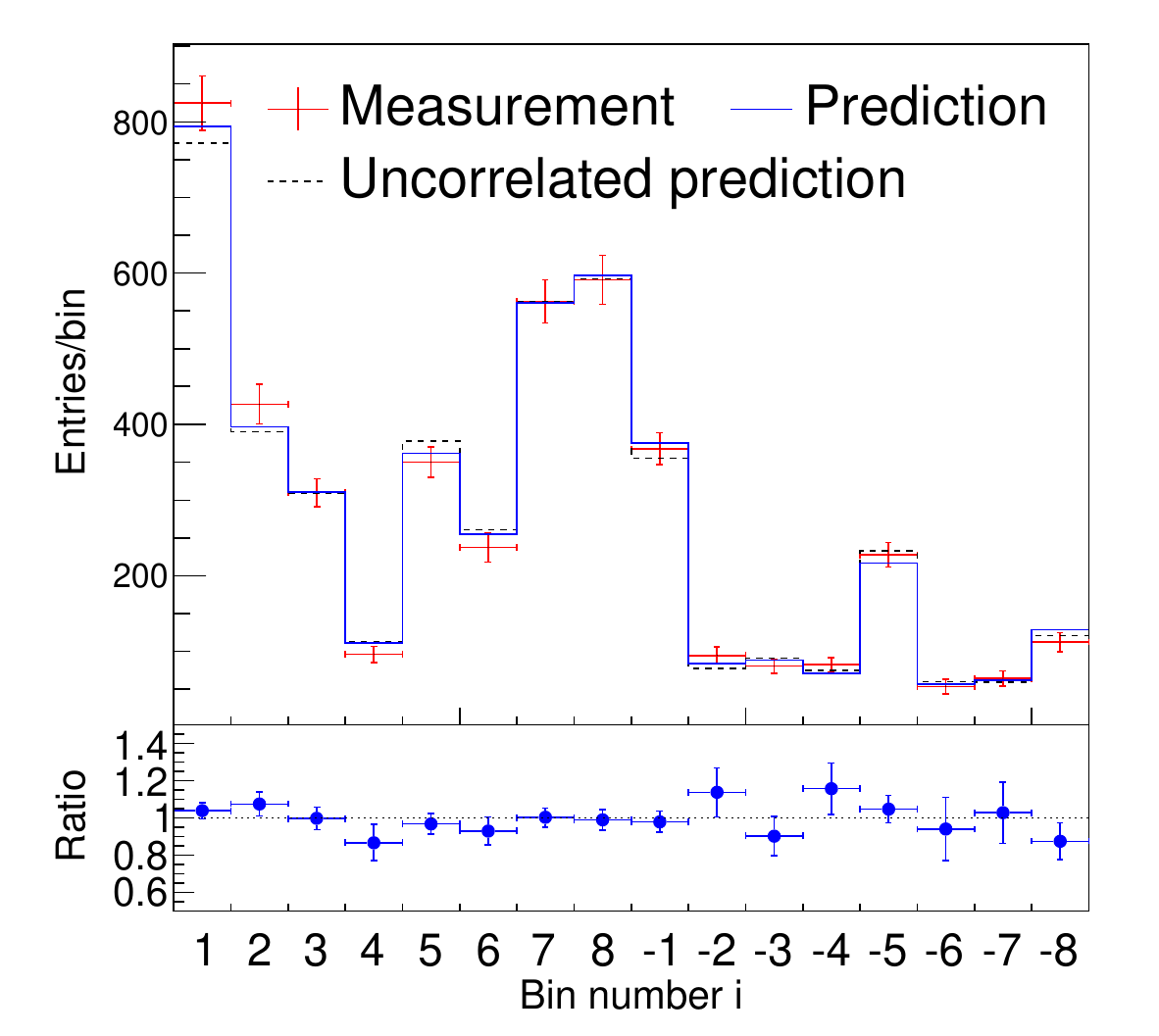}
\includegraphics[width=.48\textwidth]{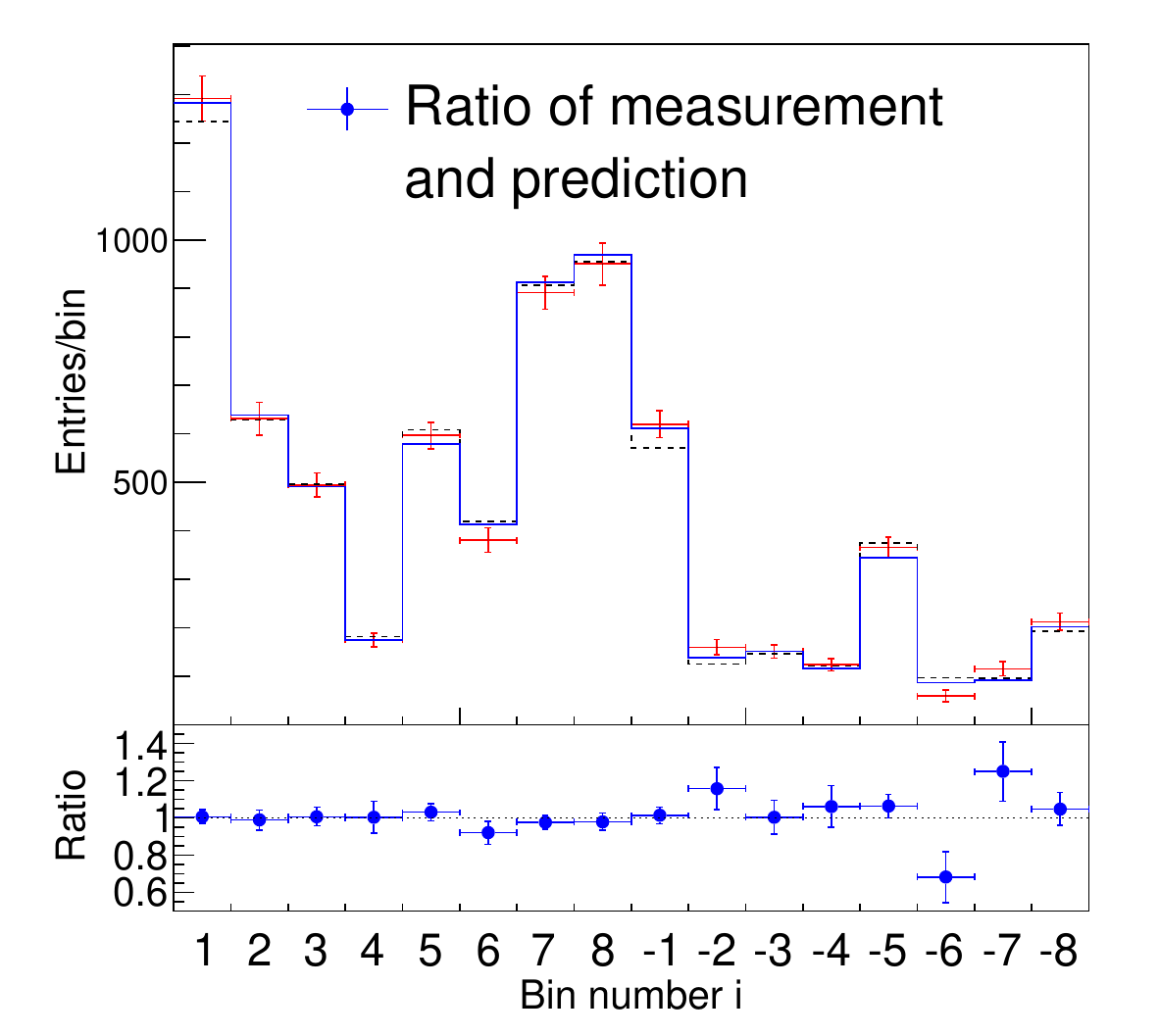}
    \caption{The $Y^{K3\pi}_i$ (left) and $Y^{K\pi\pi^0}_i$ (right) observables, with the predictions from the fit to all the BESIII inputs. Also shown are the expectations from the uncorrelated hypothesis.}
    \label{fig:kspipi_results}
\end{figure}

The fit converges with a $\chi^2/{\rm n.d.f.}$ of 61/59, indicating compatibility between the inputs.  Figure~\ref{fig:kspipi_results} shows the $Y^{K3\pi}_i$ and $Y^{K\pi\pi^0}_i$ observables with the fit results superimposed. Also shown are the predictions from the uncorrelated hypothesis, which are directly proportional to the $K_i$ values.

\begin{figure}[!ht]
\begin{center}
\includegraphics[width=.45\textwidth]{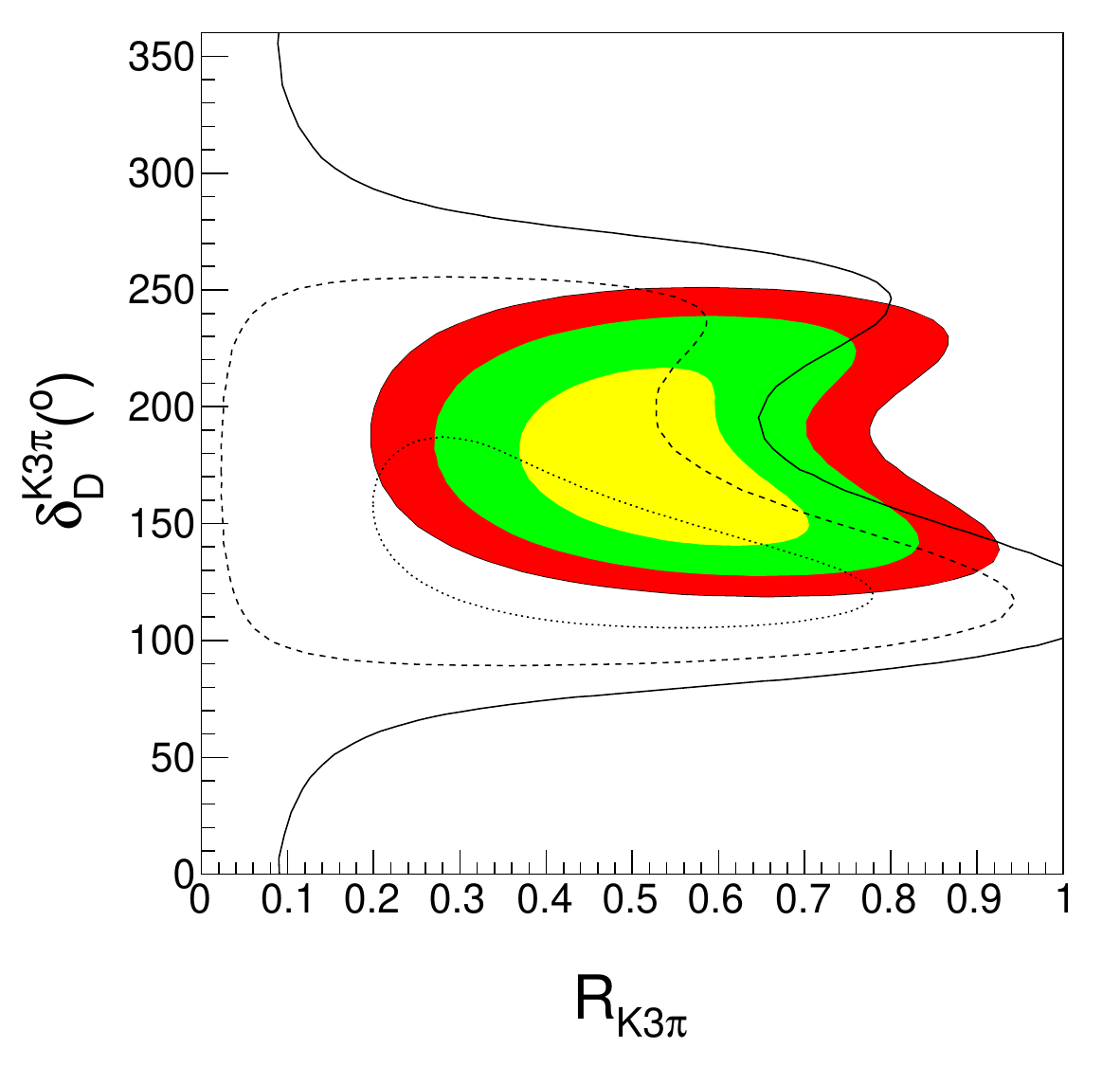}
\includegraphics[width=.45\textwidth]{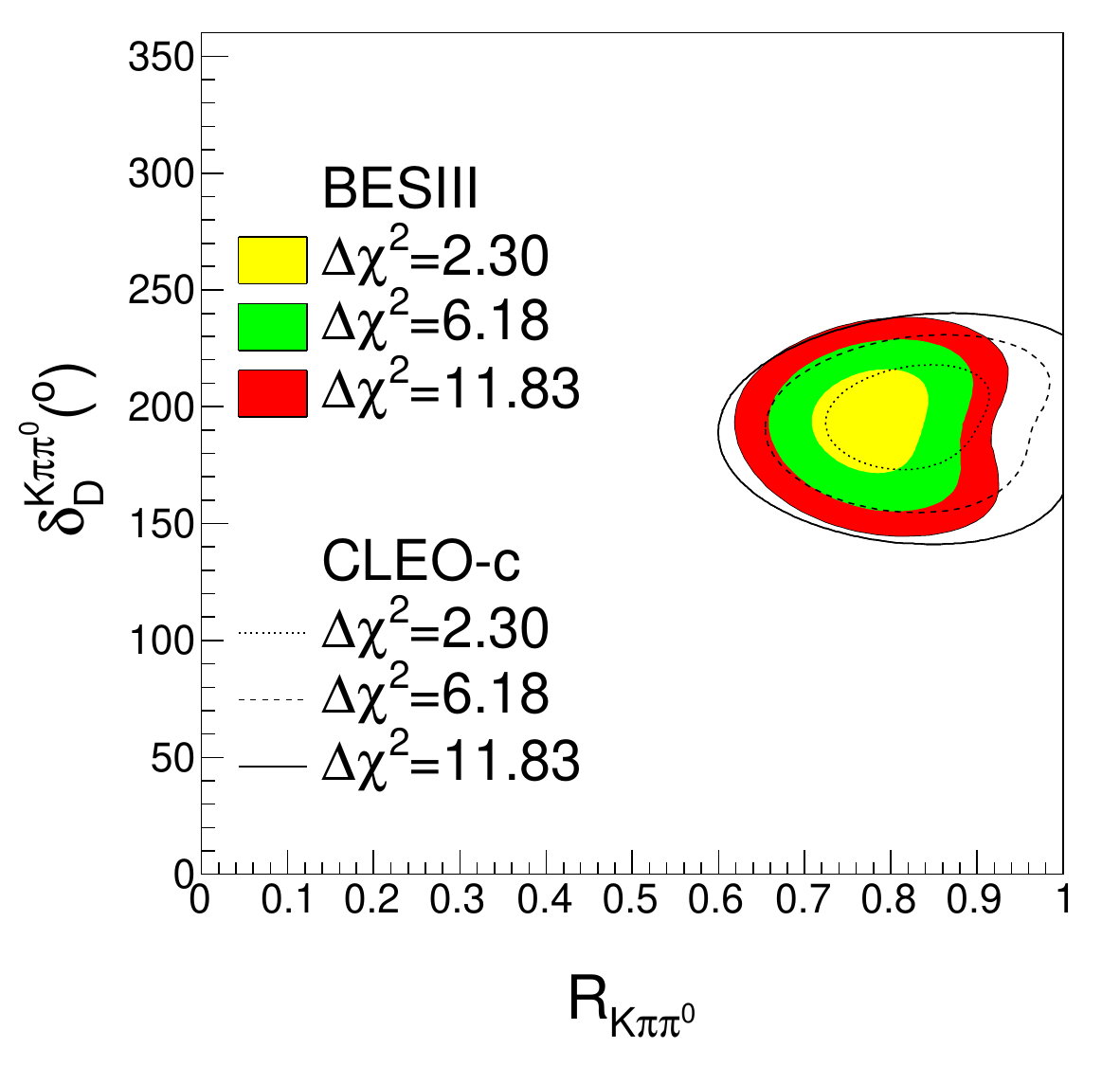}
\caption{Scans of $\Delta \chi^2$ in the global ($\Rkthreepi$, $\dkthreepi$) and ($\Rkpipio$, $\dkpipio$) parameter space, showing the $\Delta \chi^2$=2.30, 6.18, 11.83 intervals, which correspond to $68.3\%$, $95.4\%$ and $99.7\%$ confidence levels in the two-dimensional parameter space. 
Also shown are the equivalent contours determined from the CLEO-c data~\cite{Evans:2016tlp}.}
 \label{fig:parameter1_1sigma}
\end{center}
\end{figure}

The results for the hadronic parameters are given in Table~\ref{tab:parameter} (with the accompanying correlation matrix available in Appendix~\ref{sec:appendix1}), and $\Delta \chi^2$ scans  in $(\Rkthreepi,\dkthreepi)$ and $(\Rkpipio,\dkpipio)$ parameter space are presented in Fig.~\ref{fig:parameter1_1sigma}. 
The measurement of $\Rkthreepi$ is in agreement with the value of $0.46$ predicted by the LHCb models~\cite{Aaij:2017kbo}. These results are also compatible with those obtained from CLEO-c data~\cite{Evans:2016tlp}, with an improvement in the $1\sigma$ uncertainties for the coherence factors.~\footnote{It should be noted that plots in previous publications using CLEO-c data~\cite{Lowery:2009id,Libby:2014rea,Evans:2016tlp} showed contours corresponding to the intervals $\Delta \chi^2 =1,4$ and $9$. Here the CLEO-c contours have been re-drawn according to the choice of $\Delta \chi^2$ intervals adopted in this paper, which are appropriate to indicate $1\sigma$, $2\sigma$ and $3\sigma$ coverage in a two-dimensional plane.}  The region of $(\Rkthreepi,\dkthreepi)$ parameter space encompassed by the $2\sigma$ and $3\sigma$ confidence intervals is significantly more constrained.  Combined fits of the BESIII data, together with the CLEO-c and LHCb observables are reported in Appendix~\ref{sec:appendix2}.  The fit results for the auxiliary parameters are all compatible with their measured values, and in no case is their precision significantly improved by the fit.

\begin{table}[!ht]
\caption{Fitted central values for the hadronic parameters, from the global and binned analyses.   The uncertainties include both statistical and systematic contributions.}
\label{tab:parameter}
\begin{center}
\begin{tabular}{lccccccc}
\toprule
Parameter       & \hspace*{0.3cm}Global fit\hspace*{0.3cm}   & \multicolumn{4}{c}{Binned fit} \\
	&   &Bin 1 &Bin 2 & Bin 3 &Bin 4 \\
\midrule
$\Rkthreepi$    	&0.52$_{-0.10}^{+0.12}$ 	 &0.58$^{+0.25}_{-0.33}$ &0.78$^{+0.50}_{-0.21}$	&0.85$^{+0.15}_{-0.12}$ &0.45$^{+0.33}_{-0.37}$\\ \\
$\dkthreepi$  	&$\left(167_{-19}^{+31}\right)^\circ$ &$\left(131^{+124}_{-16}\right)^\circ$ &$\left(150^{+37}_{-39}\right)^\circ$  &$\left(176^{+57}_{-21}\right)^\circ$ &$\left(274^{+19}_{-30}\right)^\circ$\\ \\
$\rkthreepi$ ($\times 10^{-2}$) &5.46$\pm$0.09   &5.44$^{+0.45}_{-0.14}$ &5.80$^{+0.14}_{-0.13}$ &5.75$^{+0.41}_{-0.14}$ &5.09$^{+0.14}_{-0.14}$\\ \\
$\Rkpipio$    	&0.78$\pm$0.04  	&\multicolumn{4}{c}{0.80$\pm$0.04}		\\ \\
$\dkpipio$  	&$\left(196_{-15}^{+14}\right)^\circ$  &\multicolumn{4}{c}{$\left(200\pm 11 \right)^\circ$} \\ \\
$\rkpipio$ ($\times 10^{-2}$)  &4.40$\pm$0.11 &\multicolumn{4}{c}{4.41$\pm$0.11} \\
\bottomrule
\end{tabular}
\end{center}
\end{table}

It is instructive to estimate the intrinsic statistical precision of the sample, and the contribution of each source of systematic bias to the overall uncertainty.  Therefore, the fit is re-performed many times, with each input varied in turn according to a Gaussian distribution with width set to its assigned uncertainty, and all the other inputs fixed.  The width of the distribution of the fitted parameters is taken as an estimate of the uncertainty associated with the varying input.  The results are presented in Table~\ref{tab:inputsys}.  The most important contributions are seen to come from the finite size of the $C\!P$-tagged $D \to K^-\pi^+$ samples, the uncertainty on the $D \to \ks K^-\pi^+$ background, and on the knowledge of the $D\to\kspipi$ strong-phase and $K_i$ parameters.  The statistical uncertainty is dominant for all four measurements.

\begin{table}[!htbp]
\caption{Indicative sizes of the systematics uncertainties on the global hadronic parameters. First are listed those components associated with the detector and measurement procedure, and then those associated with the external parameters.  Also shown is the statistical uncertainty for comparison.}
\label{tab:inputsys}
\begin{center}
\begin{tabular}{lrrrr}
\toprule
 Systematics & $\Rkthreepi$ &$\dkthreepi$ & \hspace*{0.4cm}$\Rkpipio$\hspace*{0.0cm} &$\dkpipio$\\
\midrule
Size of $C\!P$-tagged $D\to K^-\pi^+$ samples  &$0.04$  &$7.0^\circ$ &$0.02$ &$6.9^\circ$\\
$K/\pi$ tracking and identification &$0.02$  &$3.8^\circ$ &$<0.01$ &$2.3^\circ$\\
$\pi^0$ reconstruction &$<0.01$  &$<0.1^\circ$ &$<0.01$ &$<0.1^\circ$\\
Impact of resonance modelling on efficiency &$<0.01$  & $2.5^\circ$ &$<0.01$ &$0.4^\circ$\\
Size of Monte Carlo samples & $0.01$  &$1.5^\circ$ &$<0.01$ &$1.3^\circ$\\
$D \to K_S^0K^-\pi^+$ background &$0.05$  &$1.0^\circ$ &$0.01$ &$4.6^\circ$\\
Fit method for signal yields &$0.02$  &$3.4^\circ$ & $<0.01$ &$1.1^\circ$\\	
\midrule
$c_i$, $s_i$ & $^{+0.01}_{-0.00}$ &$3.0^\circ$ & $<0.01$ &$\left(^{+0.6}_{-0.7}\right)^\circ$\\
$K_i$  &  $0.01$ &$\left(^{+6.7}_{-6.1}\right)^\circ$ &  $0.01$ &$\left(^{+3.1}_{-4.4}\right)^\circ$\\
$\br{}(\Dz\rightarrow S)$, with $S=\kmthreepi$ and $\kmpipio$  & $0.01$ & $\left(^{+1.7}_{-1.5}\right)^\circ$ & $0.01$ &$\left(^{+3.4}_{-2.2}\right)^\circ$\\
${\br{}(\Dz\rightarrow \bar{S})}/{\br{}(\Dz\rightarrow S)}$  & $^{+0.02}_{-0.01}$ &$2.7^\circ$ & $<0.01$ &$0.2^\circ$\\
$\br{}(\Dz\rightarrow K^-\pi)$  & $0.01$ &$\left(^{+0.8}_{-1.2}\right)^\circ$ & $<0.01$ &$\left(^{+0.9}_{-0.7}\right)^\circ$\\
$\rkpi$ &$<0.01$ &$\left(^{+0.2}_{-0.1}\right)^\circ$ &$<0.01$ &$0.2^\circ$\\
$\dkpi$  & $<0.01$ &  $<0.1^\circ$ &  $<0.01$ &  $<0.1^\circ$ \\
$x$, $y$		   &$<0.01$  &$\left(^{+1.0}_{-1.1}\right)^\circ$ &$<0.01$  &$0.5^\circ$\\
$F^+_{\pi\pi\pi^0}$ &$<0.01$ &$\left(^{+0.3}_{-0.4}\right)^\circ$ &$<0.01$ &$0.1^\circ$\\
\midrule
Statistical & $^{+0.08}_{-0.09}$ & $\left(^{+29.3}_{-18.7}\right)^\circ$ & $0.04$ &$\left(^{+10.6}_{-12.6}\right)^\circ$\\
\bottomrule
\end{tabular}
\end{center}
\end{table}

\subsection{Binned ${ D \to \kmthreepi}$ analysis}

The binned analysis proceeds in an identical manner to the global case. In this case there are 170 observables and 15 free fit parameters.
Because the binning scheme is constructed to exclude 
$D \to \ks K^-\pi^+$ background there is no uncertainty from this source.  
In Monte Carlo simulation it is found that around 90\% of decays are assigned to the correct bin.  A migration matrix, determined from simulation, is used to correct for incorrect assignments. Fits to ensembles of simulated experiments confirm that the results are unbiased and assigned reliable uncertainties.
 The measured values of the observables are presented in Table~\ref{tab:rhoresults_binned}. The accompanying correlation matrix may be found in Appendix~\ref{sec:appendix1}.


\begin{table}[!ht]
\caption{The measured binned observables for $D \to \kmthreepi$.  The first uncertainty is statistical and the second systematic.}\label{tab:rhoresults_binned}
\begin{center}
\scriptsize
\begin{tabular}{lrc c cc}
\toprule
\multicolumn{2}{l}{Observable}  &  Bin 1 &  Bin 2 & Bin 3 & Bin 4 \\ \midrule
\multicolumn{2}{l}{$\Delta^{K3\pi}_{C\!P}$} & $0.073 \pm 0.018 \pm 0.019$ & $0.065 \pm 0.023 \pm 0.019$ & $0.108 \pm 0.022 \pm 0.020$ & $0.002 \pm 0.015 \pm 0.019$\\
\multicolumn{2}{l}{$\rho^{K3\pi}_{K\pi,LS}$}      & $0.853 \pm 0.306 \pm 0.033$ & $0.207 \pm 0.161 \pm 0.008$  & $0.348 \pm 0.151 \pm 0.014$ & $0.901 \pm 0.218 \pm 0.036$ \\
\multicolumn{2}{l}{$\rho^{K3\pi}_{K\pi\pi^0,LS}$} & $0.941 \pm 0.190 \pm 0.037$ & $0.608 \pm 0.150 \pm 0.024$  & $0.485 \pm 0.283 \pm 0.019$  & $0.742 \pm 0.190 \pm 0.030$ \\
\multirow{4}{*}{$\rho^{K3\pi}_{LS}$} & Bin 1
& $0.569 \pm 0.399 \pm 0.023$ & $3.356 \pm 1.271 \pm 0.135$ & $0.459 \pm 0.604 \pm 0.018$ & $2.413 \pm 0.925 \pm 0.097$ \\
 & Bin 2 & & $0.270 \pm 0.413 \pm 0.011$ & $1.037 \pm 0.775 \pm 0.042$  & $1.072 \pm 0.668	 \pm 0.043$ \\
 & Bin 3 & & & $0.274 \pm 0.437 \pm 0.011$ & $2.518 \pm 0.926 \pm 0.101$ \\
 & Bin 4 & & & & $1.330 \pm 0.821 \pm 0.054$ \\ 
\bottomrule
\end{tabular}
\end{center}
\end{table}

The results for the fit to hadronic parameters are given in Table~\ref{tab:parameter} (with the correlation matrix in Appendix~\ref{sec:appendix1}), and $\Delta \chi^2$ scans in $(\Rkthreepi,\dkthreepi)$ space are shown in Fig.~\ref{fig:binparameter}. The fit quality, with $\chi^2/{\rm n.d.f.} = 180/155$, is satisfactory. In Appendix~\ref{sec:appendix2} may be found the results for a combined fit to the BESIII and CLEO-c data.

The amplitude models may be used to calculate predictions for the coherence factor in each bin, and the variation in strong-phase between bins~\cite{Evans:2019wza}. By making use of the measured value of $\dkthreepi$ from the global analysis it is then possible to obtain an effective prediction of the average strong-phase difference bin-by-bin, and correlated uncertainty. Following this procedure the predicted values of the coherence factors and strong-phase differences are found to be $\left(0.67,(95^{+31}_{-19})^\circ\right)$, $\left(0.85,(147^{+31}_{-19})^\circ\right)$, $\left(0.82,(188^{+31}_{-19})^\circ\right)$ and $\left(0.63,(247^{+31}_{-19})^\circ\right)$, for bins 1 to 4, respectively.  The measurements and the predictions are compatible.


\begin{figure}[!ht]
\centering
\includegraphics[width=.45\textwidth]{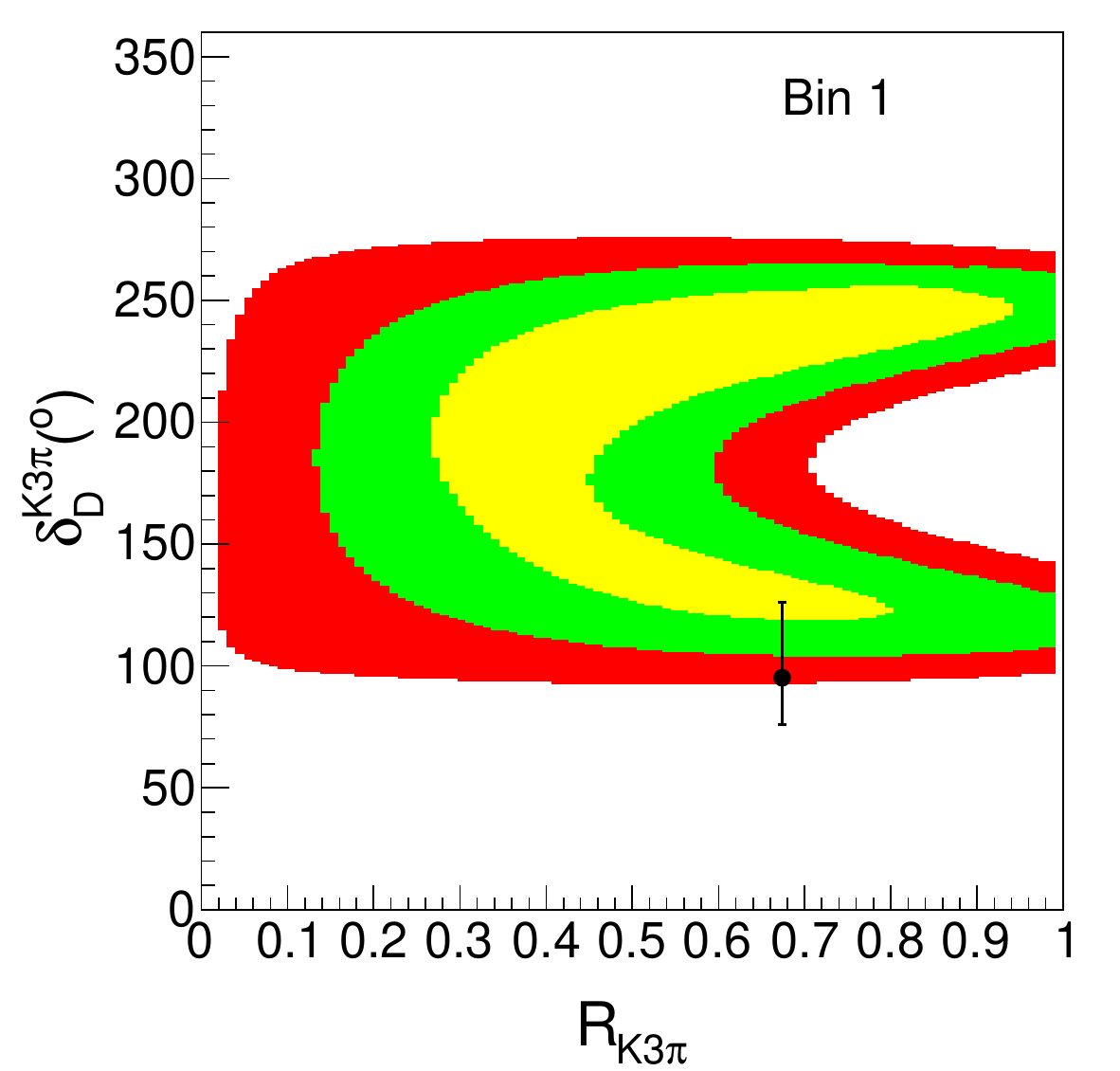}
\includegraphics[width=.45\textwidth]{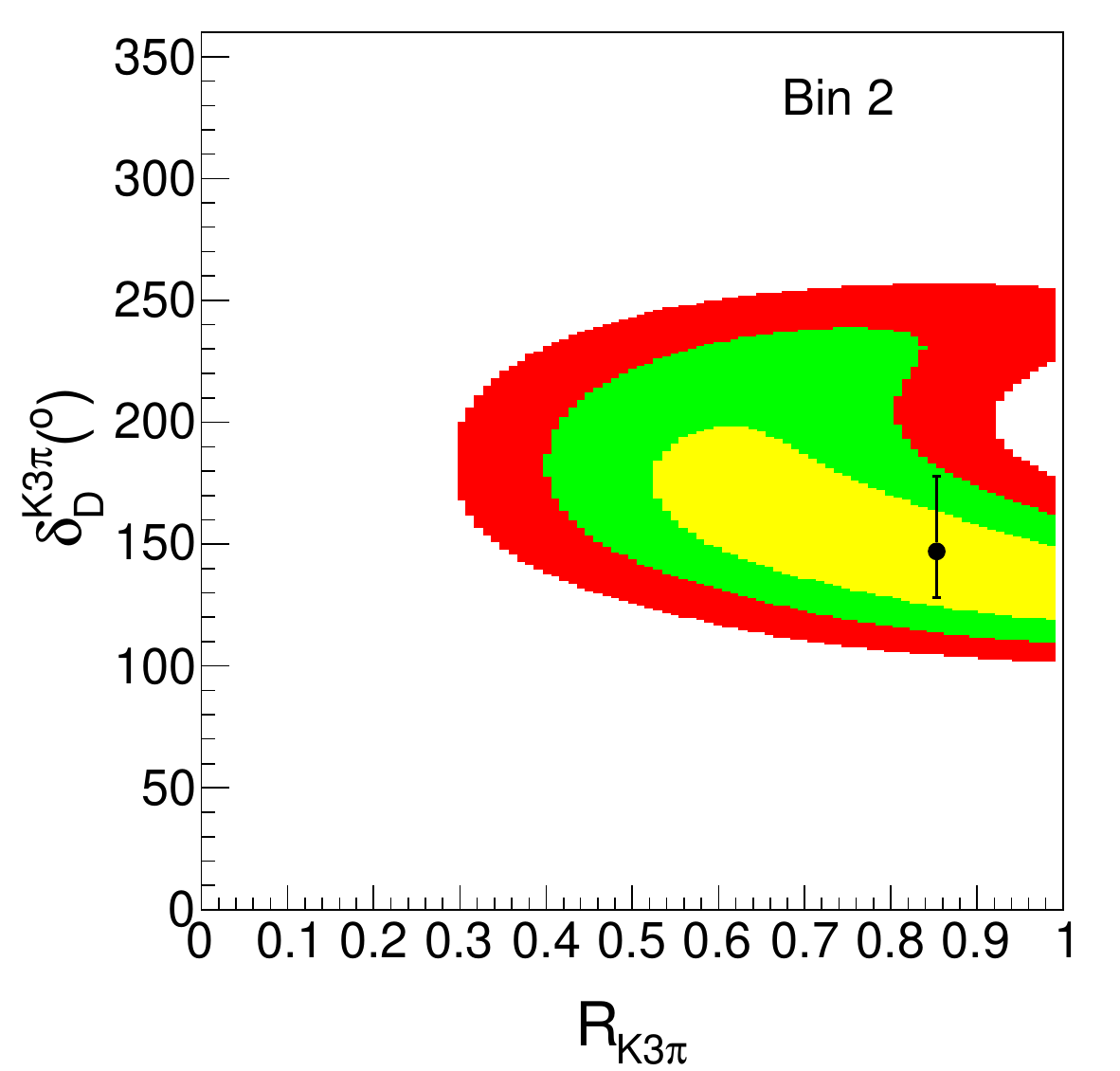}\\
\includegraphics[width=.45\textwidth]{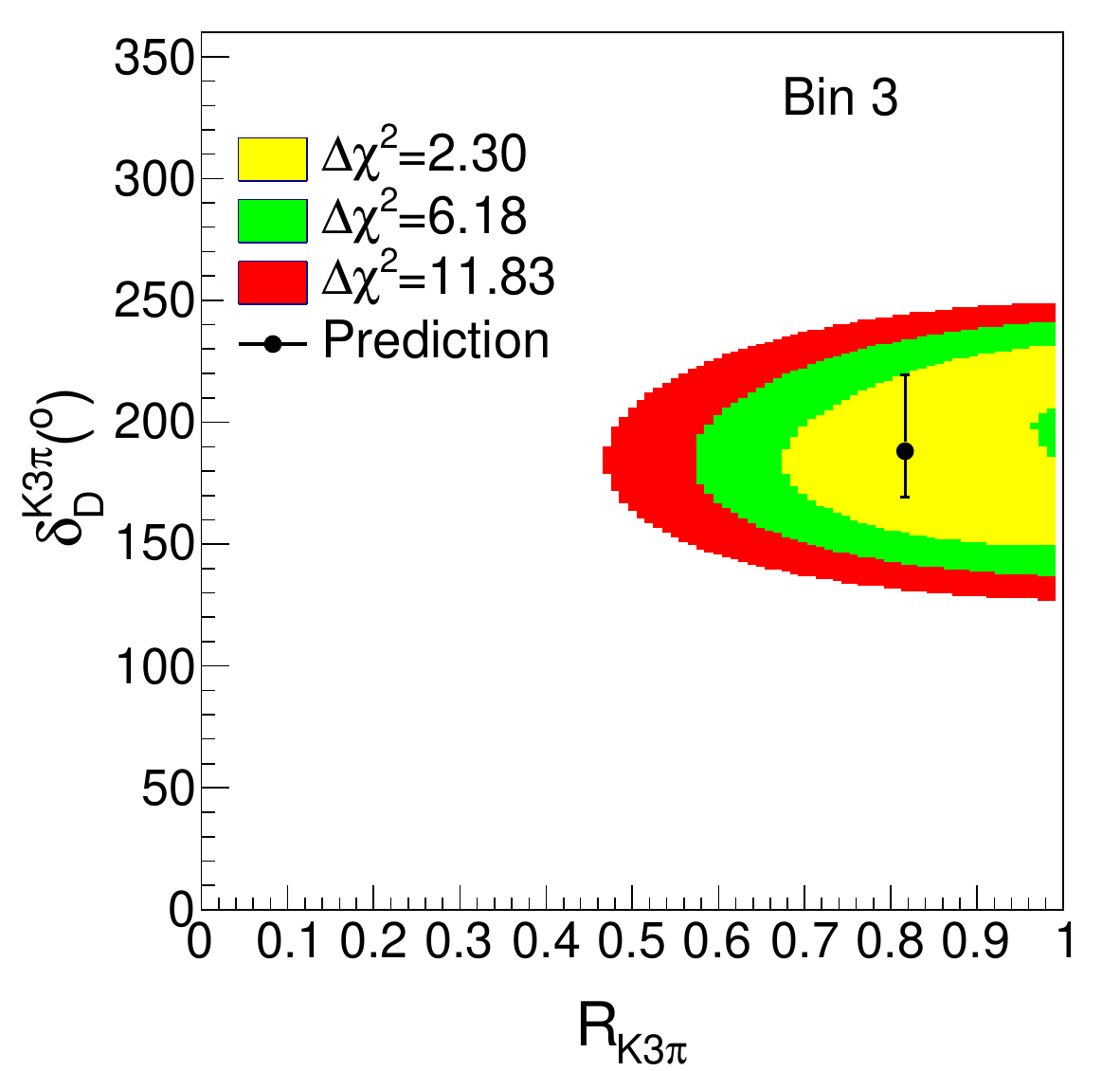}
\includegraphics[width=.45\textwidth]{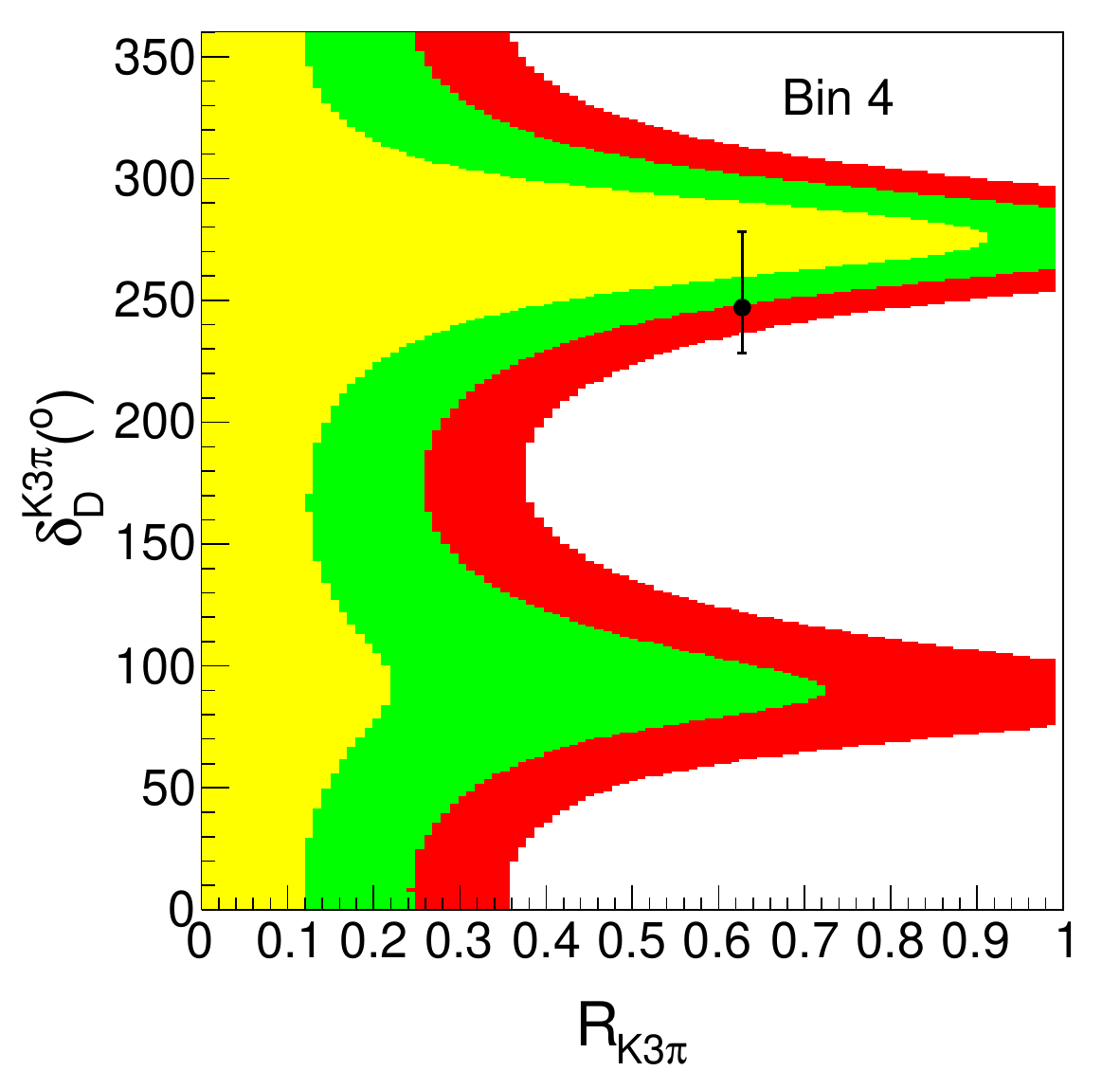}
\caption{Scans of $\Delta \chi^2$ in the four bins of the ($\Rkthreepi$, $\dkthreepi$)  parameter space, showing the $\Delta \chi^2$=2.30, 6.18, 11.83 intervals, which correspond to $68.3\%$, $95.4\%$ and $99.7\%$ confidence levels in the two-dimensional parameter space. Also indicated is the prediction from the model, where the global offset in the average strong-phase difference comes from the global result.}\label{fig:binparameter}
\end{figure}
\section{Impact of the results on the measurement of $\gamma$}
\label{sec:gamma}

Improved knowledge of the global coherence factors and average strong-phase differences in the processes $D\to \kmthreepi$ and $D \to \kmpipio$  will be valuable in future studies of $B^- \to DK^-$ decays at LHCb~\cite{Malde:2015mha,Bediaga:2018lhg} and Belle~II~\cite{Kou:2018nap}. However, neither of these channels, when used in isolation, is able to provide a standalone measurement of the Unitarity Triangle angle $\gamma$ with interesting sensitivity; rather, they bring important constraints when considered as part of the full ensemble of $B^- \to DK^-$ measurements using several $D$ decay modes.  These constraints will be tightened as a consequence of the BESIII measurements.

Subdividing the phase space of $D \to \kmthreepi$ decays into four bins allows a standalone measurement of $\gamma$ to be performed~\cite{Evans:2019wza}, and here it is possible to calculate the impact of the  measurements reported in this paper.  Within each bin of phase space, and considering all possible charge configurations, there are four decay rates that may be measured: 
\begin{eqnarray}
{\Gamma(B^{\mp}\to(K^{\pm}\pi^{\mp}\pi^{\mp}\pi^\pm)_D K^{\mp})} &  \propto &
 (r_{B})^{2} + (r_{D}^{K3\pi})^{2}  +  \nonumber \\
& & 2r_{B}r_{D}^{K3\pi}R_{K3\pi}\cos{(\delta_{B}+\delta_D^{K3\pi} \mp \gamma)}\, , \label{eq:adssuppressed} \\
{\Gamma(B^{\mp}\to(K^{\mp}\pi^{\pm}\pi^{\pm}\pi^\mp)_D K^{\mp})} &  \propto &
 1 + (r_B r_{D}^{K3\pi})^{2}  +  \nonumber \\
& & 2r_{B}r_{D}^{K3\pi}R_{K3\pi}\cos{(\delta_{B}-\delta^{K3\pi}_D \mp \gamma)}\,  \label{eq:adsfavoured}.
\end{eqnarray}
Here $r_B \sim 0.1$ is the absolute ratio of $B^- \to {\bar D^0} K^-$  to $B^- \to D^0 K^-$ amplitudes.
The phase difference between these two amplitudes is $(\delta_B - \gamma)$, where $\delta_B$ is a $C\!P$-conserving strong phase.   These expressions neglect small corrections from $D^0$-$\bar{D^0}$ oscillations, which can be included in a straightforward manner~\cite{Rama:2013voa}.  It can be seen that  the suppressed pair of decays, $B^{\mp}\to(K^{\pm}\pi^{\mp}\pi^{\mp}\pi^\pm)_D K^{\mp}$, has highest sensitivity to $\gamma$, as in this case the interference term involving this parameter enters at leading order.

An ensemble of simulated data sets is generated, each containing around 600 suppressed decays.  The sample size  is roughly equivalent  to that which is expected in the Run 1 and Run 2 data sets of LHCb, extrapolating from published results~\cite{Aaij:2016oso}.  The distribution of events between $B^-$ and $B^+$ is simulated according to Eq.~\ref{eq:adssuppressed}, with the central values of the $D$ hadronic parameters chosen to be the values measured by BESIII, and with the parameters of the $B$ decay and $\gamma = 71^\circ$ set to the central values of a recent global average~\cite{Amhis:2019ckw}.  Each data set is subjected to a $\chi^2$ fit to determine $\gamma$, the $B$-meson decay parameters, and the hadronic parameters of the $D$ decay, where the latter are constrained  according to the central values and covariances of the BESIII measurement, as expressed in the likelihood contours.   In order to quantify the contributions to the overall fit uncertainties that are induced by the BESIII measurements alone, the exercise is then repeated, but with simulated $B$-decay data sets that are 100 times larger.

\begin{figure}[!htb]
\centering
\includegraphics[width=.48\textwidth]{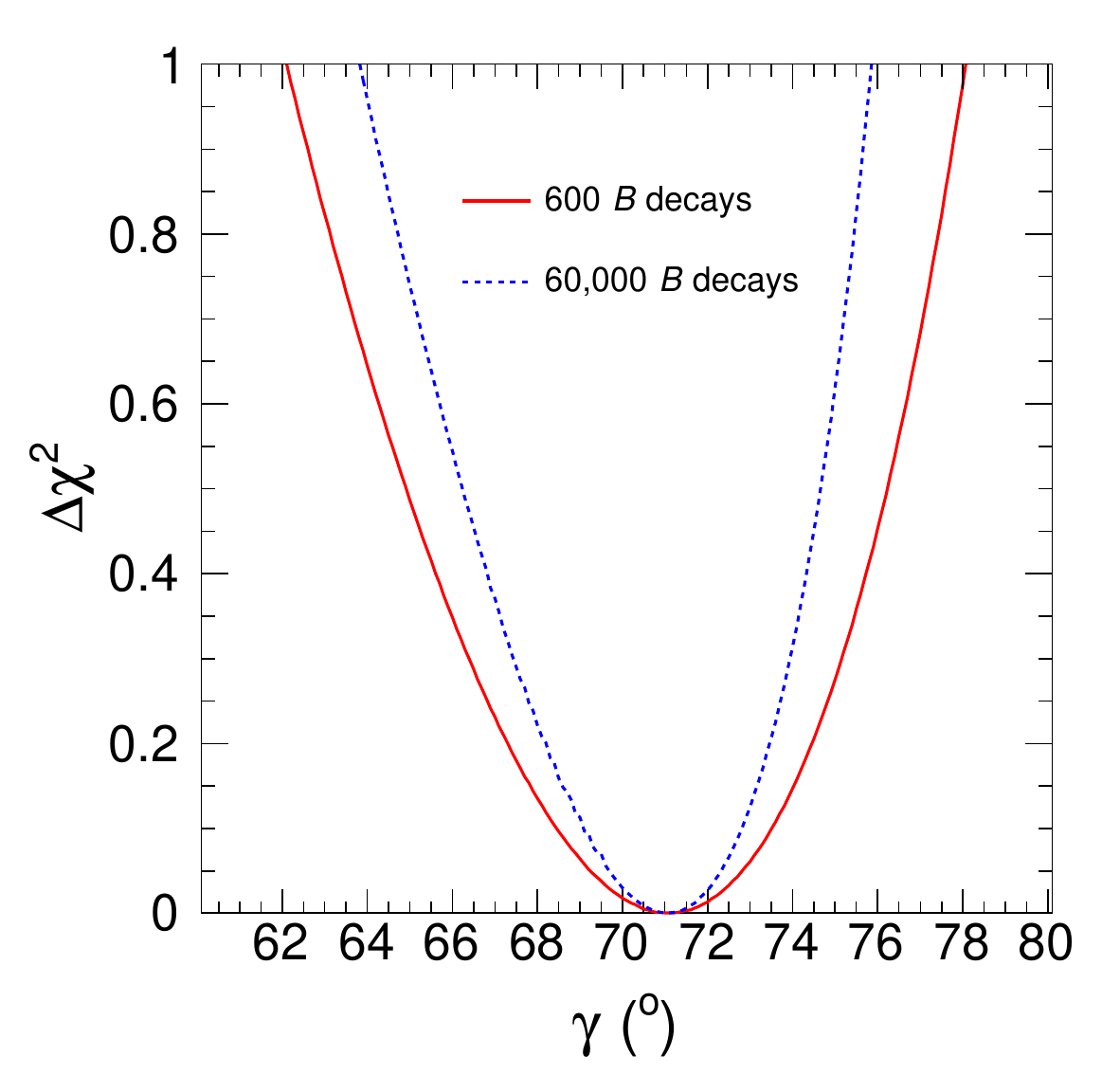}
\caption{The distribution of $\Delta\chi^2$ vs. $\gamma$ as fitted from two ensembles of simulated data sets, each data set containing 600 or 60,000 suppressed $B$ decays.}\label{fig:gammastudy}
\end{figure}

Figure~\ref{fig:gammastudy} shows $\Delta \chi^2$, the change in $\chi^2$ with respect to the minimum that these studies give, plotted as a function of $\gamma$. The minimum is centred on the input value.~\footnote{Due to a trigonometrical ambiguity two solutions are returned for $\gamma$, but only one, which is shown here, is broadly compatible with the existing constraints on the Unitarity Triangle.}  It is seen that an analysis of 600 suppressed $B$ decays, together with the knowledge of the $D$ hadronic parameters reported in this paper, will allow $\gamma$ to be determined with a precision of $\left(^{+7}_{-9}\right)^\circ$, to which the BESIII measurements contribute an uncertainty of  $\left(^{+5}_{-7}\right)^\circ$. Such a result would be only slightly less precise than the current best standalone determination of $\gamma$, which comes from an LHCb analysis of $B^- \to DK^-$, $D\to K^0_S\pi^+\pi^-$ and $D\to K^0_S K^+ K^-$ decays, and has an uncertainty of $\pm 5^\circ$~\cite{Aaij:2020xuj}.  Hence it is concluded that the BESIII results on the binned $D \to K^-\pi^+\pi^+\pi^-$ hadronic parameters are a valuable contribution to improving the knowledge of $C\!P$ violation in $b$-hadron decays.

\section{Conclusion}
\label{sec:summary}

Observables have been measured for the decays $D \to \kmthreepi$ and $D \to \kmpipio$ in a sample of quantum-correlated $D\bar{D}$ pairs.  Many of these observables exhibit significant deviations from the values expected in the absence of quantum correlations, and the ensemble of measurements allow for the coherence factor and average strong-phase difference of each decay to be determined.  In the case when the analysis integrates over the phase space of both decays, the results are
\begin{center}
\begin{tabular}{lcl}
$R_{K3\pi} \hspace*{0.17cm}= 0.52^{+0.12}_{-0.10}\,$, & & $\delta_D^{K3\pi}\hspace*{0.16cm} = \left(167^{+31}_{-19}\right)^\circ$,\\[0.2cm]
$R_{K\pi\pi^0} = 0.78 \pm 0.04\,$,  & &  $\delta_D^{K\pi\pi^0} = \left(196^{+14}_{-15}\right)^\circ$.
\end{tabular}
\end{center}
The measurements of the coherence factor are more precise than the existing world-average values, and the allowed region in the $(R_{K3\pi}, \delta_D^{K3\pi})$ and $(R_{K\pi\pi^0}, \delta_D^{K\pi\pi^0})$ planes is more restricted than is the case for the observables measured with CLEO-c data~\cite{Evans:2016tlp}.  This improved knowledge will be valuable when these channels are included with other $D$-meson decay modes in studies of $B^- \to DK^-$ decays at LHCb and Belle~II,  and will enable a more precise determination of the angle $\gamma$ of the Unitarity Triangle.  It will also be useful in interpreting $D^0$-$\bar{D}^0$ oscillation measurements performed with these multibody decays.

The analysis has been re-performed in four bins of phase space of the $D \to \kmthreepi$ decay, to yield bin-by-bin values of the coherence factor and average  strong-phase difference.  A  study of $B^- \to DK^-$ decays that makes use of such a sub-division is inherently more sensitive, and can provide a standalone measurement of $\gamma$ that is expected to be among the most precise available using a single $D$-decay mode.   The BESIII results will contribute an uncertainty of around $6^\circ$  to this measurement, which is less than that estimated to arise from the size of currently available $B$-meson decay samples.  Updated analyses with the larger data sets that BESIII expects to collect at the $\psi(3770)$ resonance in the coming years will allow this uncertainty to be significantly decreased~\cite{Ablikim:2019hff}.
\section*{Acknowledgments}

We are grateful to Alex Lenz for valuable discussions during the preparation of this paper. The BESIII collaboration thanks the staff of BEPCII and the IHEP computing center for their strong support. This work is supported in part by National Key R\&D Program of China under Contracts Nos. 2020YFA0406400, 2020YFA0406300; National Natural Science Foundation of China (NSFC) under Contracts Nos. 11625523, 11635010, 11735014, 11822506, 11835012, 11935015, 11935016, 11935018, 11961141012; the Chinese Academy of Sciences (CAS) Large-Scale Scientific Facility Program; Joint Large-Scale Scientific Facility Funds of the NSFC and CAS under Contracts Nos. U1732263, U1832207; CAS Key Research Program of Frontier Sciences under Contracts Nos. QYZDJ-SSW-SLH003, QYZDJ-SSW-SLH040; 100 Talents Program of CAS; Fundamental Research Funds for the Central Universities; INPAC and Shanghai Key Laboratory for Particle Physics and Cosmology; ERC under Contract No. 758462; German Research Foundation DFG under Contracts Nos. Collaborative Research Center CRC 1044, FOR 2359; Istituto Nazionale di Fisica Nucleare, Italy; Ministry of Development of Turkey under Contract No. DPT2006K-120470; National Science and Technology fund; STFC (United Kingdom); The Knut and Alice Wallenberg Foundation (Sweden) under Contract No. 2016.0157; The Royal Society, UK under Contracts Nos. DH140054, DH160214; The Swedish Research Council; U. S. Department of Energy under Contracts Nos. DE-FG02-05ER41374, DE-SC-0012069.

\bibliographystyle{JHEP}
\bibliography{references}

\appendix
\section{$M_{\rm BC}$ distributions} 
\label{sec:appendix0}

\begin{figure}[!h]
    \centering
    \includegraphics[width=.9\textwidth]{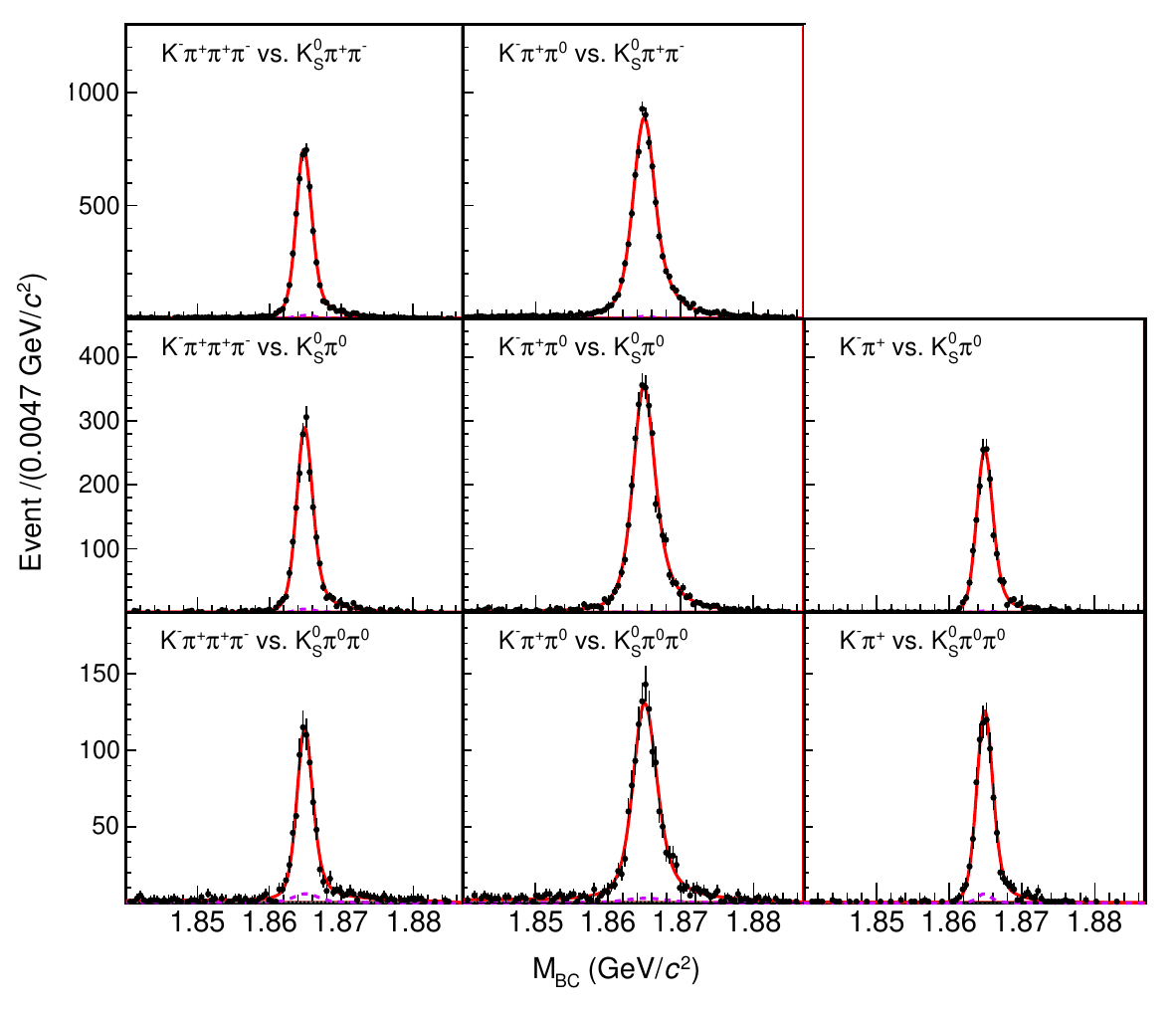}
    \caption{$M_{\rm BC}$ distributions for decays tagged by the modes $D\to \kspipi$, $D \to K^0_S\pi^0$ and $D \to K^0_{S}\pi^0\pi^0$.     The points with error bars are data; the red line indicates the total fit; 
     the dashed purple line shows the  peaking-background contributions; the combinatorial-background contribution is at too low a level to be visible.  }
    \label{fig:mbcexamples3}
\end{figure}

\begin{figure}
    \centering
    \includegraphics[width=.9\textwidth]{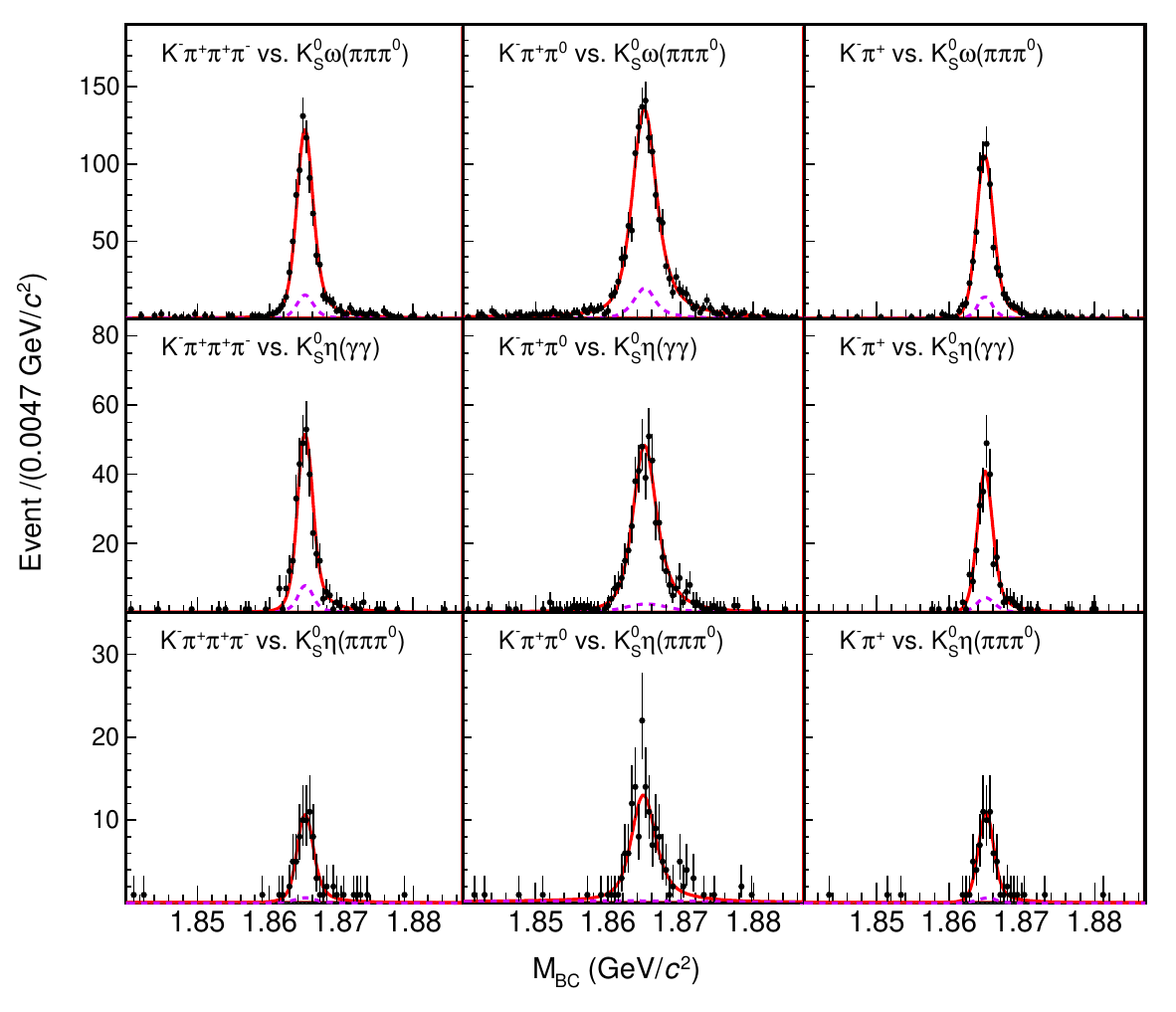}
    \caption{$M_{\rm BC}$ distributions for decays tagged by the modes $D\to K^0_S \omega(\pi^+\pi^-\pi^0)$, $D \to K^0_S \eta(\gamma\gamma)$ and $D \to K^0_S \eta(\pi^+\pi^-\pi^0)$.    The points with error bars are data;  the red line indicates the total fit; 
       the dashed purple line shows the  peaking-background contributions; the combinatorial-background contribution is at too low a level to be visible.   }
    \label{fig:mbcexamples4}
\end{figure}

\begin{figure}
    \centering
    \includegraphics[width=.9\textwidth]{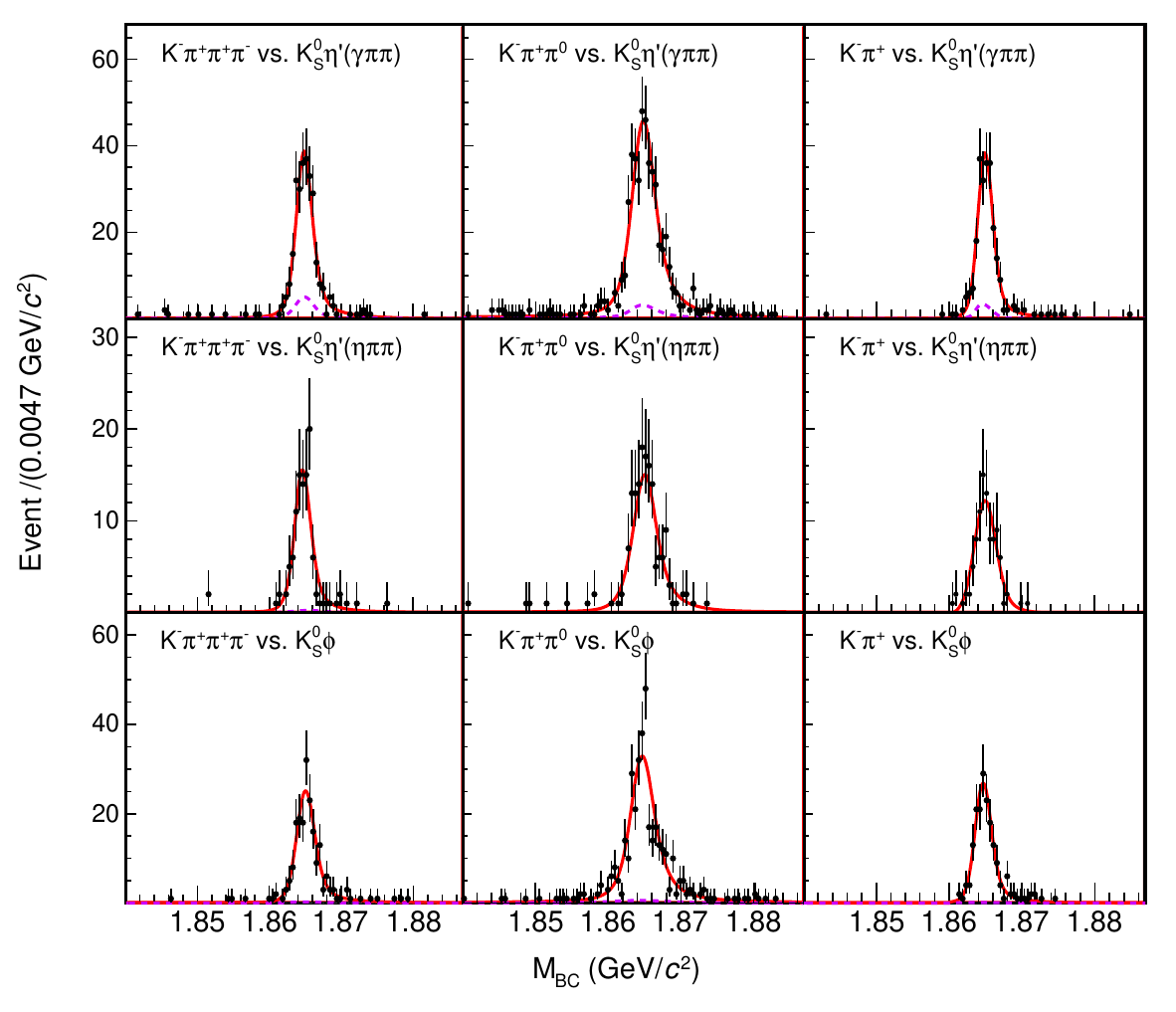}
    \caption{$M_{\rm BC}$ distributions for decays tagged by the modes $D\to K^0_S \eta'(\gamma\pi^+\pi^-)$, $D \to K^0_S \eta'(\eta\pi^+\pi^-)$ and $D \to K^0_S \phi$. The points with error bars are data;  
           the dashed purple line shows the  peaking-background contributions; the combinatorial-background contribution is at too low a level to be visible.   }
    \label{fig:mbcexamples5}
\end{figure}

Figures~\ref{fig:mbcexamples3}, \ref{fig:mbcexamples4} and \ref{fig:mbcexamples5} present the $M_{\rm BC}$ distributions for those double-tagged events not shown in the main body of the text.

\section{Correlation matrices}
\label{sec:appendix1}

Table~\ref{tab:rhocorrelation} presents the correlation matrix for the observables in the global analysis, and Table~\ref{tab:globalmatrix} gives the correlation matrix for the hadronic parameters determined in this study.  Tables~\ref{tab:binrhocorrelation} and~\ref{tab:binnedmatrix} show the corresponding matrices for the binned analysis.

\begin{table}[!htb]
    \begin{center}
    \caption{Correlation matrix for the observables in the global analysis.} \label{tab:rhocorrelation}
\begin{tabular}{ccccccc}
\toprule
 &$\Delta^{K\pi\pi^0}_{C\!P}$ & $\rho^{K3\pi}_{K\pi,LS}$ & $\rho^{K3\pi}_{K\pi\pi^0,LS}$ & $\rho^{K3\pi}_{LS}$ & $\rho^{K\pi\pi^0}_{K\pi,LS}$ & $\rho^{K\pi\pi^0}_{LS}$\\
 \midrule
$\Delta^{K3\pi}_{C\!P}$ & 0.49 & 0.00  & 0.00  &0.00 &0.00  &0.00 \\	
$\Delta^{K\pi\pi^0}_{C\!P}$ &  1.00 	     & 0.00  & 0.00  &0.00 &0.00  &0.00   \\
 $\rho^{K3\pi}_{K\pi,LS}$ &  	      &1.00        &0.01	&0.01  & 0.00  & 0.02  \\
 $\rho^{K3\pi}_{K\pi\pi^0,LS}$ & 	        &         &1.00	    &0.01 &0.00  &0.02 \\
 $\rho^{K3\pi}_{LS}$ &   	         &         &        & 1.00    & 0.00  & 0.02   \\
$\rho^{K\pi\pi^0}_{K\pi,LS}$ &   	         &         &        &  		& 1.00    &0.01 \\ 
   	     \bottomrule
\end{tabular}

\end{center}
\end{table}

\begin{table}[!htb]
    \begin{center}
    \caption{Correlation matrix for the hadronic parameters determined in the global fit.}\label{tab:globalmatrix}
\begin{tabular}{crrrrr}
\toprule
 			 &$\dkthreepi$	  &$\rkthreepi$ &$\Rkpipio$  &$\dkpipio$ &$\rkpipio$\\
\midrule
 $\Rkthreepi$	&$-$0.63  & $-$0.62  &0.03    &0.15     & 0.08 \\	
 $\dkthreepi$	& 1.00    &	   0.56  &$-$0.12 &$-$0.08  &$-$0.02\\
 $\rkthreepi$  	& 		  & 1.00  	 &0.03    &$-$0.19  &$-$0.02 \\
$\Rkpipio$   	& 		  & 		     & 1.00   &0.13     &0.03  \\
$\dkpipio$   	&  		  &  		 &  	   	  &1.00 		&0.30  \\
   	     \bottomrule
\end{tabular}

\end{center}
\end{table}

\begin{table}[!htb]
    \begin{center}
    \caption{Correlation matrix for the observables in the binned analysis. The off-diagonal elements involving the binned $\rho^{K3\pi}_{LS}$ observables, which are not shown, are all 0.01 or less.
}\label{tab:binrhocorrelation}
\small
\begin{tabular}{ccccccccccccccccccc c cc}
\toprule
& &\multicolumn{3}{c}{$\Delta^{K3\pi}_{C\!P}$} &\multicolumn{4}{c}{$\rho^{K3\pi}_{K\pi,LS}$ } &\multicolumn{4}{c}{$\rho^{K3\pi}_{K\pi\pi^0,LS}$} \\
&  &Bin 2 & Bin 3 &Bin 4 &Bin 1 &Bin 2 & Bin 3 &Bin 4 &Bin 1 &Bin 2 & Bin 3 &Bin 4 \\
\midrule
\multirow{4}{*}{$\Delta^{K3\pi}_{C\!P}$} &Bin 1 &0.46  &0.49  &0.57  &0.01  &0.00  &0.01  &0.01  &0.01  &0.01  &0.00 &0.00   \\
  &Bin 2 & 1.00   &0.43  &0.50  &0.00  &0.00  &0.00  &0.01  &0.01  &0.01  &0.00  &0.01  \\
  &Bin 3 &   & 1.00     &0.53  &0.01  &0.00  &0.01  &0.01  &0.01  &0.01  &0.01  &0.01  \\
  &Bin 4  &  &      & 1.00     &0.00  &0.00  &0.00  &0.00  &0.00  &0.00  &0.00  &0.00  \\
 \multirow{4}{*}{$\rho^{K3\pi}_{K\pi,LS}$} &Bin 1  &  &      &      & 1.00     &0.01  &0.01  &0.02  &0.02  &0.01  &0.01  &0.01  \\
&Bin 2    &  &      &      &      & 1.00     &0.00  &0.01  &0.01  &0.01  &0.00  &0.01  \\
&Bin 3    &  &      &      &      &      & 1.00     &0.01  &0.02  &0.01  &0.01  &0.01  \\
&Bin 4    &  &      &      &      &      &      & 1.00     &0.03  &0.03  &0.01  &0.02  \\
\multirow{3}{*}{$\rho^{K3\pi}_{K\pi\pi^0,LS}$}  &Bin 1  &  &      &      &      &      &      &      & 1.00     &0.03  &0.01  &0.03  \\
&Bin 2    &  &      &      &      &      &      &      &      & 1.00     &0.01  &0.02  \\
&Bin 3    &  &      &      &      &      &      &      &      &      &  1.00    &0.01  \\
      \bottomrule
\end{tabular}

\end{center}
\end{table}

\small


\begin{table}[!htb]
    \begin{center}
    \caption{Correlation matrix for the hadronic parameters determined in the binned fit.}\label{tab:binnedmatrix}
\begin{tabular}{ccrrrrrrr}
\toprule
&	&Bin 1 &\multicolumn{2}{c}{Bin 2} &\multicolumn{2}{c}{Bin 3} &\multicolumn{2}{c}{Bin 4}\\
 			& &$\dkthreepi$  &$\Rkthreepi$  &$\dkthreepi$ &$\Rkthreepi$  &$\dkthreepi$ &$\Rkthreepi$  &$\dkthreepi$\\
\midrule
\multirow{2}{*}{Bin 1} & $\Rkthreepi$	&$-$0.56  &0.33 &$-$0.12 &$-$0.82  &0.58 &$-$0.45  &0.08 \\	
& $\dkthreepi$	&1.00 &$-$0.42  &0.18  &0.51 &$-$0.79  &0.55 &$-$0.29\\
\multirow{2}{*}{Bin 2} &  $\Rkthreepi$	& &1.00 &$-$0.13 &$-$0.31  &0.42 &$-$0.41  &0.00\\	
& $\dkthreepi$	&     &	   &1.00  &0.14 &$-$0.10  &0.29  &0.27\\
\multirow{2}{*}{Bin 3} & $\Rkthreepi$	&  &   &    &1.00 &$-$0.51  &0.44 &$-$0.03 \\	
& $\dkthreepi$	&     &	   & &  &1.00 &$-$0.53  &0.35\\
Bin 4 & $\Rkthreepi$	&  &   &    &     &     & 1.00  &0.03\\	
   	     \bottomrule
\end{tabular}

\end{center}
\end{table}
\section{Combination with CLEO-c and LHCb results}
\label{sec:appendix2}

\begin{figure}[!ht]
\begin{center}
{\includegraphics[width=.45\textwidth]{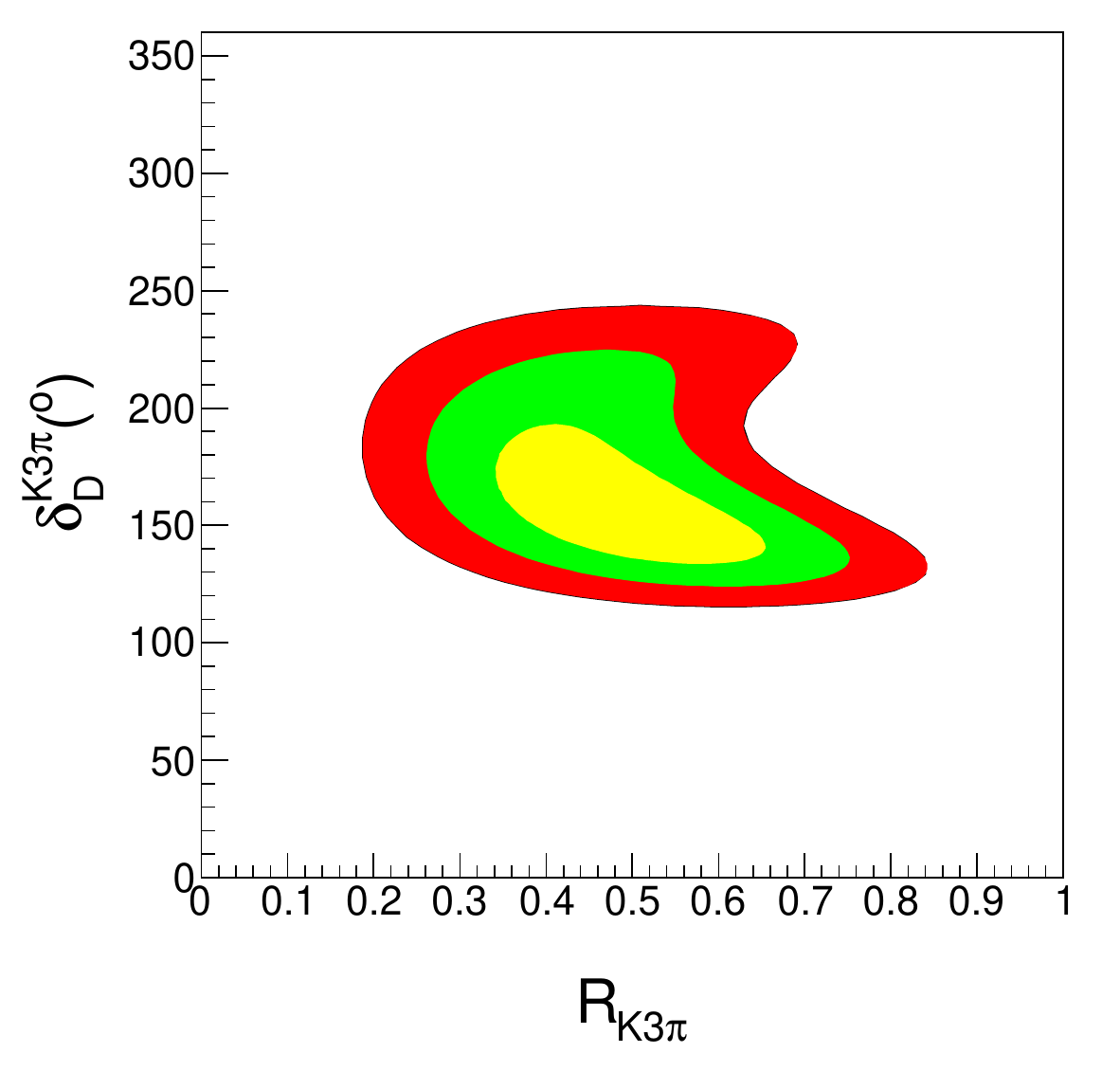}
\includegraphics[width=.45\textwidth]{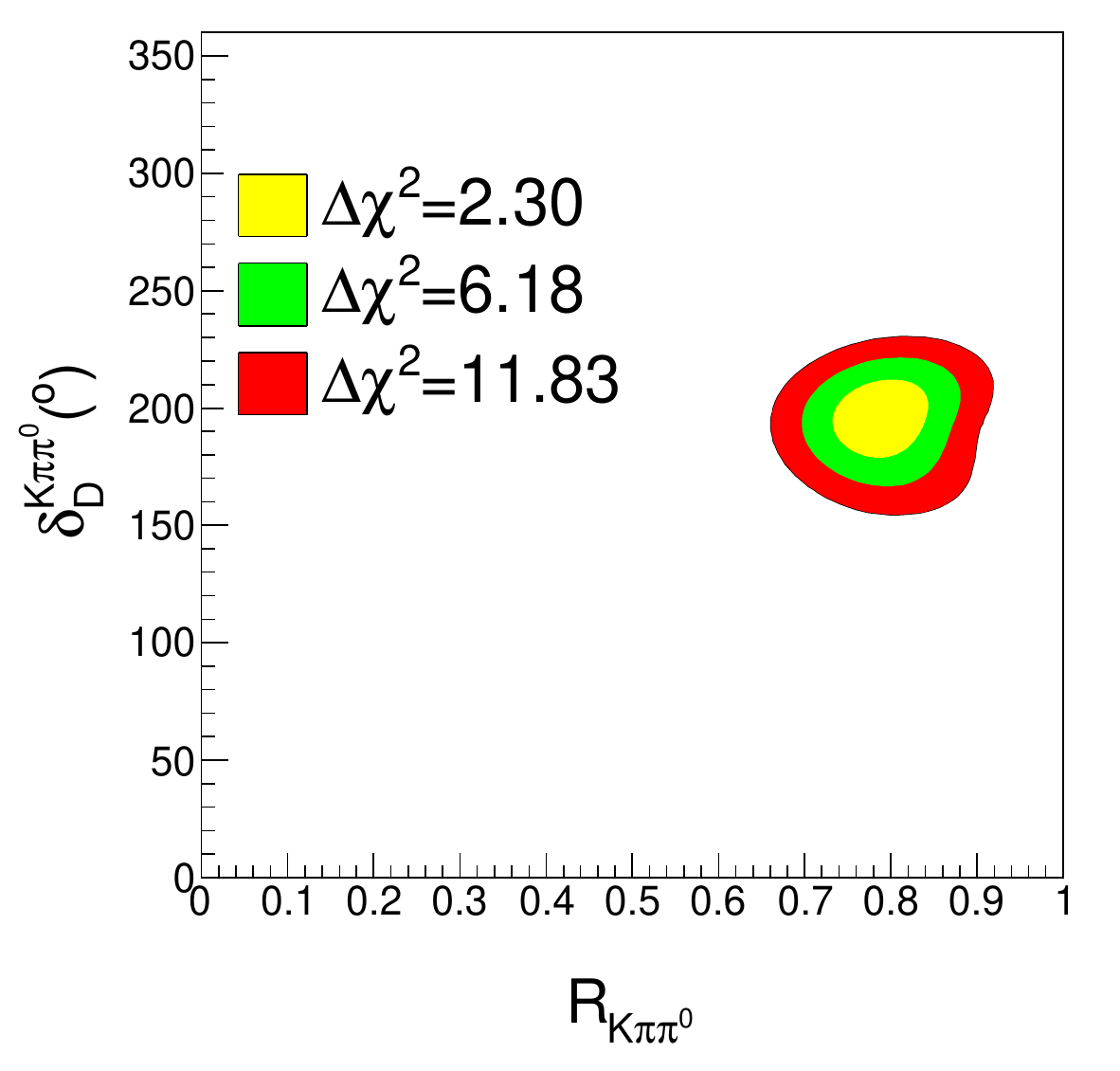}}
\caption{Scans of $\Delta \chi^2$ in the BESIII and CLEO-c global fit in the ($\Rkthreepi$, $\dkthreepi$) and ($\Rkpipio$, $\dkpipio$) parameter space, showing the $\Delta \chi^2$=2.30, 6.18, 11.83 intervals, which corresponds to the $68.3\%$, $95.4\%$ and $99.7\%$ CLs in the two-dimensional parameter space. }
 \label{fig:parameter_comb_global1}
\end{center}
\end{figure}

\begin{figure}[!ht]
\begin{center}
{\includegraphics[width=.45\textwidth]{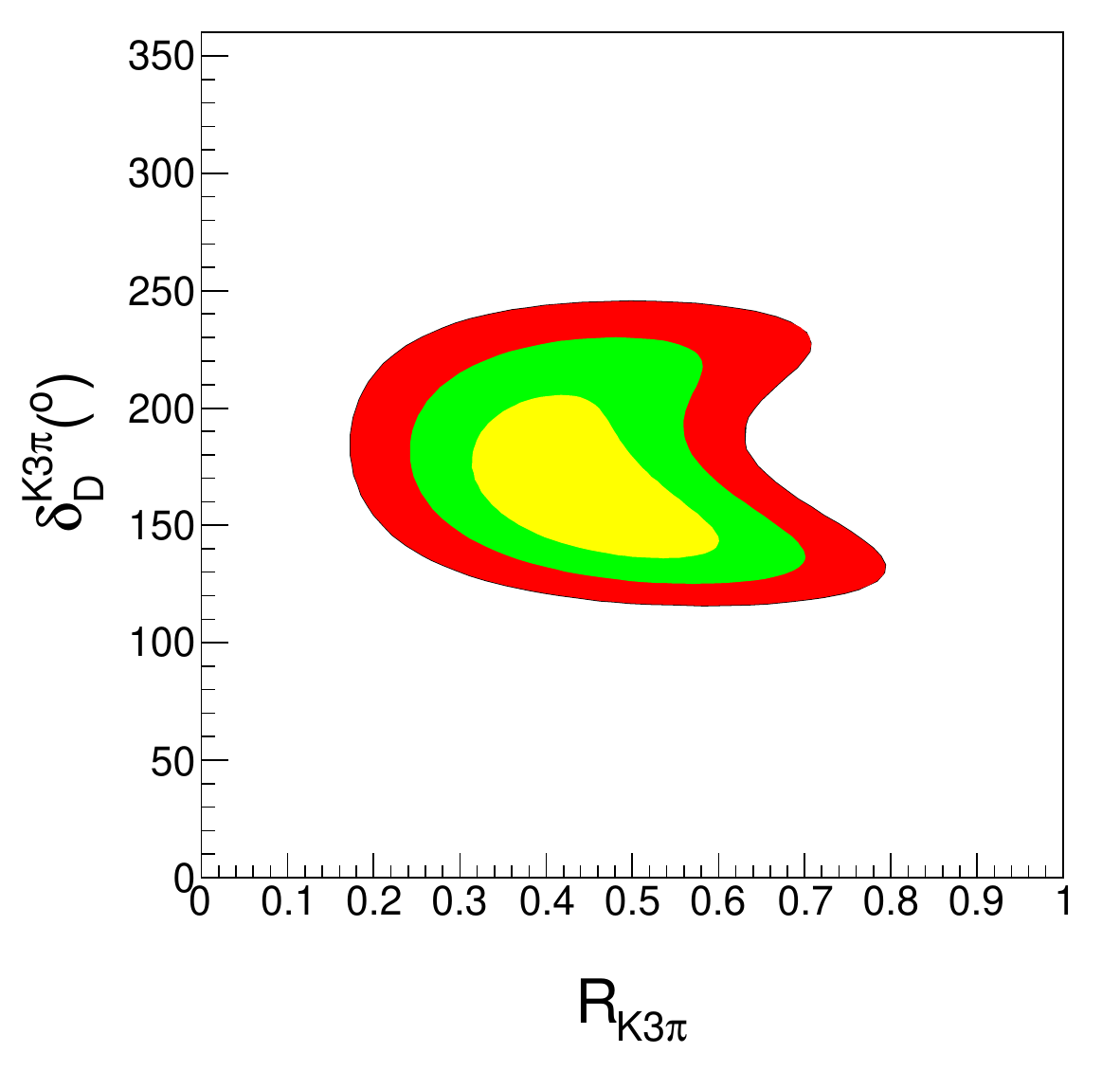}
\includegraphics[width=.45\textwidth]{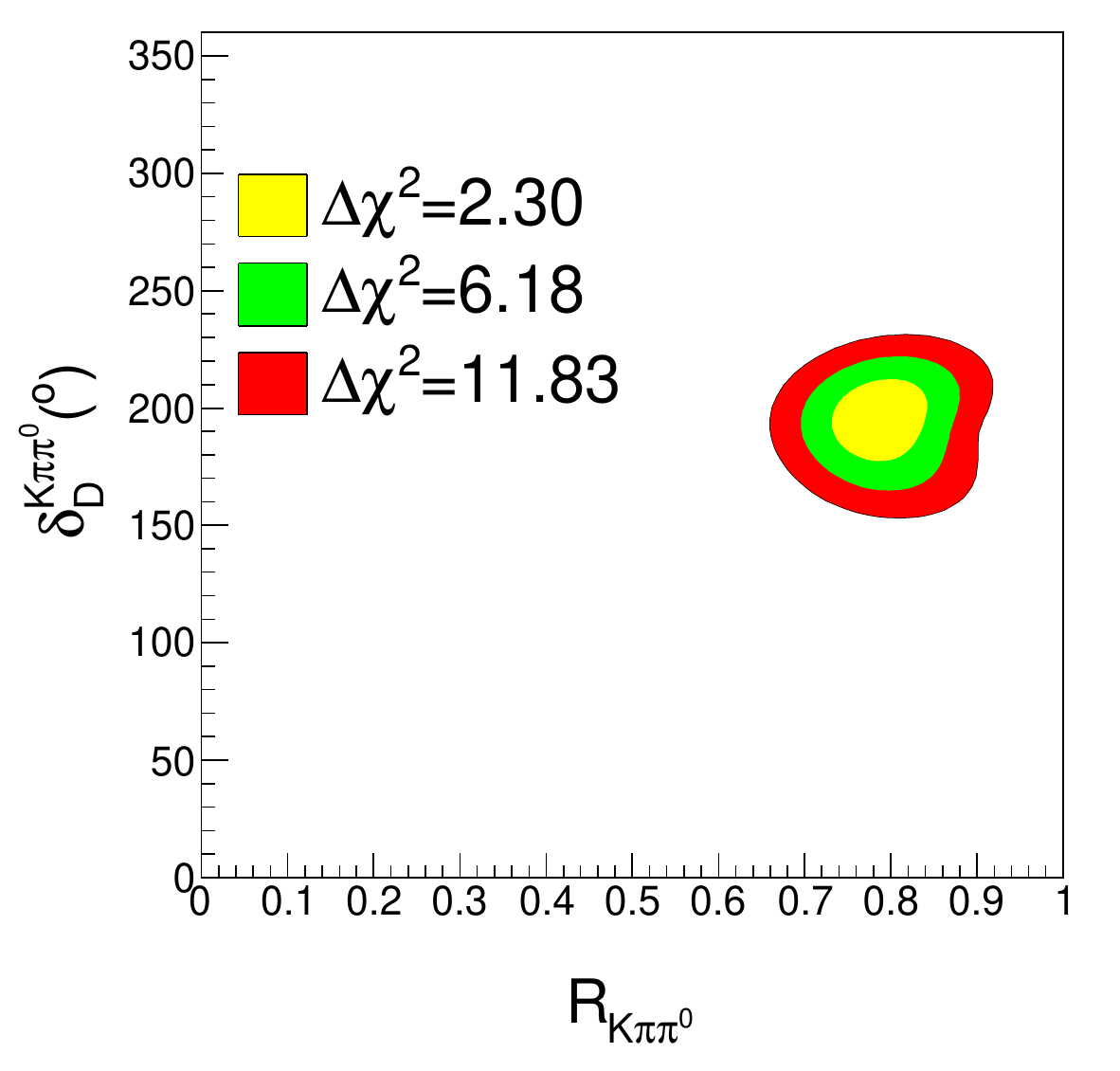}}
\caption{Scans of $\Delta \chi^2$ in the BESIII, CLEO-c and LHCb global fit in the ($\Rkthreepi$, $\dkthreepi$) and ($\Rkpipio$, $\dkpipio$) parameter space, showing the $\Delta \chi^2$=2.30, 6.18, 11.83 intervals, which corresponds to the $68.3\%$, $95.4\%$ and $99.7\%$ CLs in the two-dimensional parameter space. }
 \label{fig:parameter_comb_global2}
\end{center}
\end{figure}

A fit is performed to the global observables determined by BESIII  and those from the CLEO-c data, reported in Refs.~\cite{Libby:2014rea,Evans:2016tlp}. It is assumed that the correlations between the two sets of measurements are negligible.  The $\chi^2/{\rm n.d.f.}$ of the fit is 100/98. The results are presented in Table~\ref{tab:parameter_global_comb}, with the accompanying correlation matrix given in Table~\ref{tab:globalmatrix_comb}, and $\Delta \chi^2$ scans are shown in Fig.~\ref{fig:parameter_comb_global1}. A second fit, with a $\chi^2/{\rm n.d.f.}$ of  104/101, includes the constraints from the LHCb study of $D^0$-$\bar{D}^{0}$ oscillations~\cite{Aaij:2016rhq}.  The results are shown in Table~\ref{tab:parameter_global_comb} and Fig.~\ref{fig:parameter_comb_global2}, with the correlation matrix in Table~\ref{tab:globalmatrix_lhcb}.

\begin{table}[!ht]
\caption{Results for the global fit to the BESIII and CLEO-c data, and the BESIII, CLEO-c and LHCb data.}\label{tab:parameter_global_comb}
\begin{center}
\begin{tabular}{lcc}
\toprule
Parameter & BESIII and & BESIII, CLEO-c  \\
          & CLEO-c    & and LHCb         \\
\midrule
$\Rkthreepi$ & 0.49$_{-0.10}^{+0.11}$  & 0.44$_{-0.09}^{+0.10}$\\ \\
$\dkthreepi$  &  $\left(154_{-14}^{+22}\right)^\circ$ &$\left(161_{-18}^{+28}\right)^\circ$ \\ \\
$\rkthreepi$ ($\times 10^{-2}$) & 5.46$\pm$0.08  & 5.50$\pm$0.07\\ \\
$\Rkpipio$    &	0.79$\pm$0.04   	&	0.79$\pm$0.04	\\ \\
$\dkpipio$  &	$\left(196\pm 11\right)^\circ$  & $\left(196\pm 11\right)^\circ$\\ \\
$\rkpipio$ ($\times 10^{-2}$)  &4.41$\pm$0.11   & 4.41$\pm$0.11\\
\bottomrule
\end{tabular}
\end{center}
\end{table}

\begin{table}[!htb]
    \begin{center}
    \caption{Correlation matrix for the hadronic parameters from the global fit to the BESIII and CLEO-c data.}\label{tab:globalmatrix_comb}
\begin{tabular}{crrrrr}
\toprule
 			 &$\dkthreepi$	  &$\rkthreepi$ &$\Rkpipio$  &$\dkpipio$ &$\rkpipio$\\
\midrule
 $\Rkthreepi$	&$-$0.78  & 0.50  &0.04    &$-$0.07     & $-$0.06  \\	
 $\dkthreepi$	& 1.00    &	$-$0.34  &0.15 &$-$0.15  &$-$0.04\\
 $\rkthreepi$  	& 		  & 1.00  	 &0.02    &$-$0.11  &$-$0.02 \\
$\Rkpipio$   	& 		  & 		     & 1.00   &0.23     &0.05  \\
$\dkpipio$   	&  		  &  		 &  	   	  &1.00 		&$-$0.11  \\
   	     \bottomrule
\end{tabular}
\end{center}
\end{table}

\begin{table}[!htb]
    \begin{center}
    \caption{Correlation matrix for the hadronic parameters from the global fit to the  BESIII, CLEO-c and LHCb data.}\label{tab:globalmatrix_lhcb}
\begin{tabular}{crrrrr}
\toprule
 			 &$\dkthreepi$	  &$\rkthreepi$ &$\Rkpipio$  &$\dkpipio$ &$\rkpipio$\\
\midrule
 $\Rkthreepi$	&$-$0.75  & 0.52  &0.00    &$-$0.07     & $-$0.06 \\	
 $\dkthreepi$	& 1.00    &	  $-$0.42  &0.03 &0.17  &0.01\\
 $\rkthreepi$  	& 		  & 1.00  	 &$-$0.01    &$-$0.02  &$-$0.12 \\
$\Rkpipio$   	& 		  & 		     & 1.00   &0.19     &$-$0.01  \\
$\dkpipio$   	&  		  &  		 &  	   	  &1.00 		&0.25  \\
   	     \bottomrule
\end{tabular}
\end{center}
\end{table}

\begin{figure}[!ht]
\includegraphics[width=.45\textwidth]{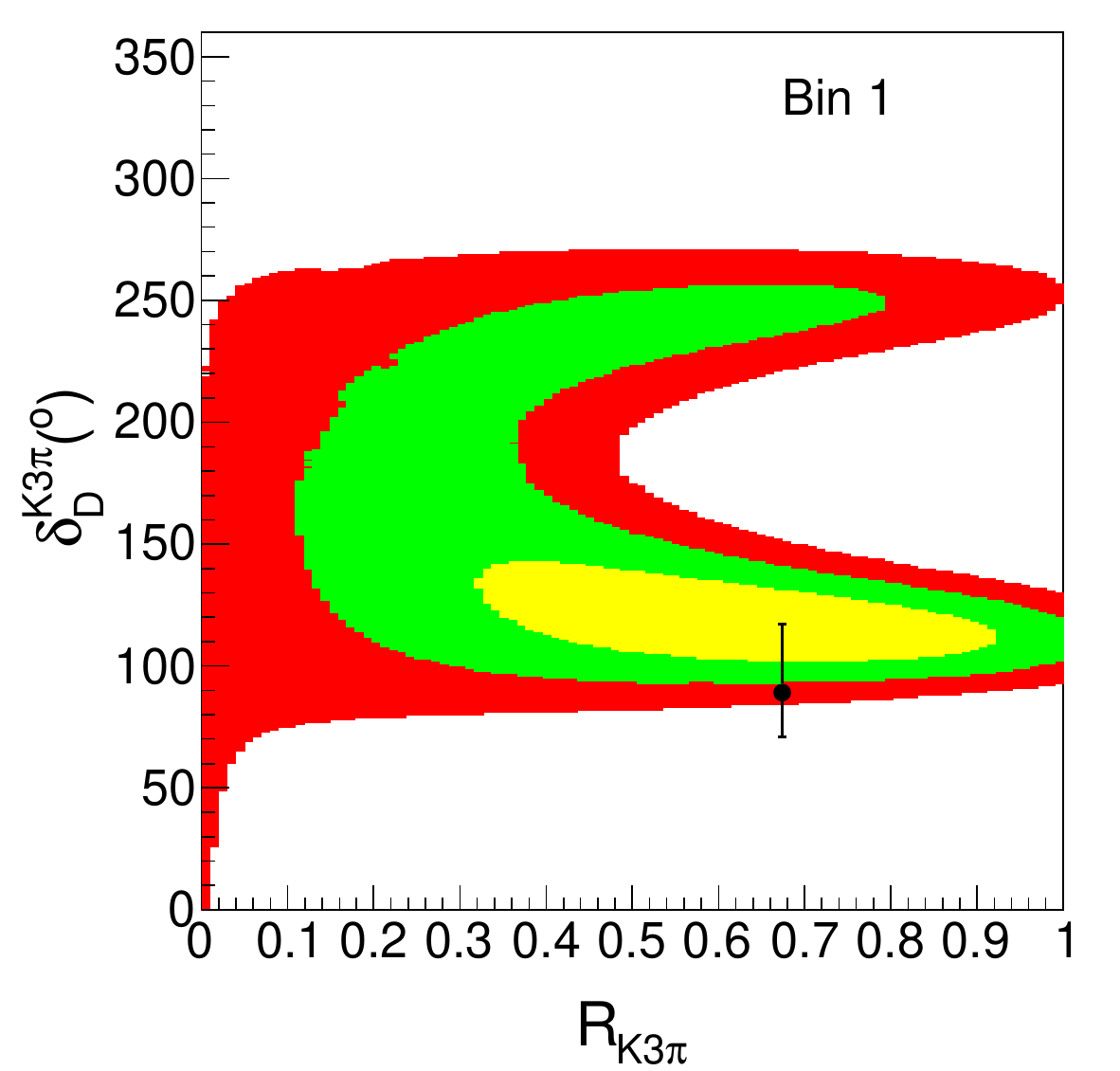}
\includegraphics[width=.45\textwidth]{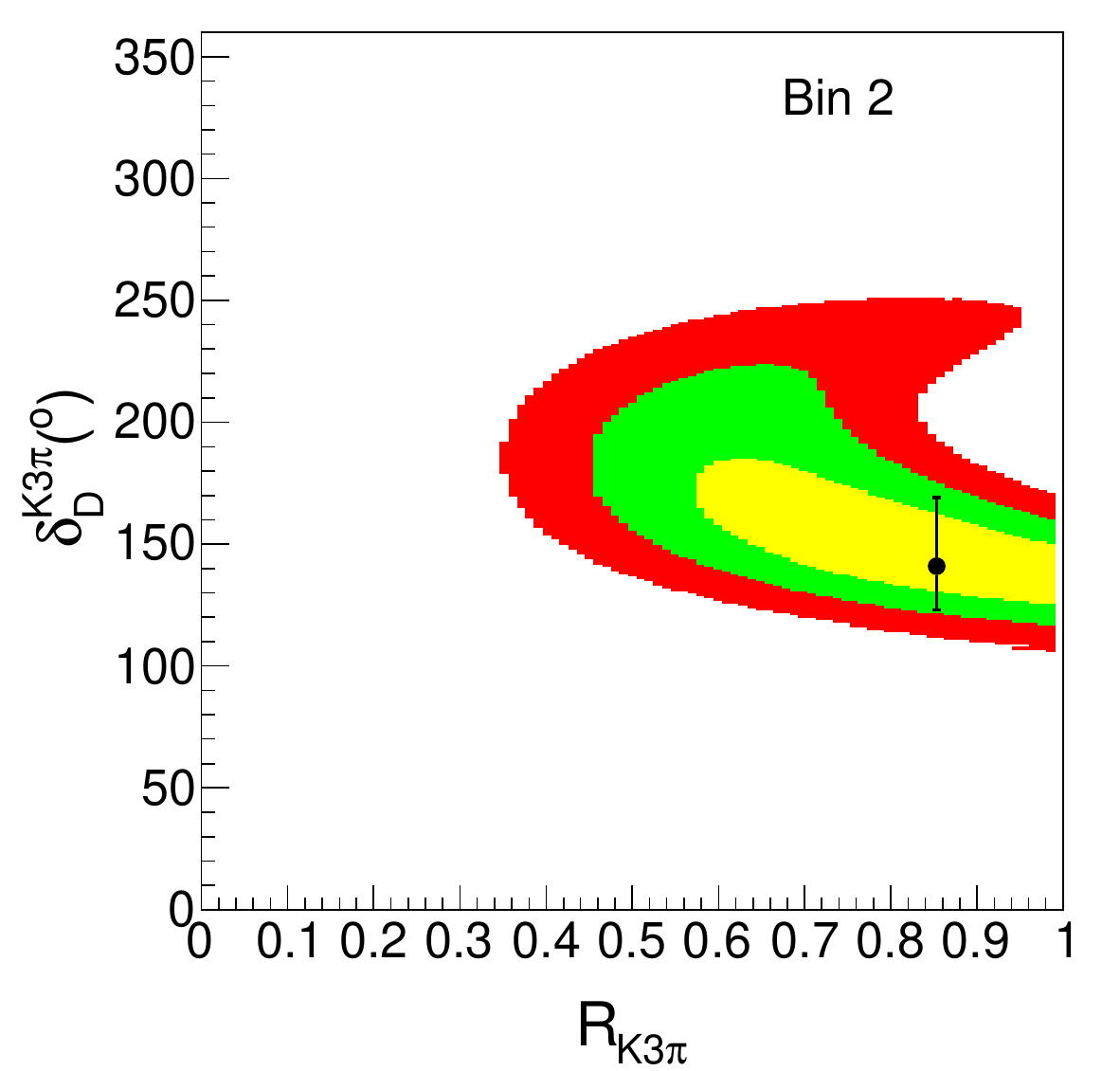}\\
\includegraphics[width=.45\textwidth]{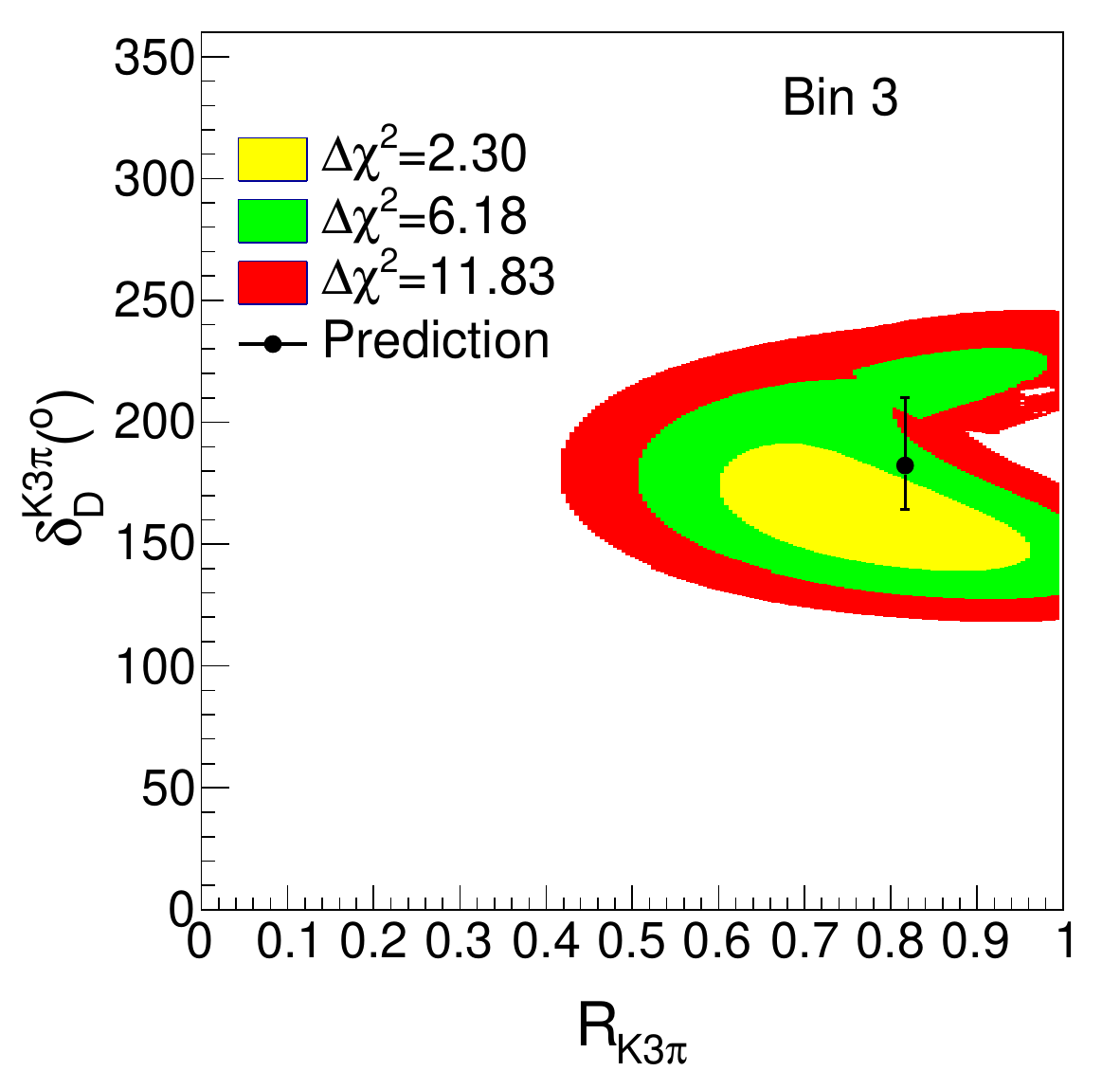}
\includegraphics[width=.45\textwidth]{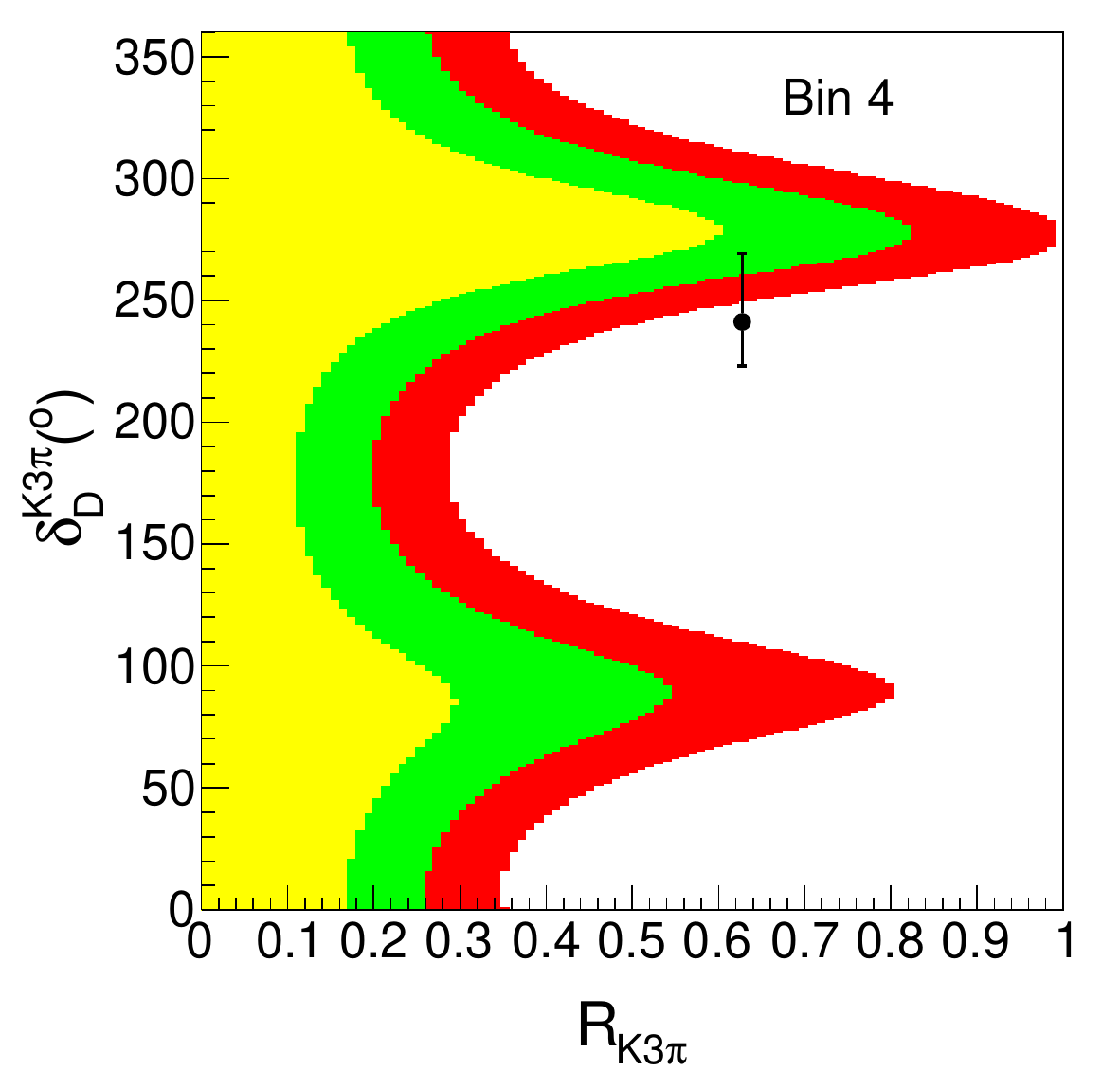}
\caption{
Scans of $\Delta \chi^2$ in the BESIII and CLEO-c binned fit in the ($\Rkthreepi$, $\dkthreepi$) and ($\Rkpipio$, $\dkpipio$) parameter space, showing the $\Delta \chi^2$=2.30, 6.18, 11.83 intervals, for the four bins of the analysis. Also indicated is the prediction from the model, where the global offset in average strong-phase difference comes from the global result. }
\label{fig:binparameter_comb}
\end{figure}

A combined fit is performed to the binned results of BESIII and CLEO-c, again assuming negligible correlations between the two experiments.  The $\chi^2/{\rm n.d.f.}$ is 356/332, and the results are presented in Table~\ref{tab:parameter_binned_comb} and Fig.~\ref{fig:binparameter_comb}.  The correlation matrix is given in Table~\ref{tab:binned_comb}.

\begin{table}[!ht]
\caption{Results for the binned fit to the BESIII and CLEO-c data.}\label{tab:parameter_binned_comb}
\begin{center}
\begin{tabular}{lcccccc}
\toprule
Parameter  & Bin 1 &Bin 2 & Bin 3 & Bin 4 \\
\midrule
$\Rkthreepi$    
&0.66$^{+0.18}_{-0.21}$ 
&0.85$^{+0.14}_{-0.21}$	
&0.78$^{+0.12}_{-0.12}$ &0.25$^{+0.16}_{-0.25}$\\ \\
$\dkthreepi$  & $\left(117^{+14}_{-9}\right)^\circ$ &
$\left(145^{+23}_{-14}\right)^\circ$  &
$\left(160^{+19}_{-20}\right)^\circ$ &
$\left(288^{+15}_{-29}\right)^\circ$\\ \\
$\rkthreepi$ ($\times 10^{-2}$)  
&5.43$\pm$0.10 &5.78$\pm$0.11   &5.76$\pm$0.10 &5.06$\pm$0.12\\ \\
$\Rkpipio$    	&\multicolumn{4}{c}{0.80$\pm$0.04}		\\ \\
$\dkpipio$& \multicolumn{4}{c}{$\left(203 \pm 11 \right)^\circ$} \\ \\
$\rkpipio$ ($\times 10^{-2}$) &\multicolumn{4}{c}{4.49$\pm$0.11} \\
\bottomrule
\end{tabular}
\end{center}
\end{table}

\begin{table}[!htb]
    \begin{center}
    \caption{Correlation matrix for the hadronic parameters from the binned fit to the BESIII and CLEO-c data.}\label{tab:binnedmatrix_comb}
\begin{tabular}{ccrrrrrrr}
\toprule
&	&Bin 1 &\multicolumn{2}{c}{Bin 2} &\multicolumn{2}{c}{Bin 3} &\multicolumn{2}{c}{Bin 4}\\
 			& &$\dkthreepi$  &$\Rkthreepi$  &$\dkthreepi$ &$\Rkthreepi$  &$\dkthreepi$ &$\Rkthreepi$  &$\dkthreepi$\\
\midrule
\multirow{2}{*}{Bin 1} & $\Rkthreepi$	&$-$0.40  &$-$0.53 &0.51 &$-$0.04  &0.17 &$-$0.02  &$-$0.04 \\	
& $\dkthreepi$	&1.00 &0.12  &$-$0.04  &0.02 &0.10  &0.01 &0.03\\
\multirow{2}{*}{Bin 2} &  $\Rkthreepi$	& &1.00 &$-$0.80 &$-$0.22  &$-$0.02 &$-$0.11  &$-$0.12\\	
& $\dkthreepi$	&     &	   &1.00  &0.21 &0.14  &0.10  &0.15\\
\multirow{2}{*}{Bin 3} & $\Rkthreepi$	&  &   &    &1.00 &$-$0.34  &0.11 &0.23 \\	
& $\dkthreepi$	&     &	   & &  &1.00 &$-$0.03  &$-$0.17\\
Bin 4 & $\Rkthreepi$	&  &   &    &     &     & 1.00  &$-$0.63\\	
   	     \bottomrule
\end{tabular}
\label{tab:binned_comb}
\end{center}
\end{table}

\end{document}